\documentclass[a4paper,fleqn,usenatbib]{mnras}

\usepackage{mathptmx}

\usepackage[T1]{fontenc}
\usepackage{ae,aecompl}

\usepackage{graphicx}	
\usepackage{amsmath}	
\usepackage{amssymb}	
\usepackage{caption}
\usepackage{epsfig} 
\usepackage{hyperref} 
\usepackage{subcaption}
\usepackage{multicol}        
\usepackage{bm}		
\usepackage{pdflscape}	
\usepackage{booktabs} 
\usepackage{natbib}
\usepackage{xfrac}
\usepackage[T1]{fontenc}
\usepackage{aecompl}
\usepackage[justification=centering]{caption}

\newcommand{\pc}{pc\,}
\newcommand{\kpc}{kpc}

\newcommand{\Myr}{Myr}

\newcommand{\kmps}{km\,s$^{-1}$\,}
\newcommand{\msol}{M$_{\odot}$}
\newcommand{\msolperyr}{M$_{\odot}$ yr$^{-1}$}

\newcommand{\lpok}{\ensuremath{\log{(P/k)}}\,}
\newcommand{\logOH}{12+$\log{\mathrm{(O/H)}}$\,}

\newcommand{\nII}{$n_{\mathrm{II}}$\,}
\newcommand{\cII}{$c_{\mathrm{II}}$\,}
\newcommand{\mH}{$m_{H}$\,}
\newcommand{\muH}{$\mu_{H}$\,}
\newcommand{\Pamb}{$P_{\mathrm{amb}}$\,}
\newcommand{\Prad}{$P_{\mathrm{rad}}$\,}
\newcommand{\Pgas}{$P_{\mathrm{gas}}$\,}
\newcommand{\ftrap}{$f_{\mathrm{trap}}$\,}
\newcommand{\rstall}{$r_{\mathrm{stall}}$\,}
\newcommand{\htr}{\ion{H}{ii} region}
\newcommand{\htrs}{\ion{H}{ii} regions}

\newcommand{\nii}{[\ion{N}{ii}]$\lambda$6584}
\newcommand{\ha}{H$\alpha$}
\newcommand{\siia}{[\ion{S}{ii}]$\lambda$6717}
\newcommand{\siib}{[\ion{S}{ii}]$\lambda$6730}
\newcommand{\siiab}{[\ion{S}{ii}] $\lambda \lambda$6717,30}

\newcommand{\U}{$\left \langle U \right \rangle$}
\newcommand{\Zsun}{$Z_{\bigodot}$\,}
\defcitealias{Goldbaum:2016aa}{G16}
\defcitealias{Dopita:2016aa}{D16}
\defcitealias{Carton:2017aa}{C17}
\defcitealias{Kewley:2002fk}{KD02}

\usepackage[usenames, dvipsnames]{color}
\usepackage[normalem]{ulem}

\newcommand{\cyan}[1]{{\textcolor{cyan}{#1}}}

\newcommand{\add}[1]{{\cyan{#1}}}

\newcommand{\aref}[1]{\hyperref[#1]{Appendix~\ref{#1}}}

\title[Effects of spatial resolution]{Quantifying the effects of spatial resolution and noise on galaxy metallicity gradients}

\author[A. Acharyya et al.]{Ayan Acharyya$^{1,2}$,\thanks{E-mail: ayan.acharyya@anu.edu.au}
	Mark R. Krumholz$^{1,2}$,
	Christoph Federrath$^{1,2}$,
	Lisa J. Kewley$^{1,2}$,
	\newauthor
	Nathan J. Goldbaum$^{3}$,
	Rob Sharp$^{1}$
	\\
	$1$Research School of Astronomy and Astrophysics, Australian National University, Canberra, ACT~2611, Australia\\
	$^2$ARC Centre of Excellence for All Sky Astrophysics in 3 Dimensions (ASTRO 3D)\\
	$^3$National Center for Supercomputing Applications, University of Illinois at Urbana-Champaign, 1205 W. Clark St., Urbana, IL 61801, USA
}

\date{Accepted XXX. Received YYY; in original form ZZZ}

\pubyear{2018}

\begin{document}
	\label{firstpage}
	\pagerange{\pageref{firstpage}--\pageref{lastpage}}
	\maketitle
	
	\begin{abstract}
		Metallicity gradients are important diagnostics of galaxy evolution, because they record the history of events such as mergers, gas inflow and star-formation. However, the accuracy with which gradients can be measured is limited by spatial resolution and noise, and hence measurements need to be corrected for such effects. We use high resolution ($\sim 20$ pc) simulation of a face-on Milky Way mass galaxy, coupled with photoionisation models, to produce a suite of synthetic high resolution integral field spectroscopy (IFS) datacubes. We then degrade the datacubes, with a range of realistic models for spatial resolution (2 to 16 beams per galaxy scale length) and noise, to investigate and quantify how well the input metallicity gradient can be recovered as a function of resolution and signal-to-noise ratio (SNR) with the intention to compare with modern IFS surveys like MaNGA and SAMI. Given appropriate propagation of uncertainties and pruning of low SNR pixels, we show that a resolution of 3-4 telescope beams per galaxy scale length is sufficient to recover the gradient to $\sim$10-20\% uncertainty. The uncertainty escalates to $\sim$60\% for lower resolution. Inclusion of the low SNR pixels causes the uncertainty in the inferred gradient to deteriorate. Our results can potentially inform future IFS surveys regarding the resolution and SNR required to achieve a desired accuracy in metallicity gradient measurements.
	\end{abstract}
	
	

	\section{Introduction}
	
	\renewcommand{\thefootnote}{*}
	Metal enrichment of the interstellar medium (ISM) is of fundamental importance in understanding the formation and evolution of galaxies. The gas-phase oxygen abundance (henceforth referred to as metallicity) distributions of galaxies contain a wealth of information about their star formation histories \citep{Maiolino:2019aa}. For example, if the stars at smaller galactocentric radii formed earlier than those in the outskirts, as expected in the inside-out galaxy formation scenario, this would lead to a negative metallicity gradient in the disc, because the star-formation going on for a longer time in the central part of the disc would enrich the ISM more than in the outskirts. Interaction with another galaxy can cause gas inflow and mixing of metal enriched gas, thus leading to shallower metallicity gradients than isolated galaxies\citep[e.g.][]{Krabbe:2008aa, Bresolin:2009aa, Rupke:2010aa, Kewley:2010aa, Rich:2012aa, Miralles:2014aa, Vogt:2015aa, Molina:2017aa, Munoz:2018aa}. Such flattening due to mergers has also been observed in numerical models \citep{Mihos:1994aa, Torrey:2012aa, Fu:2013aa, Zinchenko:2015aa, Sillero:2017aa}.
	
	It is well known that most disc galaxies exhibit a radial distribution of oxygen abundance  with a negative gradient \citep[e.g.][]{Garnett:1987aa, Kennicutt:1996aa, Garnett:1997aa, Bresolin:2002aa, Kennicutt:2003aa, Bresolin:2007aa}, and this has been further established by recent studies with large numbers of samples, for both nearby galaxies \citep[e.g.][]{Moustakas:2010aa, Rupke:2010ab, Cecil:2014aa, Sanchez-2014aa, Sanchez-Menguiano:2016aa, Molina:2017aa, Belfiore:2017aa, Poetrodjojo:2018aa, Sanchez-Menguiano:2018aa} and those at redshifts $z \sim 1-2$ \citep[e.g.][]{Molina:2017aa}. The steepness of the gradient appears to be correlated with a number of galaxy properties, e.g., the metallicity gradient - effective radius relation of discs \citep{Diaz:1989aa}, and the correlation between metallicity gradients and galaxy morphology \citep[e.g.][]{Zaritsky:1994aa, Martin:1994aa, Sanchez:2014aa}. \citet{Ho:2015aa} provide a benchmark metallicity gradient of $\sim -0.4$ dex/$R_{\mathrm{25}}$\footnote{Radius of the 25$^{\mathrm{th}}$ magnitude/arcsec$^2$ isophote in B-band.} for local field star-forming galaxies, suggesting co-evolution of the stellar and gas disc in local galaxies. \mbox{\citet{Wuyts:2016aa}}, on the other hand, report a lack of significant correlation between metallicity gradients and other global galaxy properties in a sample of $\sim$180 galaxies. Moreover, it is not well established if the shape of the metallicity profile in galaxies is universal \citep[e.g][]{Ho:2015aa} or depends on the stellar mass \citep[e.g.][]{Belfiore:2017aa, Sanchez-Menguiano:2018aa, Mingozzi:2020aa}. These disagreements, and the strength of the conclusions that one can potentially draw, show that it is important to measure metallicity gradients accurately.

	However, measured metallicity gradients can be affected by the limitations of the instruments used and the quality of the spectroscopic observations, such as spatial resolution and signal-to-noise ratio (SNR). Resolution, noise, and similar effects are important because metallicity diagnostics are based on ratios of emission lines with intensities that have a complex, non-linear relationship with the true, underlying metallicity. Redistribution of flux from inner to outer regions of a galaxy due to beam smearing, can introduce systematic biases in the line ratio profiles and therefore the metallicity gradient. The non-linearity of the metallicity diagnostics exacerbates this bias and the errors in the inferred metallicity. For example, recent studies find that metallicity gradients appear shallower at lower spatial resolution, irrespective of the metallicity indicator used \citep[e.g.][]{Yuan:2013aa, Mast:2014aa, Poetrodjojo:2019aa}.
		
	Previous attempts to quantify this effect have employed either smearing of high spatial resolution observations with coarser point spread functions (PSFs) \citep{Yuan:2013aa} or application of PSF convolution to model discs \citep{Carton:2017aa}. The former method is ideal, in that one starts from real data and then degrades it. However, it is severly limited by data availability: in order to carry out a systematic study of the likely errors as a function of intrinsic metallicity gradient, resolution, and SNR, one would need to start from a catalogue of measurements with both very high spatial resolution and very high SNR, covering a range of intrinsic metallicity gradients. At present no such catalogue is available. The latter approach has only been employed in ``toy-models" of disc galaxies so far, which assume a smooth radial variation of the star formation rate (SFR) surface density and consequently of the nebular emission line profiles. Neither approach allows one to consider effects like beam-smearing between neighbouring regions of very different star formation rate densities or mean stellar population ages, or mixing together light from {\htr}s with very different densities or ionisation parameters. A complete study of these effects using physically realistic disc models along with metallicity diagnostics commonly used in observations is still lacking. The aim of this paper is to provide quantitative results from such a study.
	
	We produce synthetic integral field spectroscopy (IFS) observations from high-resolution simulations of disc galaxies that resolve the phase structure of the interstellar medium and include realistic stellar feedback. We then perform spectral line fitting and metallicity measurements in order to understand how well we can reproduce the intrinsic metallicity gradient. Simulations have the unique advantage that we have complete control over the input physics and hence know the ``true" values of the physical properties. Converting these simulations to mock observations and treating the synthetic data cubes similar to an observed data cube allows us to directly compare theory with observations, and to explore the effects of observational limitations such as finite SNR and resolution. Such comparisons, by treating simulations and observations equally, are essential to understand the effect of instrument properties and observational parameters on metallicity gradients.
	
	This paper is organised as follows. In \autoref{sec:models} we briefly describe the simulations and the steps involved in translating them to synthetic IFU data cubes, including the sub-grid modelling of H~\textsc{ii} regions. \autoref{sec:observables} details the analysis pipeline we use to derive metallicities from the synthetic data. We present results in \autoref{sec:results}, followed by discussion of their implications in \autoref{sec:disc}. Finally, we summarise our work and draw conclusions in \autoref{sec:sum}.
	
	We use a solar oxygen abundance \logOH = 8.77 based on the local galactic concordance scale \citep{Dopita:2016aa} throughout the paper unless otherwise stated. We assume a standard flat $\Lambda$ cold dark matter cosmology with $H_0$ = 70 km s$^{-1}$ Mpc$^{-'1}$ and matter density $\Omega_M$ = 0.27 wherever required for our calculations.
	
	\section{Methods for production of synthetic observations}\label{sec:models}
	
	\begin{figure*}
		\centering
		\includegraphics[scale=0.384, trim=0.0cm 0.0cm 0.0cm 0.0cm,clip=true,angle=0]{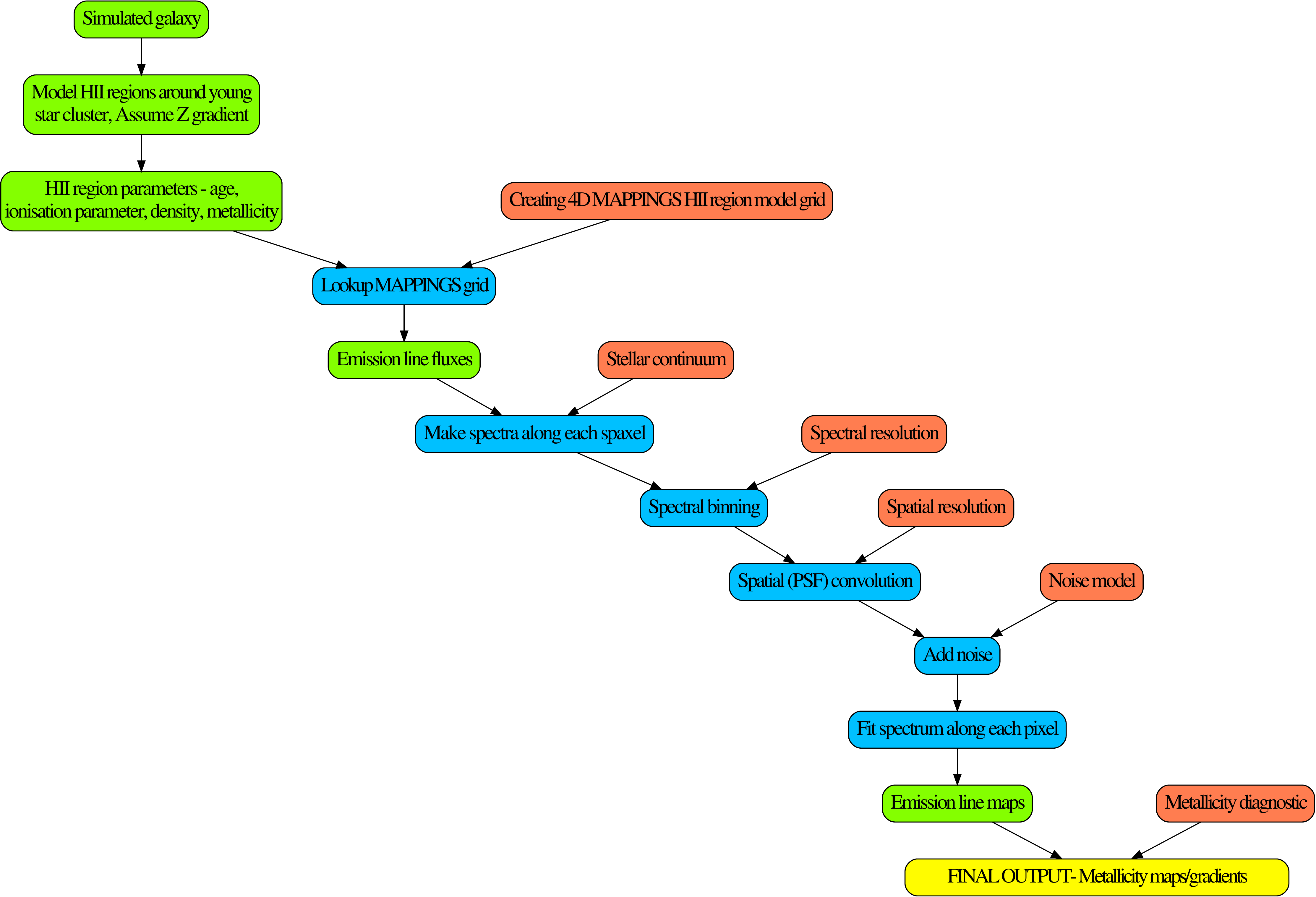}
		\caption{Flowchart for the methodology followed in this paper. The red blocks denote inputs and assumptions, blue blocks show the key steps involved, and the intermediate outputs are in the green blocks.
		}
		\label{fig:flowchart}
	\end{figure*}

	\autoref{fig:flowchart} summarises the procedure we use to produce our synthetic data cubes and then extract metallicity information from them. We explain the data generation method in this section and the analysis method in the following section. To produce our synthetic data, we start with the simulations (\autoref{sec:enzo}) and then model H~\textsc{ii} regions around each young star particle (\autoref{sec:H2R}). We then construct a MAPPINGS photoionisation {\htr} model grid (\autoref{sec:grid}) and use it to assign nebular line luminosities to each pixel of our simulated galaxy. To produce the full position-position-velocity (PPV) cube, we add the nebular fluxes to the stellar continuum derived from Starburst99 (\autoref{sec:cube}). Thereafter, we convolve the datacube with a smoothing kernel to simulate finite observational resolution (\autoref{sec:convolution}) and spectrally bin the datacube, followed by addition of noise (\autoref{sec:noise}) to simulate sky emission combined with instrumental effects.
	
	\subsection{Galaxy simulations}\label{sec:enzo}
\begin{figure*}
	\centerline{
		\def\arraystretch{0.1}
		\setlength{\tabcolsep}{0.7pt}
		\begin{tabular}{lll}
			\includegraphics[trim=0.0cm 0.0cm 5.9cm 0.0cm, clip, height=0.25\textheight]{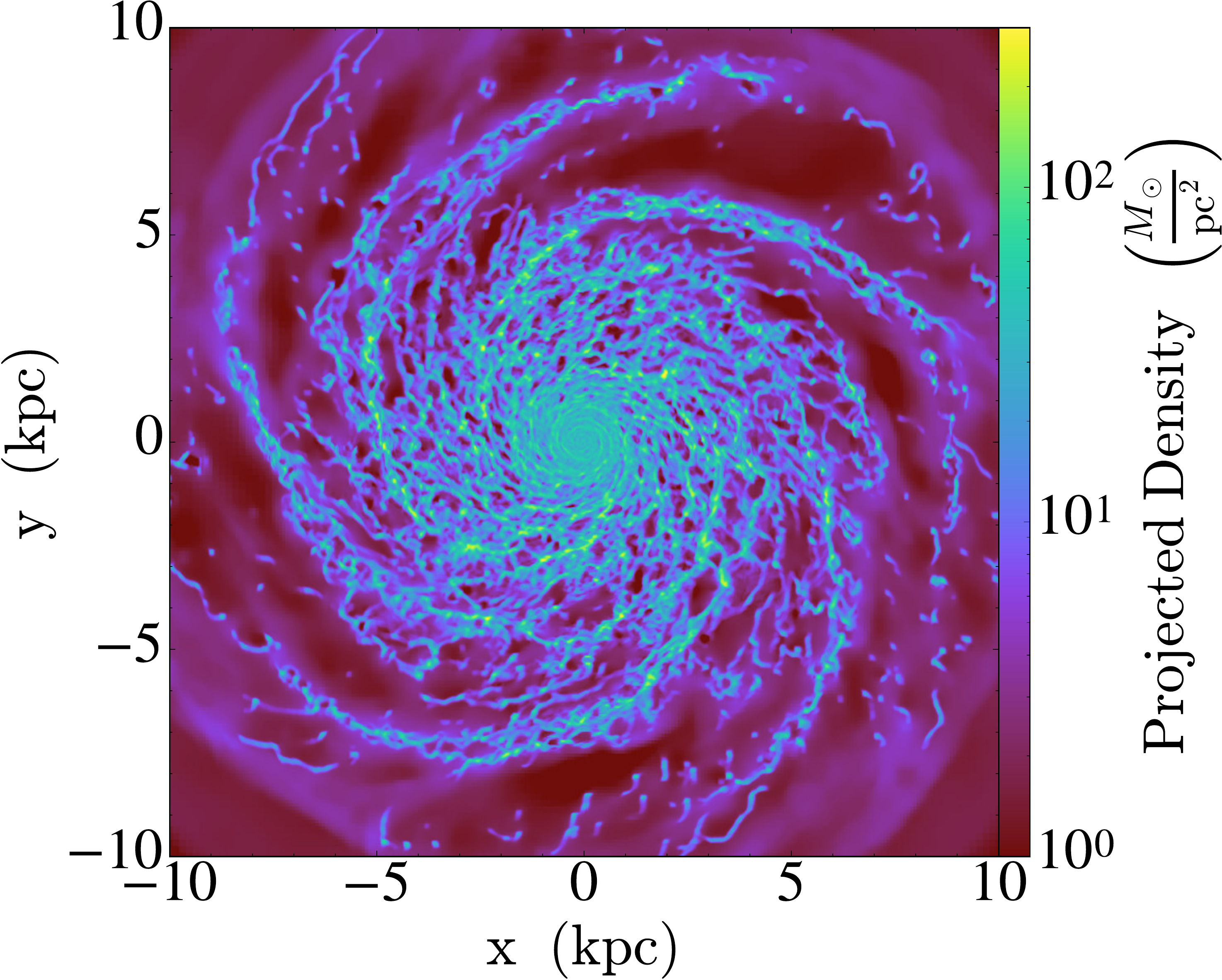} &
			\includegraphics[trim=4.1cm 0.0cm 0.0cm 0.0cm, clip, height=0.25\textheight]{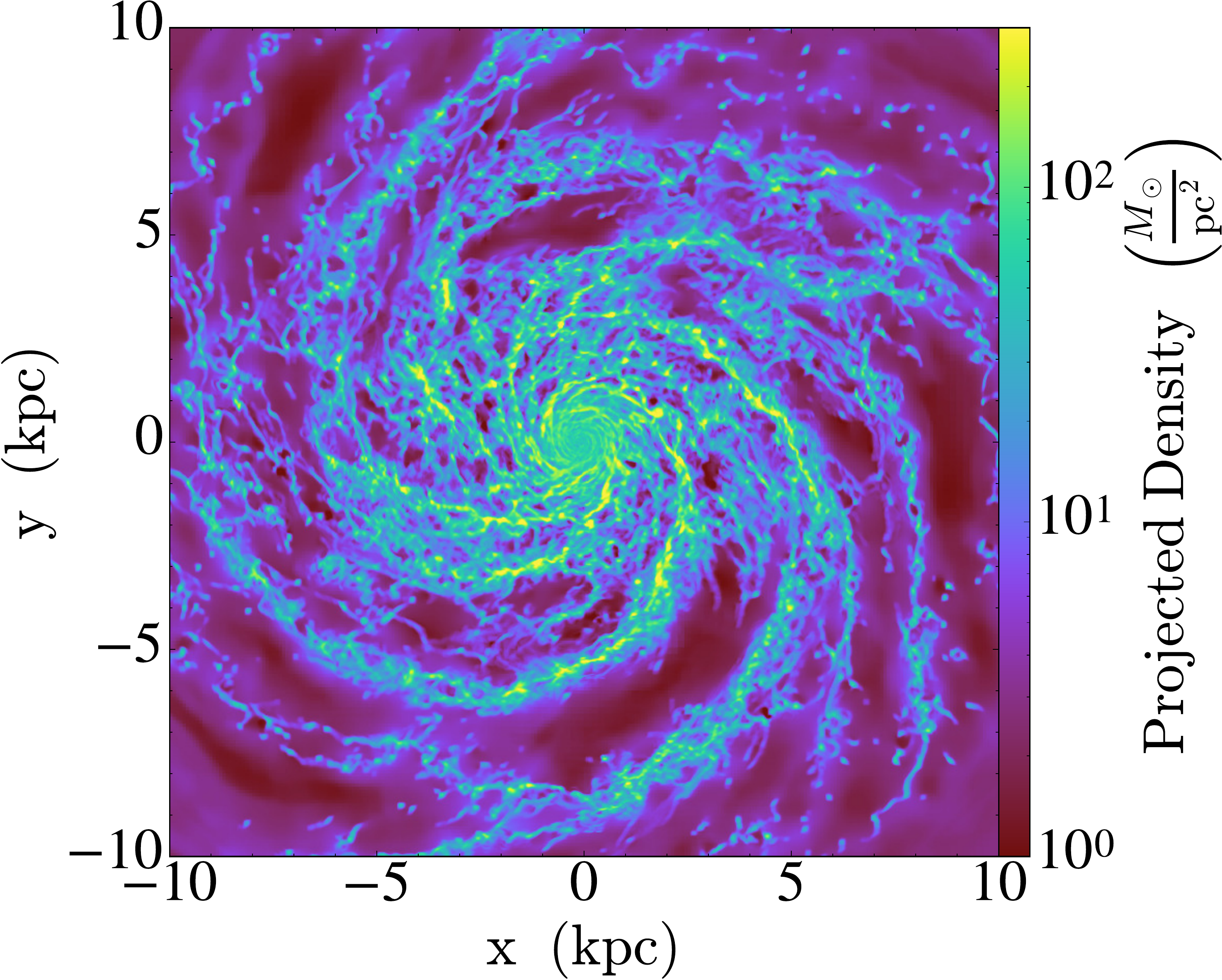} &
	\end{tabular}}
	\caption{Comparison of the projected gas density of the low gas fraction ($f_g = 0.1$; left) and fiducial gas fraction ($f_g = 0.2$; right) \citetalias{Goldbaum:2016aa} simulations at the final times in the simulations.
	} 
	\label{fig:enzo}
	\end{figure*}

	We use the isolated Milky Way (MW) type disc galaxy simulations of \citet[][G16 hereafter]{Goldbaum:2016aa}, which are publicly available, in order to generate synthetic observables. The halo mass of 10$^{12}${\msol} and stellar mass of 10$^{10}${\msol} have been chosen to loosely resemble that of the MW. The bulge-to-disc ratio is 0.1 \citep{Kim:2013aa}, slightly less bulge-dominated than the MW \citep[$\sim 0.3$][]{Bland-Hawthorn:2016aa}, but not very different. The simulations include hydrodynamics, self-gravity, star formation, photoionisation and supernova feedback, a live stellar disc (i.e., rather than using a fixed stellar potential as is common in some galaxy simulations, in \citetalias{Goldbaum:2016aa} the stellar potential is generated by stars that move self-consistently and thus are free to develop spiral structure, scatter off structures in the gas, etc.), and stellar plus dark matter halo. Their high spatial resolution (20 \pc) means that even if we are unable to resolve individual \ion{H}{ii} regions, we have only few ($\sim$ 10) {\htr}s within each cell. The simulations include photoionisation heating using a Str\"{o}mgren volume approximation, but do not include full ionising radiative transfer. Consequently, we need to further use photoionisation models to appropriately model the expansion of the {\htrs} (see \autoref{sec:H2R}) and the emission line fluxes from them. \citetalias{Goldbaum:2016aa} carry out simulations with three different gas fractions $f_g$ (defined as ratio of gas to total baryonic mass within the disc): low gas fraction (LGF, $f_g$ = 0.1), fiducial ($f_g$ = 0.2), and high gas fraction (HGF, $f_g$ = 0.4). For each case we use the final time slice from the simulations. We show a comparison between the projected gas densities of the simulations with different gas fractions in \autoref{fig:enzo} to demonstrate the higher gas content in the fiducial gas fraction simulation (right) than the low gas fraction scenario (left). The current MW gas fraction ($\sim$15\%) is between the LGF and fiducial cases of the simulated galaxy. We compare results obtained from the fiducial and LGF versions of the \citetalias{Goldbaum:2016aa} simulations in \aref{sec:ap}; in the main text we will focus on the fiducial case as that has a star formation rate ($\sim$2 {\msolperyr}) comparable to that of the Milky Way. Our tests show that the results for the LGF case are qualitatively identical. 
	
	The \citetalias{Goldbaum:2016aa} simulations and their successors using similar methodology \citep{Fujimoto:2018aa, Fujimoto:2019aa} are particularly well-suited for this experiment because they have been subjected to extensive observational comparisons and show very good agreement with all diagnostics on $\gtrsim 100$ pc and larger scales. In particular, the simulations show excellent agreement with the observed mass spectrum and structural properties of giant molecular clouds, and with the spatially resolved Kennicutt-Schmidt relation \citep[see][for details]{Fujimoto:2019aa}; thus the spatial distribution of star formation within them provides a good approximation to the MW.
	
	\subsection{Modelling HII regions}\label{sec:H2R}
	The first step in producing the synthetic observations is to identify the young ($\leq$ 5 \Myr) star particles in the simulations. We only include the young star particles in our calculations because the ionising luminosities of stellar populations drop rapidly at ages above 5 {\Myr}; for the stellar population models we adopt in this work (see below), $>97\%$ of the total ionising flux is emitted in the first 5 Myr. 
	
	Star particles that are in close proximity can potentially lead to merging of {\htrs} resulting in a super-bubble being jointly driven by the host stellar associations. In order to treat the overlapping {\htrs} we merge star particles in every 40 {\pc}$^3$\footnote{Although the resolution of the \citetalias{Goldbaum:2016aa} simulations is 20 {\pc} at the highest refinement level, we chose 40 {\pc} as the base spatial resolution for this study for computational ease.} cell by summing their luminosities and masses. We assign the luminosity-weighted mean of position, velocity and age of the individual star particles to the merged particle. Henceforth, any reference to the star particles denotes the merged star particle in every cell. 
	
\begin{figure*}
	\centering
	\includegraphics[trim=0.0cm 0.0cm 0.0cm 0.0cm, clip, height=0.25\textheight]{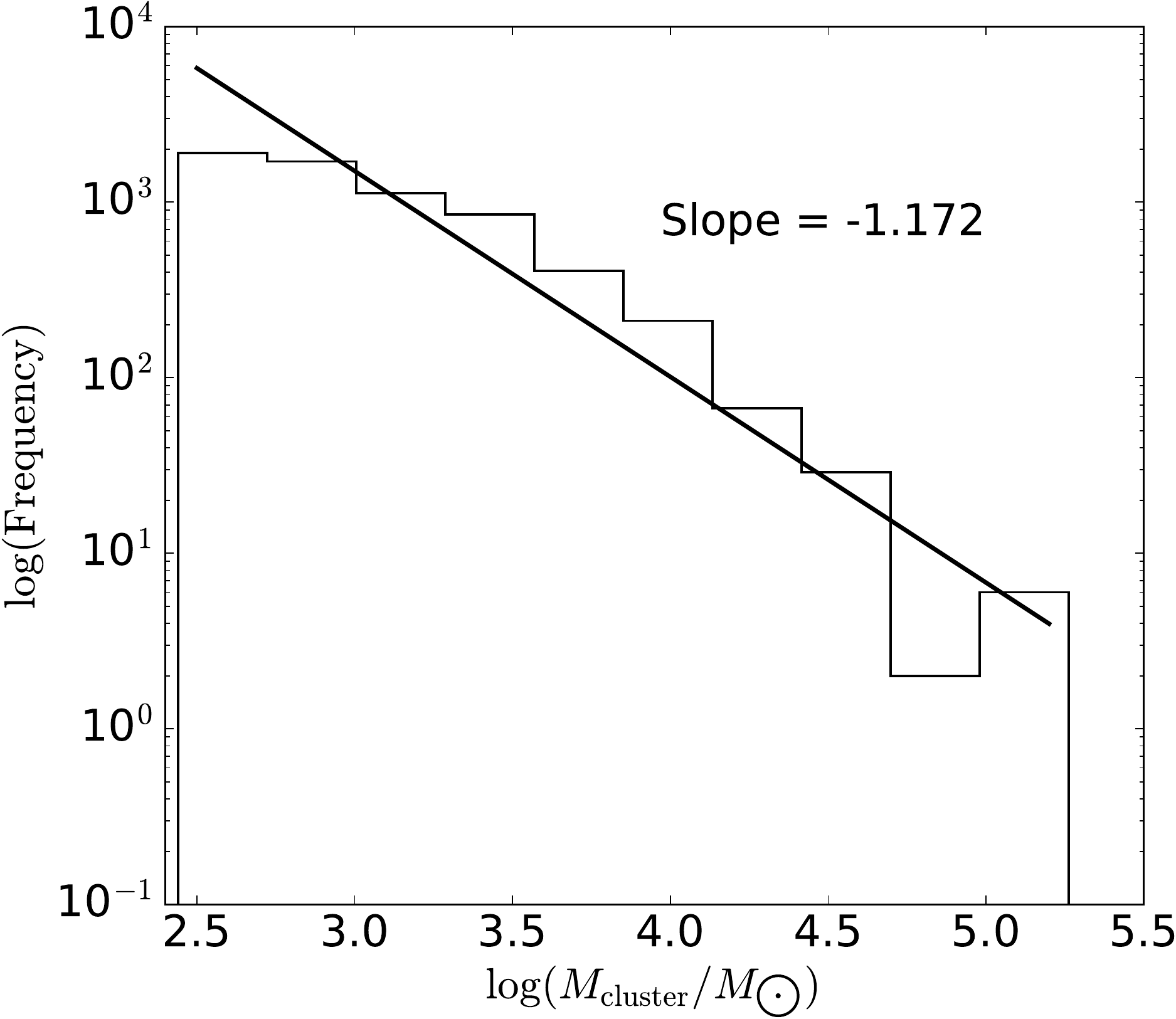}
	\hspace{10 pt}
	\includegraphics[trim=0.0cm 0.0cm 0.0cm 0.0cm, clip, height=0.25\textheight]{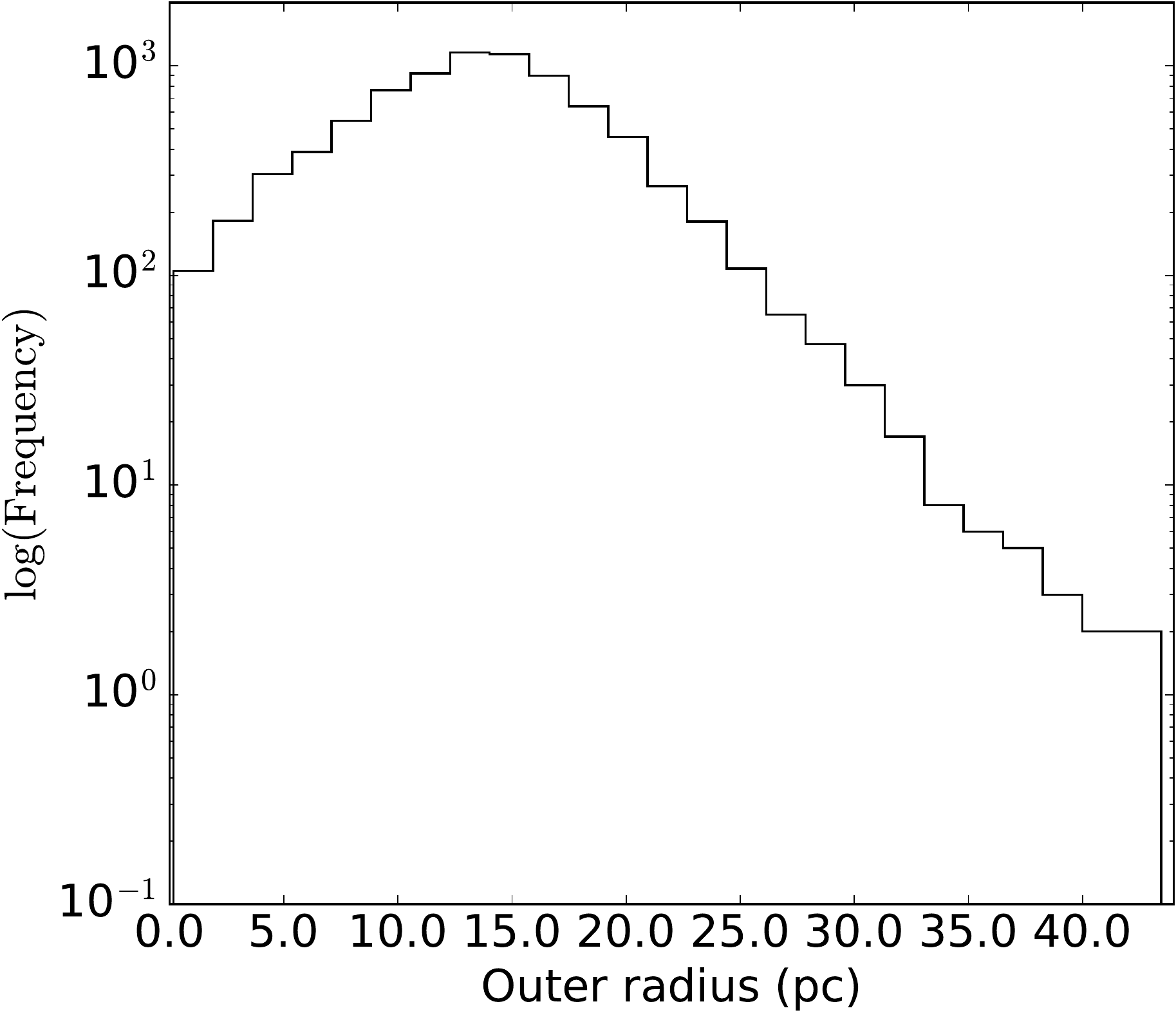}

	\caption{Distribution functions for star cluster masses (\textit{left}) and outer radii (\textit{right}) of {\htrs} for the fiducial gas fraction ($f_g = 0.2$) \citetalias{Goldbaum:2016aa} simulations.
	}
	\label{fig:histograms}
	\end{figure*}

	To verify that our procedure results in a reasonable distribution of cluster sizes, in \autoref{fig:histograms} we show the resulting cluster mass function (CMF), where a ``cluster'' here refers to one of our merged particles. The CMF is close to a power-law distribution $dN/d\log M \propto M^{-1}$ from $300$ $\mathrm{M}_\odot$ (the mass of individual star particles in the \citetalias{Goldbaum:2016aa} simulation) to $\approx 10^5$ $\mathrm{M}_\odot$. This is in good agreement with the CMFs frequently observed in spiral galaxies -- see the review by \citet{Krumholz:2019aa} for a summary of recent measurements. However, to ensure that our results are not overly-dependent on the merging procedure, in \aref{sec:mergeHII} we demonstrate that an alternative approach -- not merging stars at all (so that each cluster is $\approx 300$ $M_\odot$) produces nearly identical results for the inferred metallicity gradient compared to when the stars are merged. For this reason, in the remainder of the main text we will focus only on the case of merging within 40 {\pc}$^3$ cells, yielding the CMF shown in \autoref{fig:histograms}. Given our list of merged clusters, we follow the method of \cite{Verdolini:2013aa} to compute for each star particle a series of quantities as follows:

	\paragraph*{Bolometric and ionising luminosity:}
	We run fixed mass ($M =$ 10$^6$ {\msol}) Starburst99 \citep{Leitherer:1999aa} models with a \citet{Kroupa:2001vn} Initial Mass Function (IMF) to obtain the bolometric and ionising luminosity for star clusters aged 0 to 5 \Myr, with a uniform linear spacing of 0.1 \Myr. The IMF uses exponents 1.3 and 2.3 for mass ranges 0.1\,{\msol} $\leq M \leq$ 0.5\,{\msol}  and  0.5\,{\msol} $\leq M \leq$ 120\,{\msol} respectively. We then assign a luminosity to each star particle in the simulation by linearly interpolating the Starburst99 outputs to the star particle age, and scaling linearly to the star particle mass. We note that this effect implicitly neglects stochastic sampling of the IMF, which can be significant in clusters with masses $\lesssim 10^{3.5}$ {\msol} \citep[e.g.,][]{da-Silva:2012aa}. However, given our CMF (\autoref{fig:histograms}), such clusters contribute somewhat less than half of the total ionising photon budget. Consequently, neglecting stochastic sampling has relatively modest effects as long as a typical observational beam contains enough total young stars that we expect it to contain a significant number of clusters with masses $\gtrsim 10^{3.5}$ $M_\odot$. This is the case for the resolution of all surveys to date with the exception of MUSE/PHANGS, and thus we will not worry about this complication further here (Arora et al., in preparation).
	
	\paragraph*{Stall radius:}
	We define \rstall as the maximum radius up to which an {\htr} (assumed spherical, isotropic and constant density) can expand, until its internal pressure (sum of gas pressure \Pgas and radiation pressure \Prad) reaches a state of equilibrium with the ambient pressure \Pamb, taken to be the pressure in the computational cell that contains the star particle. The stall radius is defined implicitly by the set of equations \citep{Krumholz:2009aa, Verdolini:2013aa}
	\begin{eqnarray}
	\label{eq:Pamb}
	P_{\rm amb} = P_{\rm gas} + P_{\rm rad} \\
	P_{\rm rad} = \frac{\psi S \epsilon_0 f_{\rm trap}}{4 \pi r_{\rm stall}^2 c } \\
	P_{\rm gas} = \mu_H n_{\rm II} m_H c_{\rm II}^2 
	\end{eqnarray}
	where $\psi = L/S\epsilon_0$ is the ratio of the star cluster's total bolometric luminosity to ionising power, $S$ is the total number of ionising photons injected per unit time and $\epsilon_0$ = 13.6~eV. \ftrap is the factor by which radiation force is enhanced by trapping the energy within the expanding shell. The mean mass per H nucleus \muH is given by $ 1 + Y/X + Z/X \approx 1.4$, for Solar composition ($X=0.73$, $Y=0.25$, $Z=0.02$) where $X$, $Y$, and $Z$ are the mass fractions of hydrogen, helium and all other elements, respectively. 
	The mass of a hydrogen atom, the sound speed in the {\htr} and the speed of light are denoted by \mH, \cII and $c$, respectively. 
	To fully specify the system, we also need to know the number density of H nuclei \nII in the {\htr}, which we derive by assuming ionisation balance:
	\begin{eqnarray}
	\phi S = \frac{4}{3}\pi r_{\rm stall}^3 \alpha_B n_{\rm II} n_e \\
	n_e = \left(1+ \frac{Y}{4X}\right)n_{\rm II} \\
	\Rightarrow n_{\rm II} = \sqrt{\frac{\phi S}{\frac{4}{3}\pi r_{\rm stall}^3 \alpha_B (1+\frac{Y}{4X})}}
	\label{eq:nII}
	\end{eqnarray}
	where $\phi$ is the fraction of ionising radiation not being absorbed by the dust, and $\alpha_B = 2.59\times10^{-13}$ cm$^{-3}$s$^{-1}$ is the recombination coefficient (assuming electron temperature of 10$^4$ K). Substituting the above expressions into \autoref{eq:Pamb} gives
	\begin{eqnarray}
	\label{eq:rstall_eqn}
	P_{\rm amb} = \frac{a}{r_{\rm stall}^\frac{3}{2}} + \frac{b}{r_{\rm stall}^2}
	\end{eqnarray}
	where $a = 0.65 m_H c_{\rm II}^2 \sqrt{\phi S}$ and $b =\psi S \epsilon_0 f_{\rm trap}/4\pi c$ are constants. We solve \autoref{eq:rstall_eqn} using a 1D Newton's method (initial guess provided by the analytically-obtained solution to \Pamb = \Pgas) to get the values of \rstall for each {\htr} in the simulation. We take $\psi = \mbox{3.2}$, $\phi = \mbox{0.73}$, and $f_{\rm trap} = \mbox{2}$, following \cite{Verdolini:2013aa}.
	
	\paragraph*{Instantaneous radius and density:}
	To compute the instantaneous radius $r$ of our {\htrs}, we use the analytic approximation provided by \cite{Krumholz:2009aa}, which interpolates between the similarity solutions describing the expansion of gas pressure driven ($r_{\rm gas}$) or radiation pressure driven ($r_{\rm rad}$) {\htrs}.
	\begin{eqnarray}
	r = r_{\rm ch} \left(x_{\rm rad}^{(7-\kappa_\rho)/2}+x_{\rm gas}^{(7-\kappa_\rho)/2}\right)^{2/(7-\kappa_\rho)} \\
	x_{\rm rad} = \left(\frac{4-\kappa_\rho}{2}\tau^2\right)^{1/(4-\kappa_\rho)} \\
	x_{\rm gas} = \left(\frac{(7-2\kappa_\rho)^2}{4(9-2\kappa_\rho)}\tau^2\right)^{2/(7-2\kappa_\rho)}
	\end{eqnarray}
	where $\tau = \frac{t}{t_{\mathrm{ch}}}$, $\kappa_{\rho}$ is the index of the power-law density distribution of the gas into which the {\htr} is expanding (for which we adopt $\kappa_\rho = 0$ in this work), and $r_{\mathrm{ch}}$ and $t_{\mathrm{ch}}$ are the characteristic spatial and time scales as described by equations (4) and (9) of \cite{Krumholz:2009aa}, given by (for the case $\kappa_\rho = 0$)
	\begin{eqnarray}
	r_{\rm ch} = \frac{\alpha_B}{12\pi \phi}\left(\frac{\epsilon_0}{\kappa_B T_{\rm II}}\right)^2 f_{\rm trap}^2\frac{\psi^2S}{c^2}\\
	t_{\rm ch} = \sqrt{\frac{4\pi}{3}\frac{\rho_0 c}{f_{\rm trap}L}r_{\rm ch}^{4}}.
	\end{eqnarray}	
	Here $\rho_0$ is the ambient density of the material into which the {\htr} is expanding and $r_{\rm ch}$ is the characteristic radius where the radiation pressure and gas pressure are equal, with the gas pressure being dominant at $r>r_{\mathrm{ch}}$ and radiation pressure dominating the $r<r_{\mathrm{ch}}$ regime, and $t_{\rm ch} = r_{\rm ch}/c_{\rm II}$, where $c_{\rm II}$ is the ionised gas sound speed. The \textit{lesser} of $r$ and \rstall is assigned as the outer radius $r_{\mathrm{out}}$ of the \ion{H}{ii} region. We provide the distribution of radii ($r_{\mathrm{out}}$) for all the {\htrs} in the simulation in \autoref{fig:histograms}. We show in \autoref{sec:size} that the relationship between pressure and size for our simulated HII regions agrees within the uncertainties with that observed by Galactic {\htrs} by \citet{Tremblin:2014aa}. Once the radius is known, we can immediately compute the \htr~density from ionisation balance, following \autoref{eq:nII} with $r_{\rm stall}$ replaced by $r_{\rm out}$: $n_{\rm II} = \left[3\phi S/4\pi r_{\rm out}^3 \alpha_B (1 + Y/4X)\right]^{1/2}$. At this stage, we have the age, density, and outer radius of every young {\htr} in the simulation. 

	It should be noted that our calculation of the radius and density (and, below, the volume-averaged ionisation parameter) implicitly assumes that H~\textsc{ii} regions have constant internal density. We emphasise that we make this approximation only for the purposes of computing the dynamics, and that our calculation of line emission uses the more common isobaric approximation. The primary effect of assuming constant density is to somewhat lower the total recombination rate compared to that which would occur in regions with internal density variations, which in turn leads to a slight overestimate of the expansion rate. However, this is only a tens of percent level effect \citep{Bisbas:2015aa}.
	
	\subsection{Comparison of size distribution of HII regions with literature}\label{sec:size}
\begin{figure}
		\centering
		\includegraphics[width=1.0\linewidth]{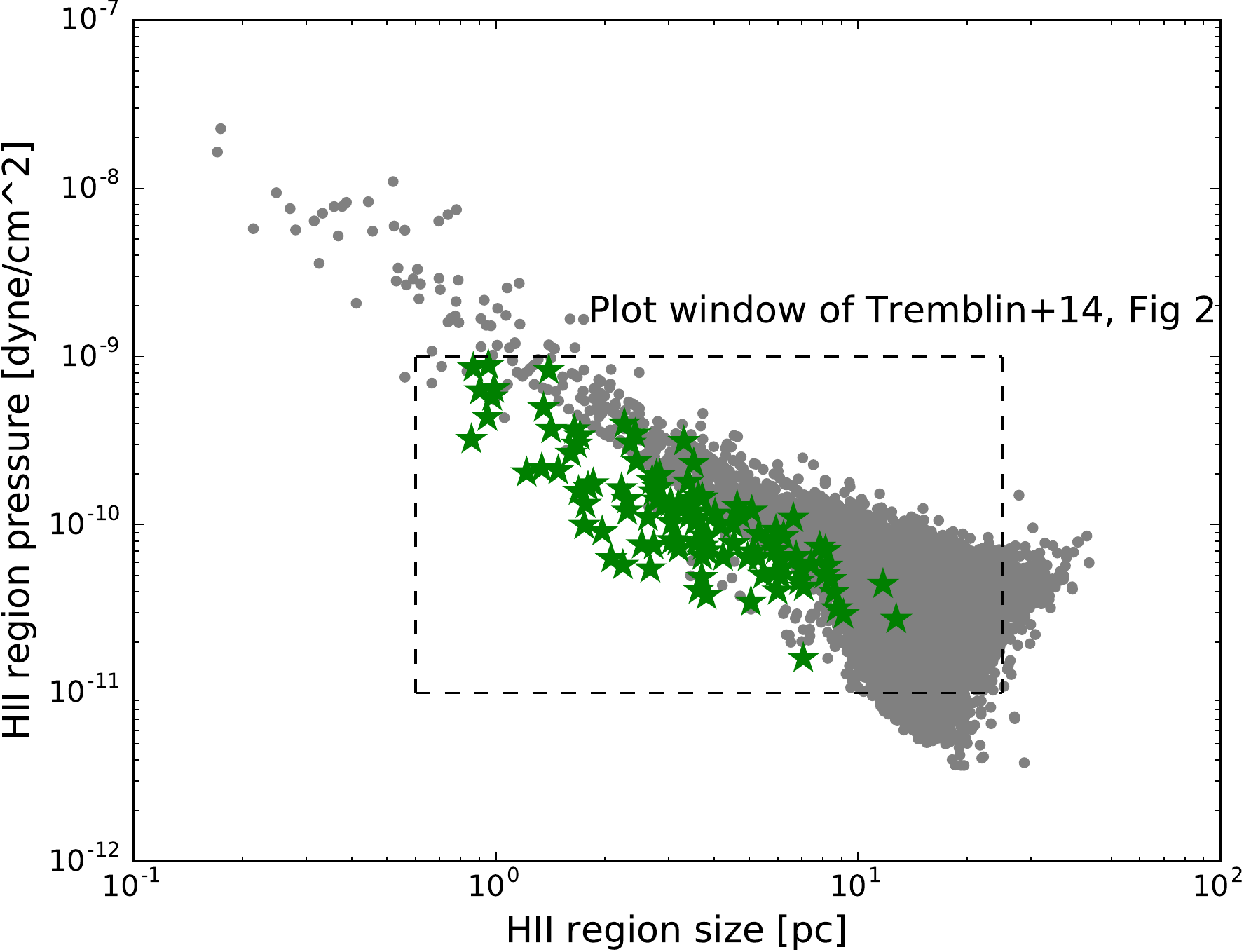}
		\caption{Pressure-radius parameter space of our model {\htrs} (\textit{gray circles}) compared with observed Galatic {\htrs} from \citet{Tremblin:2014aa} (\textit{green stars}). The models shown here are based on a radially varying $\Omega$, which, as demonstrated in Section 1.1 of the Supplementary Online Material, does not produce any qualitative difference from our fiducial $\Omega$=0.5 models.
		}
		\label{fig:size_dist}
	\end{figure}

	It is difficult to make a direct comparison of our model {\htr} sizes with that of observed {\htrs}, particularly for extra-galactic studies, due to spatial resolution limitations of the observations. Numerous extra-galactic studies report a {\htr} size distribution peaking between $\sim$50 - 100 pc, which is generally the resolution limit at such distances. \citet{Tremblin:2014aa} however study Galactic {\htrs} through radio recombination lines, which enables them to probe smaller length scales. Completeness remains an issue, since they have finite sensitivity, but we can minimise the problem by not comparing the absolute size distributions (which are sensitive to completeness) of our simulated HII regions to the data, but instead compare the pressure-radius relation. This should be less sensitive to completeness, but nonetheless captures the physics of HII region expansion. We make this comparison in \autoref{fig:size_dist}, which shows that a majority of our models lie reasonably well within the parameter space spanned by the observations. However, we emphasise that a direct comparison of size distribution between the two is difficult to carry out without accounting for sensitivity and completeness issues in detail - which is beyond the scope of this paper.
	
	\subsection{Constructing the 4D MAPPINGS grid}\label{sec:grid} 
		
	Once we have the physical size and density for each {\htr} in the simulation, we need to compute a grid of {\htr} models in order to be able to read off the nebular emission line fluxes. The strength of emission lines originating from a {\htr} depends on the shape of its driving radiation field, density structure, ionisation structure and chemical composition. The driving radiation field in turn, depends on the age and mass distribution of the stars in the stellar association that produces the {\htr}. Hence, we need to compute a four dimensional grid of age, metallicity $Z$, volume-averaged ionisation parameter \U~and density $n_{\mathrm{II}}$. We know the age of the star cluster from the simulation and the density from our modelling (\autoref{sec:H2R}). Next, we need to determine the ionisation parameter and metallicity of each {\htr}.
	\paragraph*{Ionisation parameter:}
	The ionisation parameter $U$ is defined as ratio of photon to hydrogen number density. In order to determine the volume-averaged ionisation parameter \U, we introduce a parameter $\Omega$ \citep[following][]{Yeh:2012aa} as a measure of the relative volume occupied by the stellar wind cavity to the ionised bubble:
	\begin{eqnarray}
	\Omega = \frac{r_i^3}{r_{\rm out}^3 - r_i^3}
	\end{eqnarray}
	where $r_i$ is the inner radius of the ionised shell. A high value of $\Omega$ corresponds to an {\htr} that is wind-dominated and low $\Omega$ implies domination by either radiation or gas pressure. Throughout the main text of the paper, we assume a fiducial value of $\Omega=0.5$, as suggested by \citet{Yeh:2012aa} for radiation-confined dust-limited {\htr} shells (their Figure 1). \cite{Yeh:2012aa} also point out that {\htrs} within the central 500 pc of M83, NGC 3256, NGC 253 typically have $\Omega\sim 0.5$ (their Figure 8 and 9). We discuss the implications of this choice in \autoref{sec:omega}. Assuming the gas density \nII to be constant throughout, the total number of ionising photons per unit time passing through a shell at a distance $r$ is given by \citep{Draine:2011}
	\begin{eqnarray}
	Q(r) = Q_0  \left[1-\left(\frac{r}{r_s}\right)^3 + \left(\frac{r_i}{r_s}\right)^3\right]
	\label{eq:Om}
	\end{eqnarray}
	where $Q_0$ is the rate of emission of ionising photons from the driving star cluster and $r_s$ is the Stromgen radius given  by:
	\begin{eqnarray}
	r_s = \left(\frac{3Q_0}{4\pi \alpha_B n_{\rm II}^2}\right)^{1/3}
	\label{eq:rs}
	\end{eqnarray}
	The local ionisation parameter $U$ at a distance $r$ from the stellar source is
	\begin{eqnarray}
	U = \frac{Q(r)}{4\pi r^2 c n_{\rm II}}
	\label{eq:U}
	\end{eqnarray}
	which, when volume-averaged over the entire {\htr}, relates to \nII, $Q_0$ and $\Omega$ as
	\begin{eqnarray}
	\left\langle U \right\rangle = \left(\frac{81\alpha_B^2n_{\rm II}Q_0}{256\pi c^3}\right)^{1/3}\left((1+\Omega)^{4/3}-\left(\frac{4}{3} + \Omega\right)\Omega^{1/3}\right).
	\label{eq:Uavg}
	\end{eqnarray}
	Relaxing the assumption of constant density would lead to a slightly lower mean ionisation parameter, since it would cause greater recombination losses, but given that density variations in H~\textsc{ii} regions are only at the tens of percent level \citep{Bisbas:2015aa}, this is a minor effect.
	
	\paragraph*{Metallicity:}
	The \citetalias{Goldbaum:2016aa} simulations do not have an intrinsic metallicity ($Z$) gradient because they do not track the chemical evolution of the gas. None of the current simulations that build up a metallicity gradient self-consistently over cosmological time scales have high enough spatial resolution for this study. We choose \citetalias{Goldbaum:2016aa} simulations because their structure is realistic on scales smaller than the beam of any survey we intend to imitate. Since the purpose of this paper is to study the effect of spatial resolution on metallicity gradients, we therefore `paint' a metallicity gradient on the \citetalias{Goldbaum:2016aa} galaxy. Having control of the metallicity gradient gives us the advantage of being able to vary it so that we can study the effect of resolution on a broad range of gradients. We assume smooth radial exponential profiles for $Z$ and choose a value of the gradient $\nabla_r \log Z$ and the central metallicity $Z_0$. Thus, for every {\htr} at a given radius $r$ we can compute its metallicity. In the light of recent works having reported azimuthal variations in metallicty gradients \citep{Ho:2018aa, Kreckel:2019aa}, we note that a smooth radial gradient may not be the most realistic representation. However, we choose this profile for simplicity and to quantify the effects of spatial resolution and noise with fewer free parameters.
	
	Having defined the parameter space that describes our {\htrs}, we next use the MAPPINGS v5.1 code \citep{Sutherland:2013aa} to run a series of photoionisation models to sample it. The parameter space sampled is as follows. The models are computed at metallicity values of 0.05, 0.1, 0.2, 0.3, 0.5, 1.0, 2.0, 3.0 and 5.0 in units of solar metallicity \Zsun \footnote{We have verified that this grid is sufficiently well-sampled that interpolation errors do not dominate our error budget by running the pipeline we describe below on a case with no spatial smoothing or noise. When we do so we find that we are able to recover metallicities to within a few percent, which is the accuracy of our metallicity diagnostics, so that interpolation is not contributing significantly to the error budget.} and six uniformly spaced ages, ranging from 0 to 5 \Myr~(both boundaries included). We use four $\log{(\langle U \rangle)}$ values ranging from $-4$ to $-1$ and six $\log{(n_{\mathrm{II}})}$ values uniformly spaced in the range $-1$ to 6 (in units of cm$^{-3}$). The MAPPINGS code requires as input the metallicity $Z$ and total rate of ionising photons emission rate $Q_0$, together with a series of quantities at the {\htr} inner edge: the ionisation parameter $U_i$, pressure $P$, and temperature $T$. We determine these quantities from the parameters making up the 4D grid -- $Z$, $\langle U \rangle$, $n_{\mathrm{II}}$ and age -- as follows. Substituting the $\Omega$, \U and \nII values into \autoref{eq:Uavg}, \autoref{eq:rs} and \autoref{eq:Om} we solve for $Q_0$ and $r_i$. We then derive the ionisation parameter at the inner edge of the {\htr} ($U_i$) by substituting $r_i$ in \autoref{eq:U}. The pressure in the photoionised gas \lpok is derived from \nII using \lpok = $\log{(n_{\mathrm{II}})} + \log{T}$ where $T$ is the electron temperature inside the {\htr}, which for the purposes of computing the input pressure we assume to be constant at 10$^4$ K.\footnote{The actual temperature profile is computed self-consistently by MAPPINGS; we only adopt a temperature here for the purposes of turning our mean density into a starting pressure.} The quantities $Z$, $U_i$, \lpok, $T$ and $Q_0$ are provided to MAPPINGS to generate each {\htr} model. For each MAPPINGS model we record the line flux for the {\nii}, {\ha}, {\siia} and {\siib} \,lines. We then assign line luminosities to each simulation {\htr} by linearly interpolating the line flux on our grid of $Z$, \U, \nII, age values.
	
	\subsection{Constructing the synthetic IFS cube}\label{sec:cube}
	The next step in our synthetic data construction procedure is to produce a synthetic integral field spectrograph (IFS) data cube by adding the nebular line fluxes output from MAPPINGS on top of an underyling stellar continuum provided by Starburst99 \citep{Leitherer:1999aa}. The Starburst99 model parameters are described in \autoref{sec:H2R}. We ensure that both MAPPINGS and Starburst99 models are based on the same stellar parameters. First, we run Starburst99 models for a range of ages from 0 to 5 \Myr. Then, we compute the driving stellar continuum radiation field for each {\htr} in the simulation by interpolating the spectra output by Starburst99 linearly in age and rescaling to the cluster mass. The interpolation in age is performed at a finely sampled resolution of 0.1 \Myr. We sample the continuum at a resolution of 10\,\kmps within a $\pm$ 500\,\kmps window around the central wavelengths of emission lines, and at 20\,\kmps outside that window. The above wavelength sampling for the continuum is sufficient because the Starburst99 continuum itself has poor spectral resolution.
	
	Next, we add the emission line fluxes for each {\htr}, computed in \autoref{sec:grid}, as Gaussians with a velocity dispersion $\sigma_v$ = 15 \kmps centred on the rest wavelength of each line, Doppler-shifted by the velocity of the cell in which the {\htr} is located; for the purposes of computing the Doppler shift, we assume that the galaxy is being viewed face-on. We assume $\sigma_v$ = 15 \kmps as a reasonable estimate of thermal and turbulent broadening \citep{Krumholz:2016aa}, although this depends on local turbulent properties of the individual galaxy observed \citep{Federrath:2017aa, Zhou:2017aa}. Our synthetic data cubes have a rest-frame wavelength range of 6400 - 6783 \AA. This range is motivated by the fact that we only need the emission lines {\ha}, {\nii}, {\siia} and {\siib} for the metallicity diagnostic (\autoref{sec:met}) we employ, all of which lie within the said range.
	
	A sufficiently high spectral resolution continuum, such as those in observed data, is marked with stellar absorption features - particularly the Balmer trough near the {\ha} line. Observers adopt sophisticated methods to fit that trough in order to measure the {\ha} emission line flux. However, for our work a simplistic continuum fitting routine is sufficient because the low resolution continuum we use does not include the aforementioned trough. In reality, a more complicated fitting algorithm may be able to measure the {\ha} line with greater accuracy, which our method is currently incapable of doing. However, because we consider only young ($< 5$ {\Myr}) stellar populations, the equivalent width (W) of the {\ha} absorption is expected to be W$\sim 2-3$ \AA  \citep{Groves:2012aa}, whereas that of the {\ha} emission in the bright pixels of our simulation is W$\sim40$ \AA. Thus, the Balmer absorption is a small fraction of the emission line strength and failing to account for it would only affect the flux at a few percent level, which is smaller than the uncertainties introduced by other steps, and hence can be ignored.
	
	To produce the ideal IFS cube, we construct a grid spanning 30\,kpc$\times$30\,kpc$\times$380\,\AA \,with each spaxel being 40\,pc$\times$40\,pc$\times$10\,{\kmps}, and sum the nebular and stellar light from all the {\htr}s in each spatial pixel. Note that the \citetalias{Goldbaum:2016aa} simulations have a spatial resolution of 20\,\pc at the maximum refinement level. However, we assume a base cell-size of 40\,\pc in the interest of minimising computation cost. The choice for the base resolution does not affect our results, as long as it is considerably smaller than our smallest point spread function (PSF) size ($\sim$ 200\,\pc). The end result of this step is an ideal (10\,\kmps spectral resolution, 40\,\pc spatial resolution, noise-free) 3D IFS datacube.
	
	The next set of steps are aimed at incorporating instrument effects into the ideal cube. First, we rebin the spectral dimension of this cube down to a resolution of 30 km s$^{-1}$, comparable to the resolutions of modern IFS systems. We then proceed to coarsen the spatial resolution of the cube, a process we describe in the next section.

	\begin{figure*}
	\centerline{
		\def\arraystretch{0.1}
		\setlength{\tabcolsep}{1.0pt}
		\begin{tabular}{rl}
			\includegraphics[trim=0.0cm 0.0cm 4.7cm 0.0cm, clip, height=0.3\textheight]{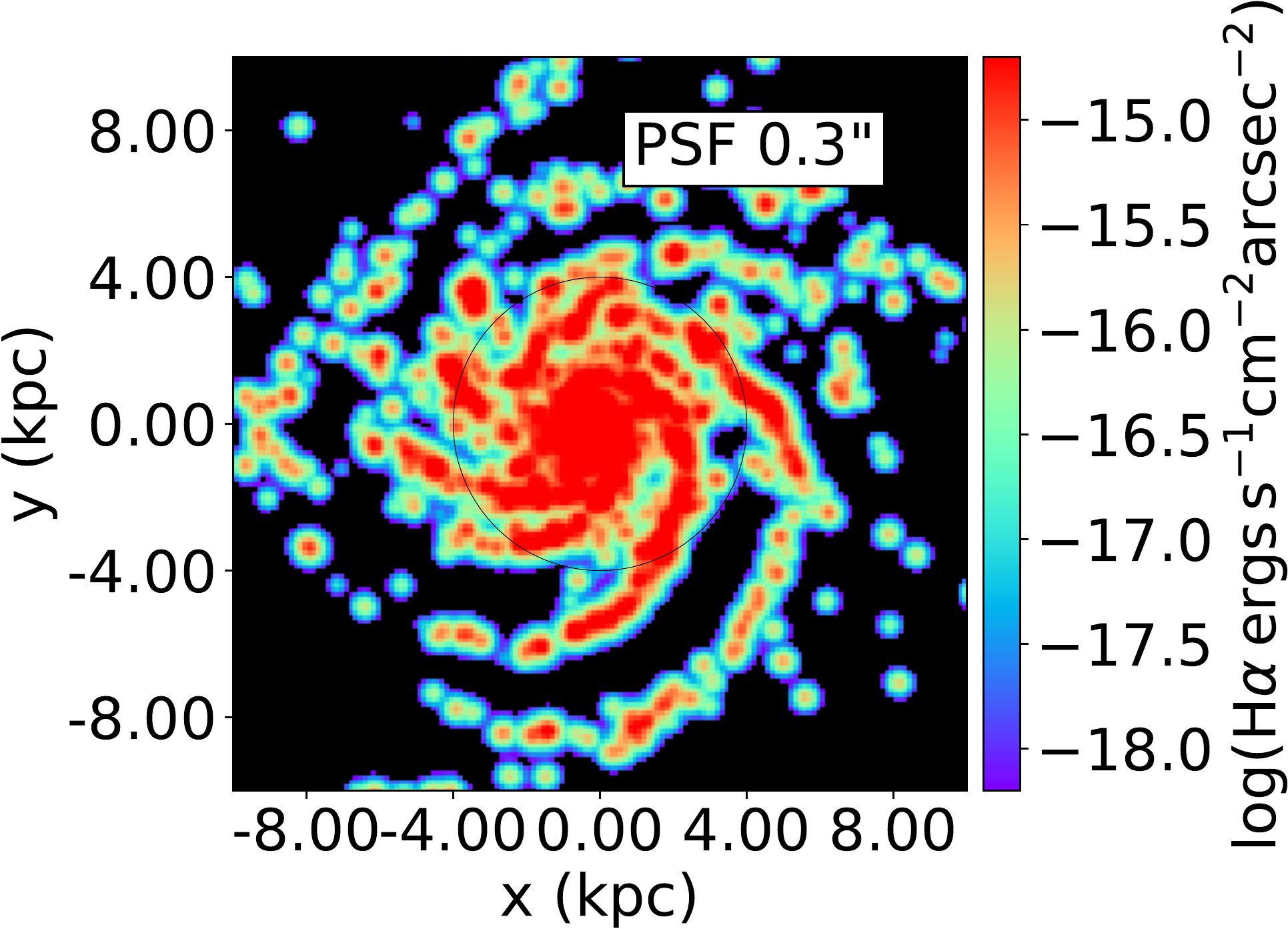} &
			\includegraphics[trim=3.5cm 0.0cm 0.0cm 0.0cm, clip, height=0.3\textheight]{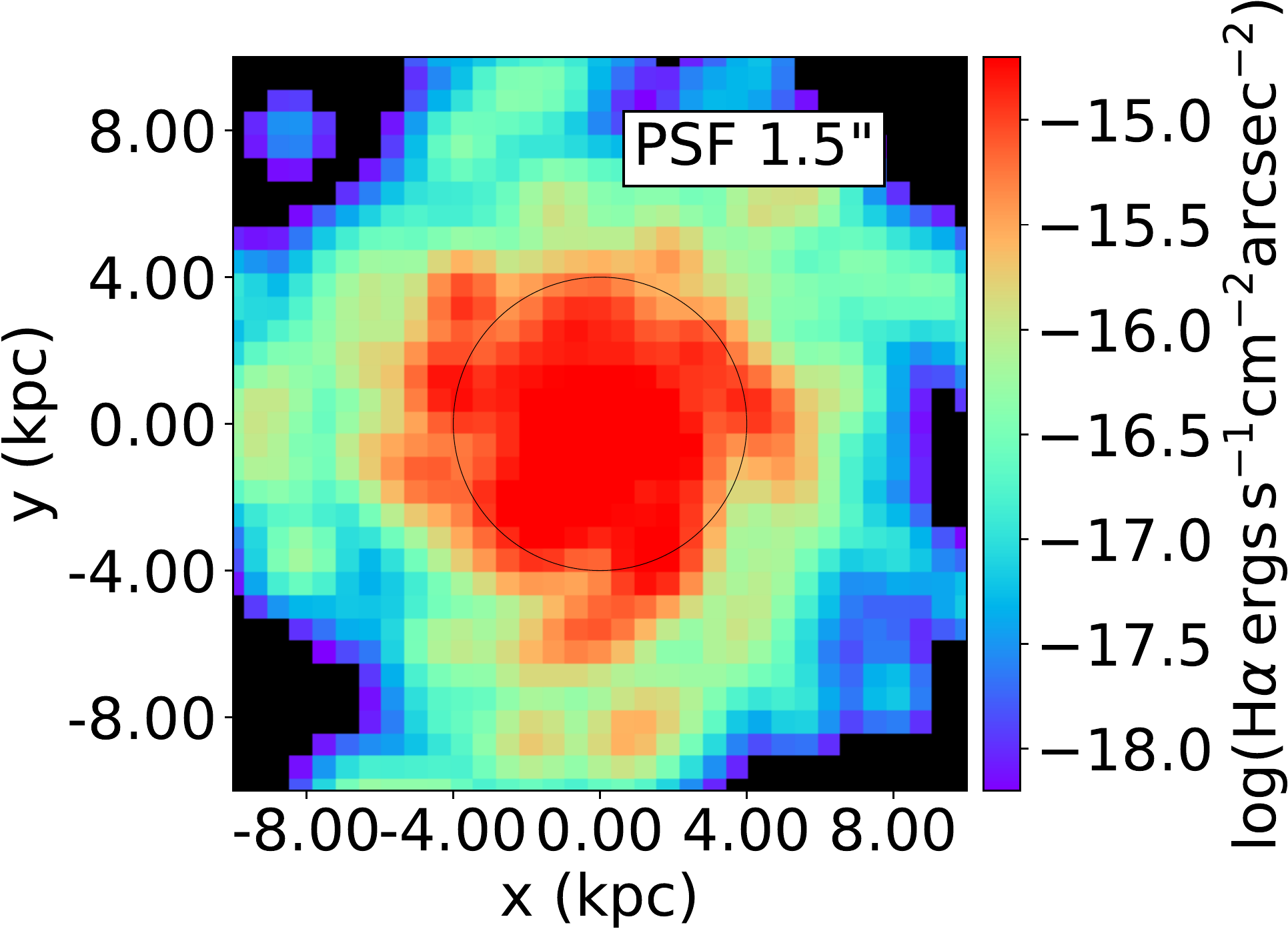}
		\end{tabular}}
		\caption{{\ha} maps produced by using a small PSF (left; FWHM = 0.3" = 0.25 {\kpc}, on source frame of galaxy) and large PSF (right; FWHM = 1.5" = 1.25 {\kpc}, on source frame), in order to highlight the effect of convolution with a Moffat profile. The thin black circle in each map marks the scale radius of the disc, which in this case is 4 {\kpc}. Both these cases are free of noise, in order to isolate the effect of spatial smoothing.
		} 
		\label{fig:compare_res}
	\end{figure*}

	\subsection{Simulating finite spatial resolution}\label{sec:convolution}
	We place our simulated galaxy at a redshift $z$ = 0.04, which is typical of the SAMI survey \citep{Green:2018aa}.\footnote{The choice of the distance is immaterial because we discuss our results in the reference frame of the galaxy, i.e. our models span a wide range of PSFs expressed as number of beams per scale length of the galaxy} Based on this distance, we convert the instrument PSF to a physical spatial resolution element, defined as the full width half maximum (FWHM) of the beam in the source plane of the galaxy. We then spatially convolve each wavelength slice with a normalised 2D Moffat kernel \citep{Moffat:1969aa}. The Moffat profile is defined by a width $\sigma$ and power index $\beta$, which are related as follows:
	\begin{eqnarray}
	\sigma = \frac{\rm FWHM}{2 \sqrt{2^{\frac{1}{\beta}} - 1}},
	\end{eqnarray}
	where FWHM is the full width at half maximum for the kernel. \cite{Trujillo:2001} report that the above analytical form provides the best fit to the PSF predicted from atmospheric turbulence theory for $\beta = 4.7$, and we use this value of $\beta$ throughout this paper. To reduce computational cost, we truncate the Moffat kernel at $5\sigma$, which incorporates 99.99\% of the total power. After the convolution, we resample each slice to 2 spaxels per beam, i.e., we choose our pixel scale to be half of the FWHM, in order to mimic typical IFS datacubes. We ensure that both the convolution and resampling procedures conserve flux to machine precision. \autoref{fig:compare_res} shows a comparison between H$\alpha$ maps when smoothed with a 0.15" (left) and 1" (right) wide PSF. No noise has been included in these cases, in order to highlight the effect of spatial smoothing.
	
	\begin{figure*}
	\centerline{
		\def\arraystretch{0.1}
		\setlength{\tabcolsep}{1.0pt}
		\begin{tabular}{rl}
			\includegraphics[trim=0.0cm 0.0cm 4.7cm 0.0cm, clip, height=0.3\textheight]{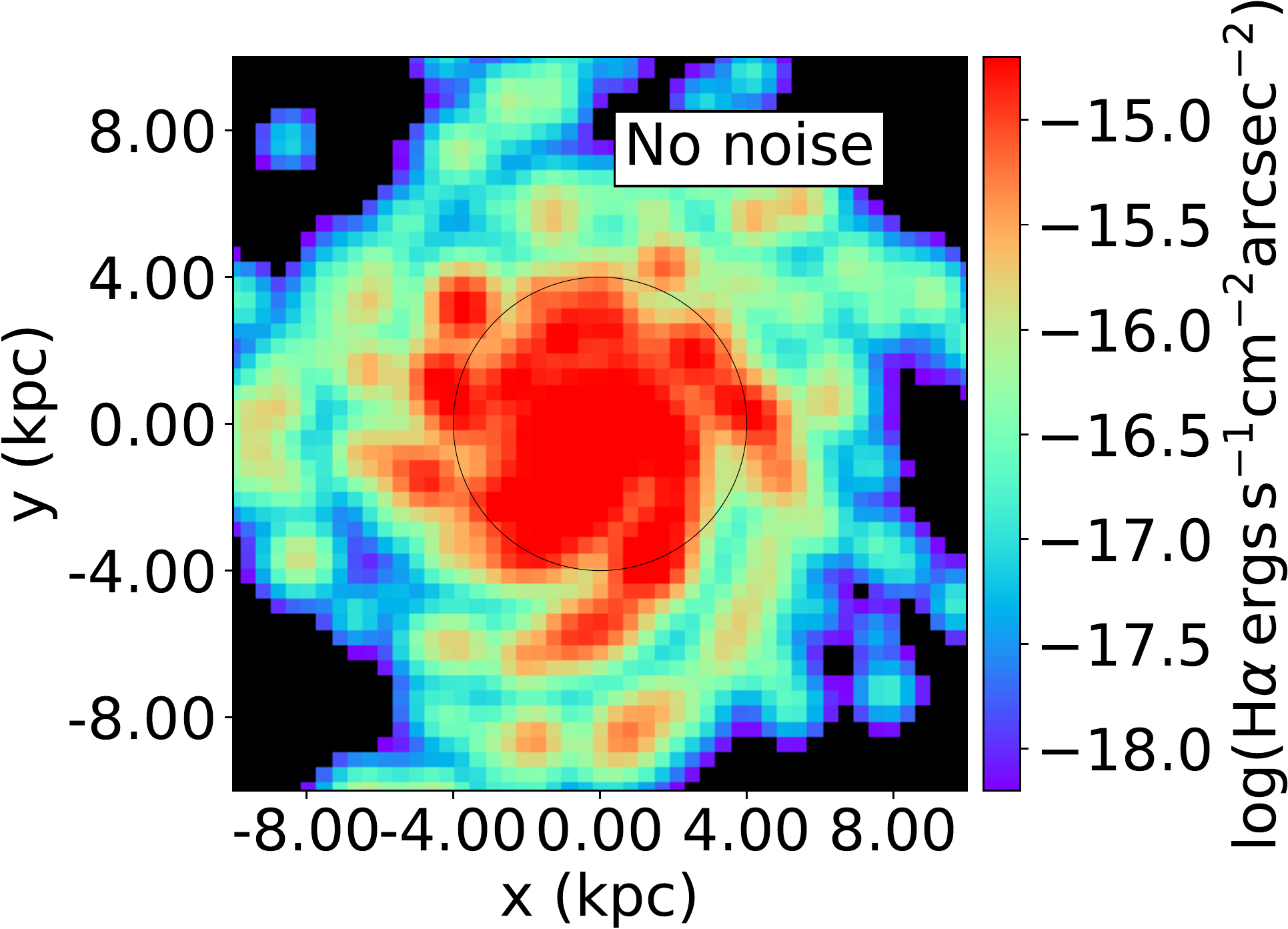} &
			\includegraphics[trim=3.5cm 0.0cm 0.0cm 0.0cm, clip, height=0.3\textheight]{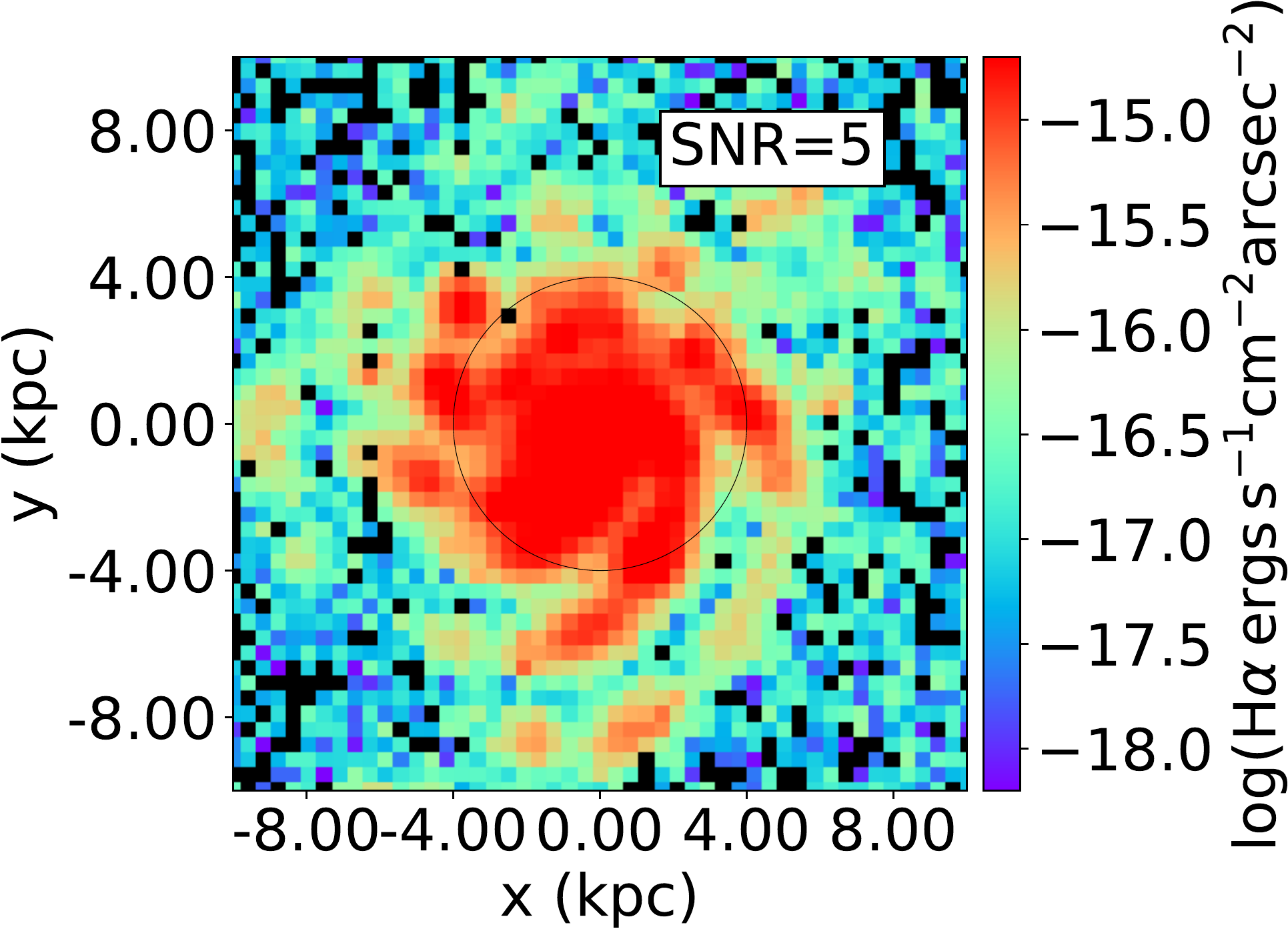}
		\end{tabular}}
		\caption{Same as \autoref{fig:compare_res}, but this time comparing between noisy (SNR=5; right) and without noise (left) case. Both images use a 1" PSF. The signal on the outskirts of the noisy image is relatively more affected than the core region. This is expected because the outer regions of the galaxy have fewer {\htrs} and hence poorer signal.
		} 
		\label{fig:compare_noise}
	\end{figure*}

	\subsection{Simulating noise}\label{sec:noise}
	\begin{figure}
	\centerline{
		\def\arraystretch{0.0}
		\setlength{\tabcolsep}{0.0pt}
		\begin{tabular}{l}
			\includegraphics[trim=0.0cm 0.0cm 0.0cm 0.0cm, clip, height=0.2\textheight]{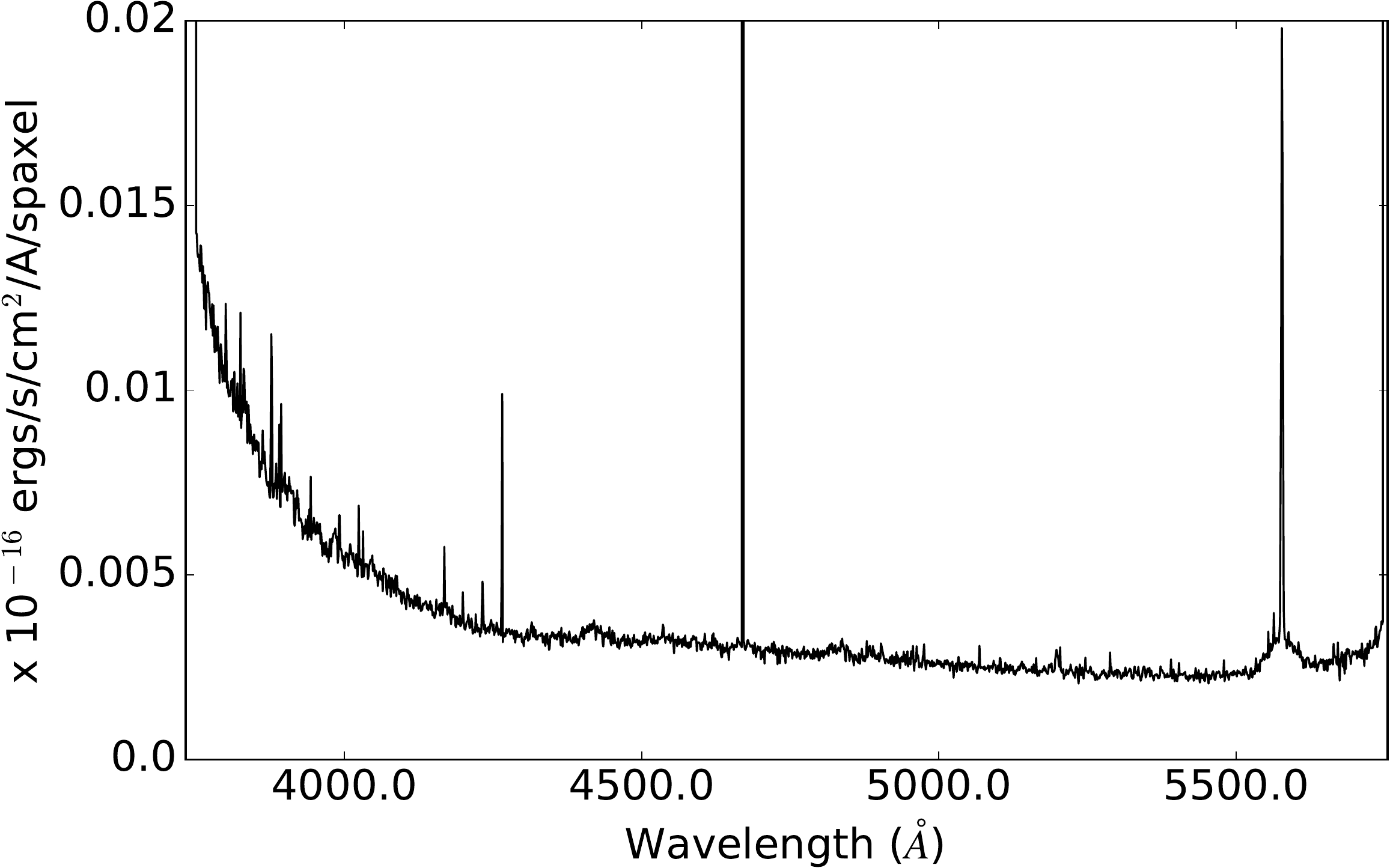} \\
			\includegraphics[trim=0.0cm 0.0cm 0.0cm 0.0cm, clip, height=0.2\textheight]{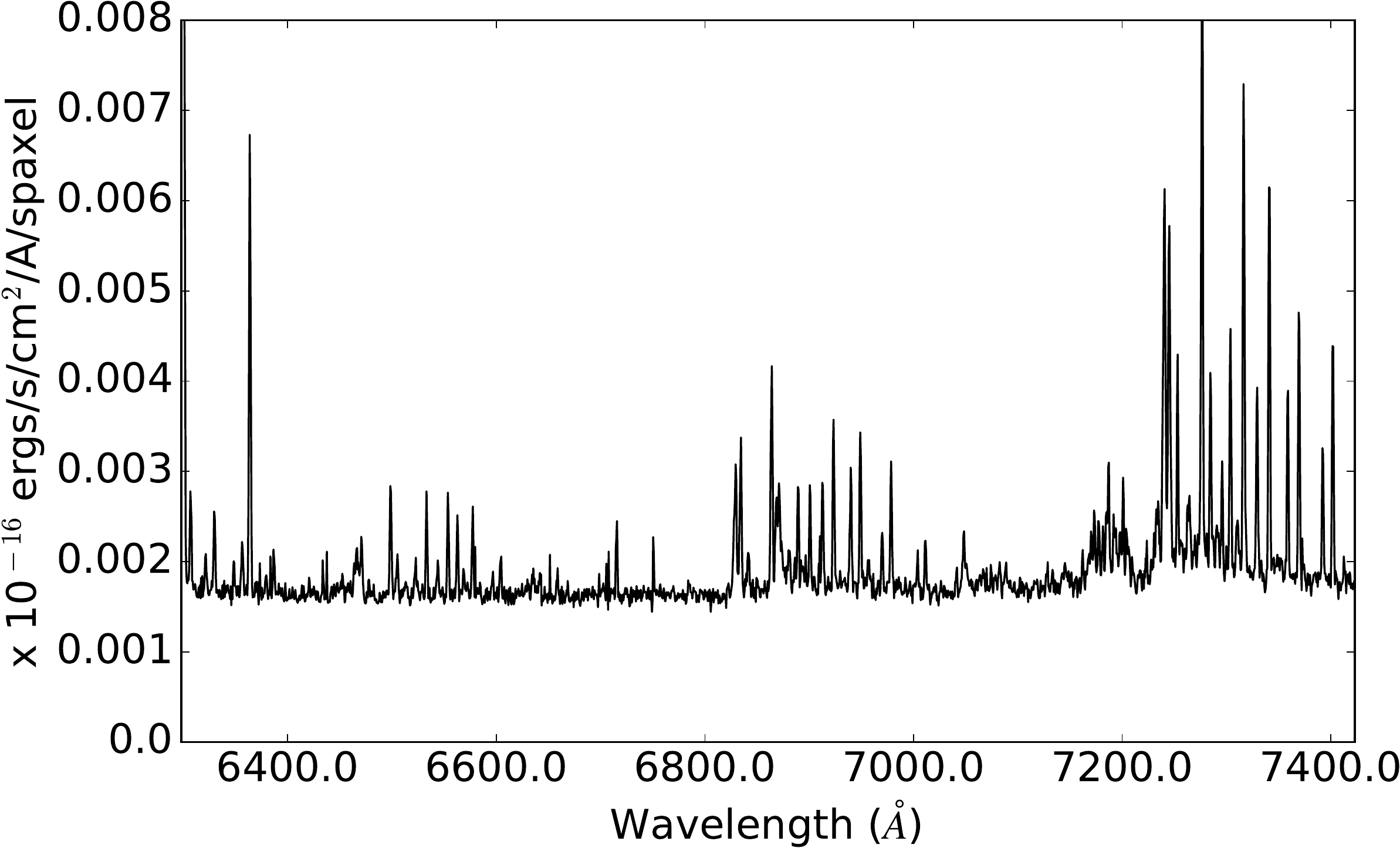}
	\end{tabular}}
	\caption{SAMI noise spectra for blue (top) and red (bottom) wavelengths. This includes noise contribution from the sky background as well as instrumental read-noise. We use this wavelength-dependent noise template to implement our noise model (\autoref{sec:noise}). 
	} 
	\label{fig:sami_noise}
	\end{figure}
	Sky noise (including contribution from OH airglow and continuum) has a non-negligible impact on the otherwise surface-brightness limited ground-based IFS observations, and thus it is important that our noise model capture this properly. Read-noise is often not the dominant source of noise in IFS data but Poisson noise, however, is important because it depends on the signal. In order to account for all these different noise sources, we use realistic wavelength-dependent noise, with the wavelength dependence taken from an observed sky frame of SAMI observations \citep{Croom:2012aa, Allen:2015aa, Green:2018aa, Scott:2018aa}. \autoref{fig:sami_noise} shows the SAMI noise spectra we used for both the blue (top) and red (bottom) channel. The blue channel noise spectra is relevant only for the additional analyses we conducted that is discussed in \aref{sec:ap}. Being a representative variance array from the outer regions of a SAMI observation, this noise spectrum captures a sky-noise limited dataset with some contribution from the detector read-noise (we include Poisson noise next). We apply this noise spectrum at every position in the synthetic IFS cube, with a normalisation level chosen as follows. For every wavelength slice of the IFS cube, we define 
	\begin{equation}
	f_{\mathrm{sky}} (\lambda) = \frac{I_{\mathrm{sky}}(\lambda)}{N_{\mathrm{sky}}}
	\end{equation}
	as the sky noise contribution at that wavelength, where $I_{\mathrm{sky}}(\lambda)$ is the sky intensity as a function of wavelength, and $N_{\rm sky}$ is the sky intensity at the wavelength of the {\nii} line, i.e., $f_{\rm sky}$ is normalised to have unit intensity at the centre of the {\nii} line. We then take the noise-less {\nii} emission line map (\autoref{sec:fitting}) and compute $S_{\mathrm{gal}}$ -- the mean intensity in an annulus between 90\% and 110\% of the scale length $r_{\mathrm{scale}}$~=~4~kpc of the galaxy. The effective (half-light) radius $r_{\mathrm{e}}$ for the galaxy is 3~kpc i.e. $r_{\mathrm{e}}$ = 0.75$r_{\mathrm{scale}}$. $S_{\mathrm{gal}}$ is used as the normalisation value for the noise across the whole field of view (FoV). In order to produce an observation with a specified signal to noise ratio (SNR), we take the zero noise spectrum at every spatial pixel for every wavelength slice and add a random number drawn from a Poisson distribution with standard deviation
	\begin{equation}
	N = f_{\mathrm{sky}} (\lambda) \times \frac{S_{\mathrm{gal}}}{\mathrm{SNR}}
	\end{equation}
	This allows for a signal-dependent Poisson component in our noise, which is otherwise absent in the SAMI noise spectrum we used. Thus, by definition, the mean SNR within an \mbox{(0.9 - 1.1)$\,\times\,r_{\mathrm{scale}}$} annulus of the {\nii} emission line map should be equal to the input SNR. The SNR measured in the datacubes is generally a bit smaller ($\sim$80\% of the input SNR) due to the additional uncertainty introduced by the continuum subtraction process (\autoref{sec:fitting}), but this effect is not large. Throughout the paper we quote SNR values as SNR per spaxel. \autoref{fig:compare_noise} shows a comparison between H$\alpha$ maps without noise (left) and with noise added (right, SNR=5).
	
	\section{Extracting observables}\label{sec:observables}
	\begin{figure*}
	\centerline{
		\def\arraystretch{2.0}
		\setlength{\tabcolsep}{0.15pt}
		\begin{tabular}{llll}
			\includegraphics[trim=0.0cm 0.9cm 4.7cm 0.0cm, clip, height=0.173\textheight]{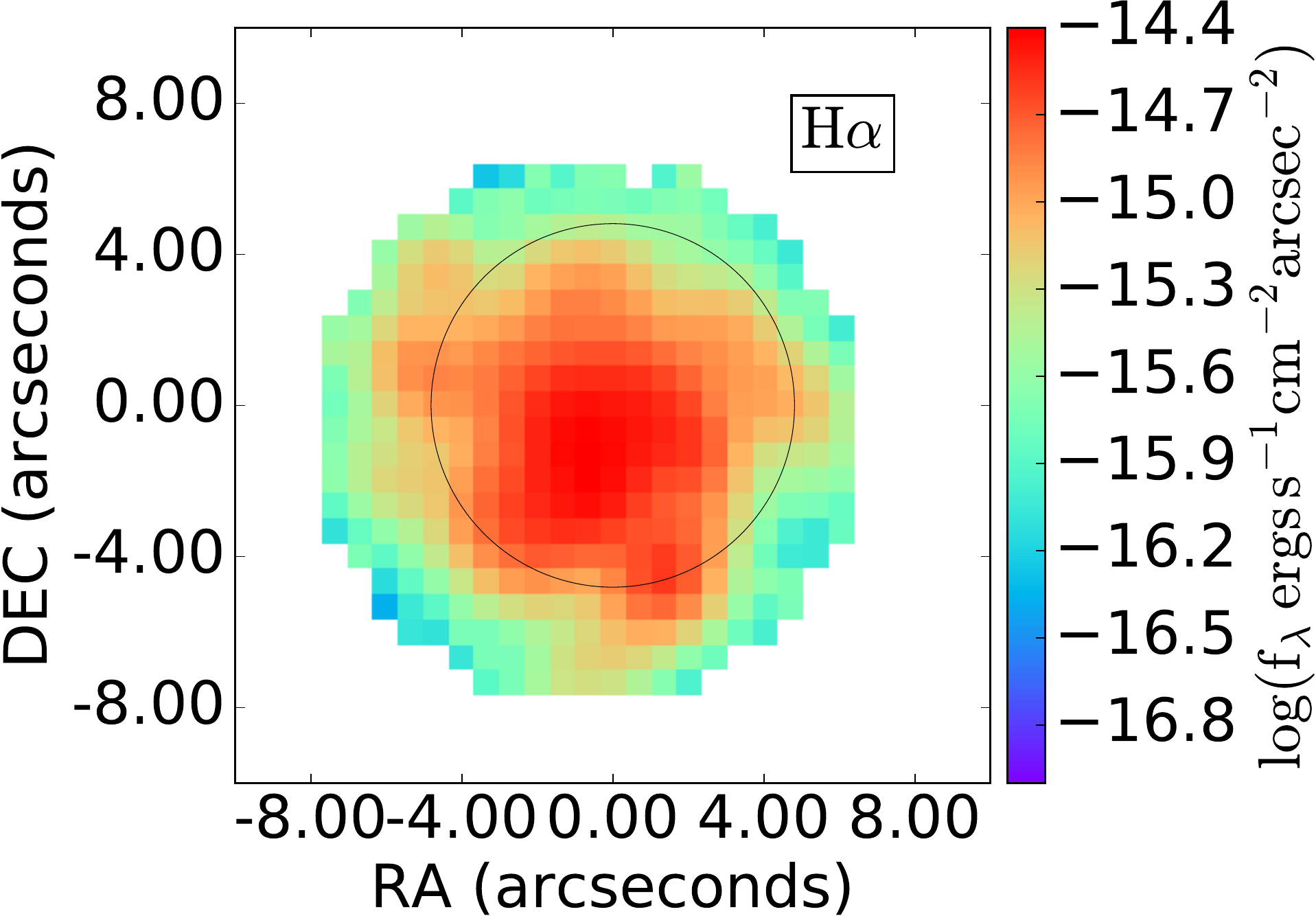} &
			\includegraphics[trim=3.3cm 0.9cm 4.7cm 0.0cm, clip, height=0.173\textheight]{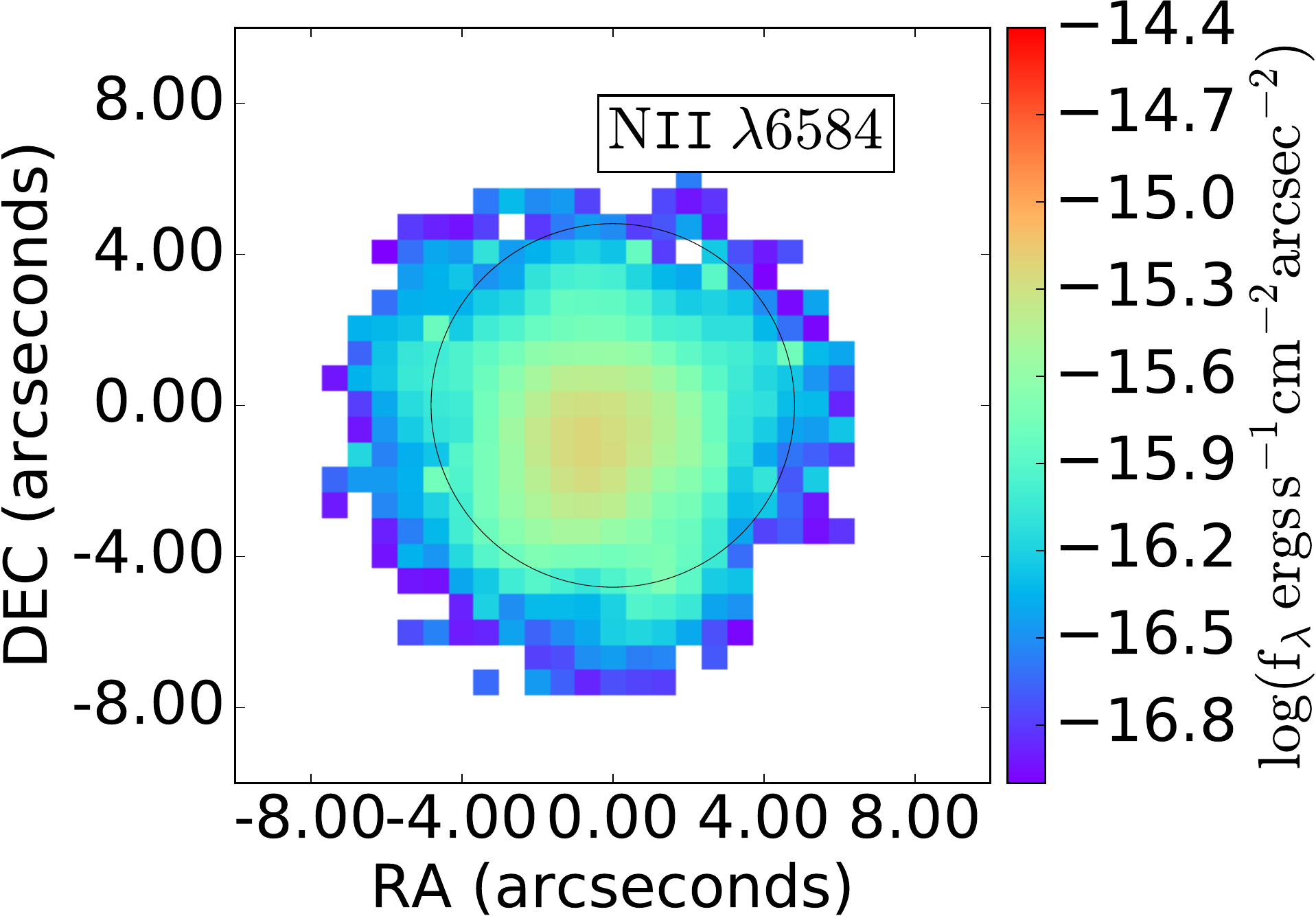} &
			\includegraphics[trim=3.3cm 0.9cm 4.7cm 0.0cm, clip, height=0.173\textheight]{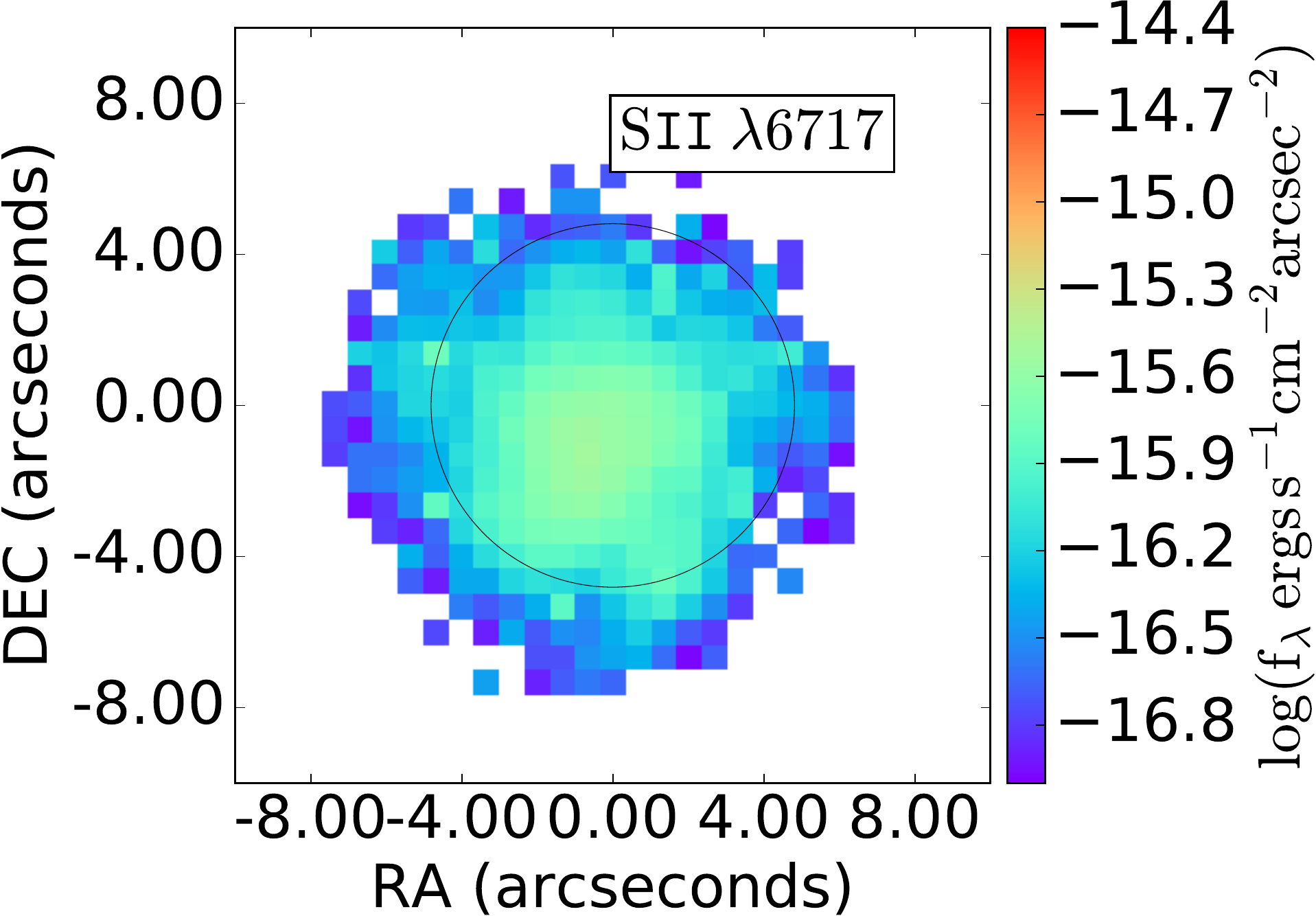} &
			\includegraphics[trim=3.3cm 0.9cm 0.0cm 0.0cm, clip, height=0.173\textheight]{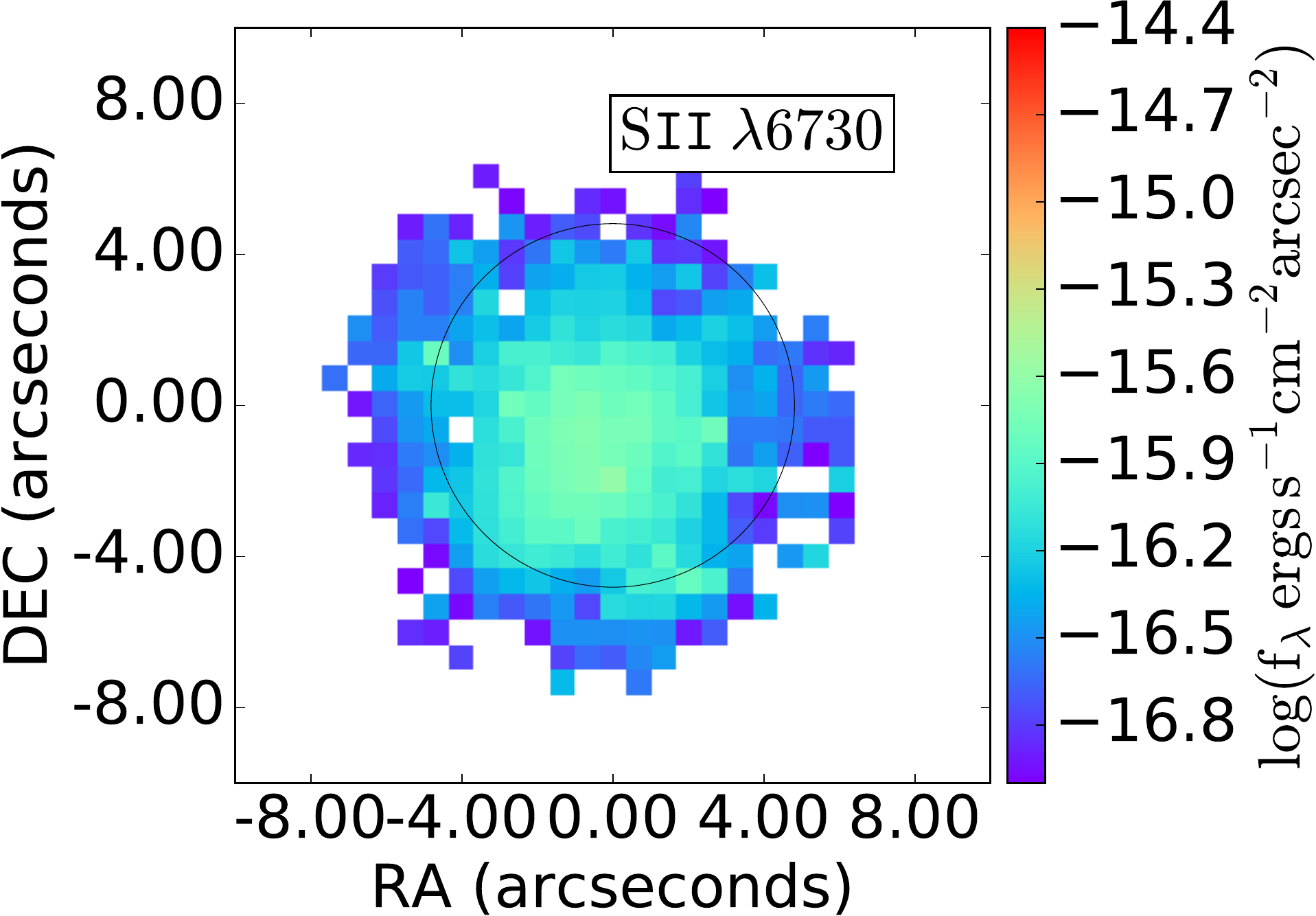} 			
			\\
			\includegraphics[trim=0.0cm 0.0cm 4.6cm 0.0cm, clip, height=0.185\textheight]{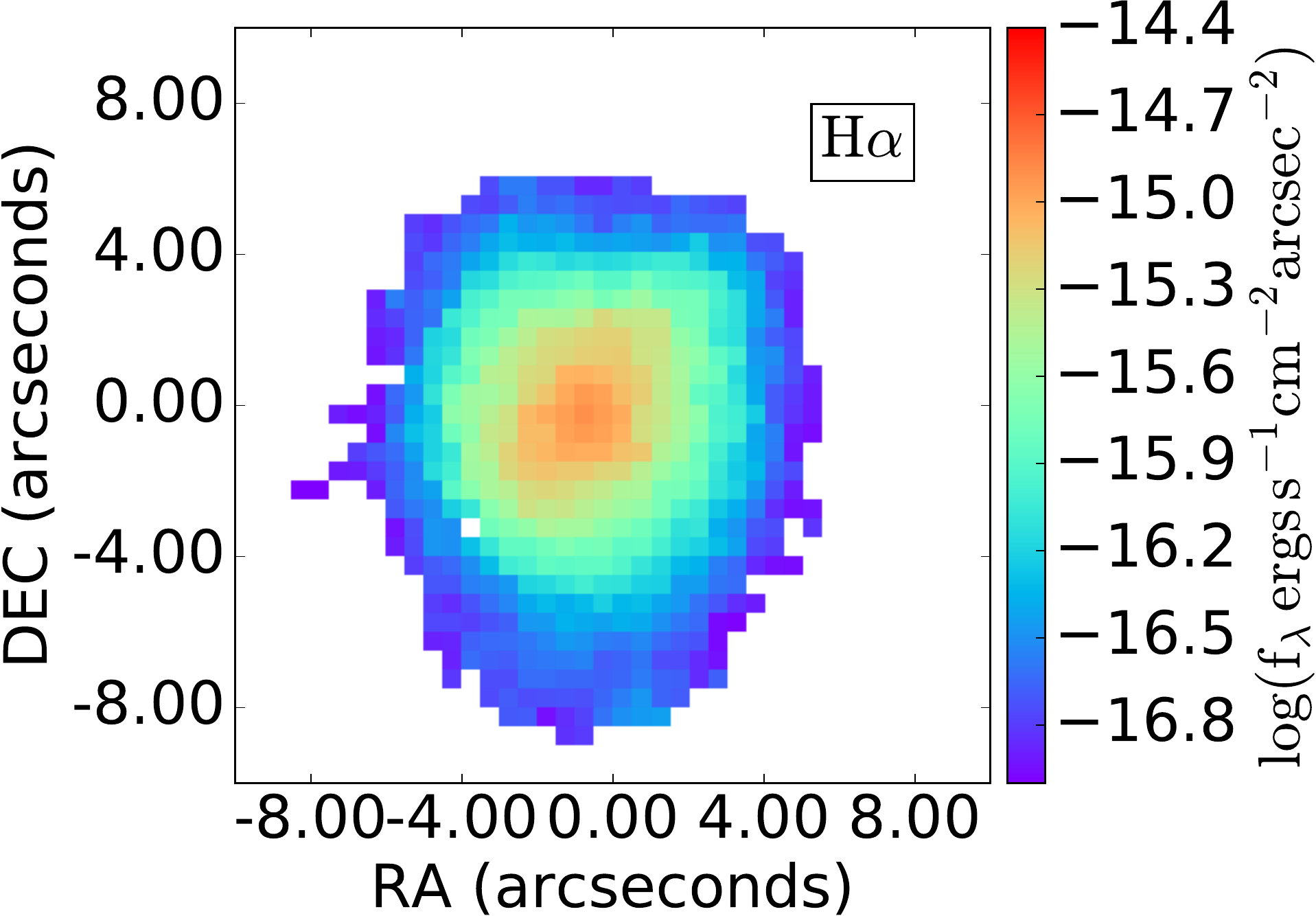} &
			\includegraphics[trim=3.3cm 0.0cm 4.6cm 0.0cm, clip, height=0.185\textheight]{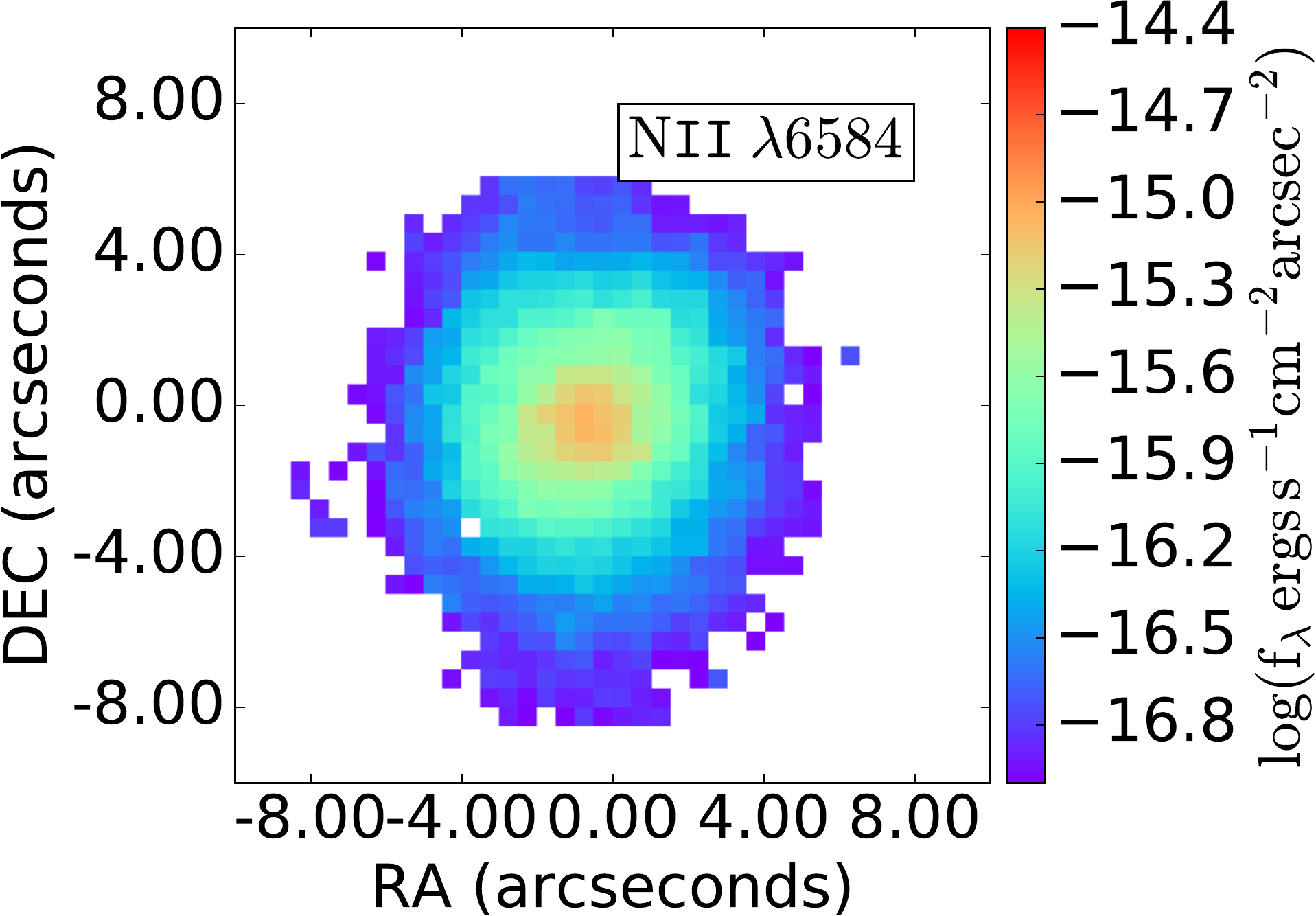} &
			\includegraphics[trim=3.3cm 0.0cm 4.6cm 0.0cm, clip, height=0.185\textheight]{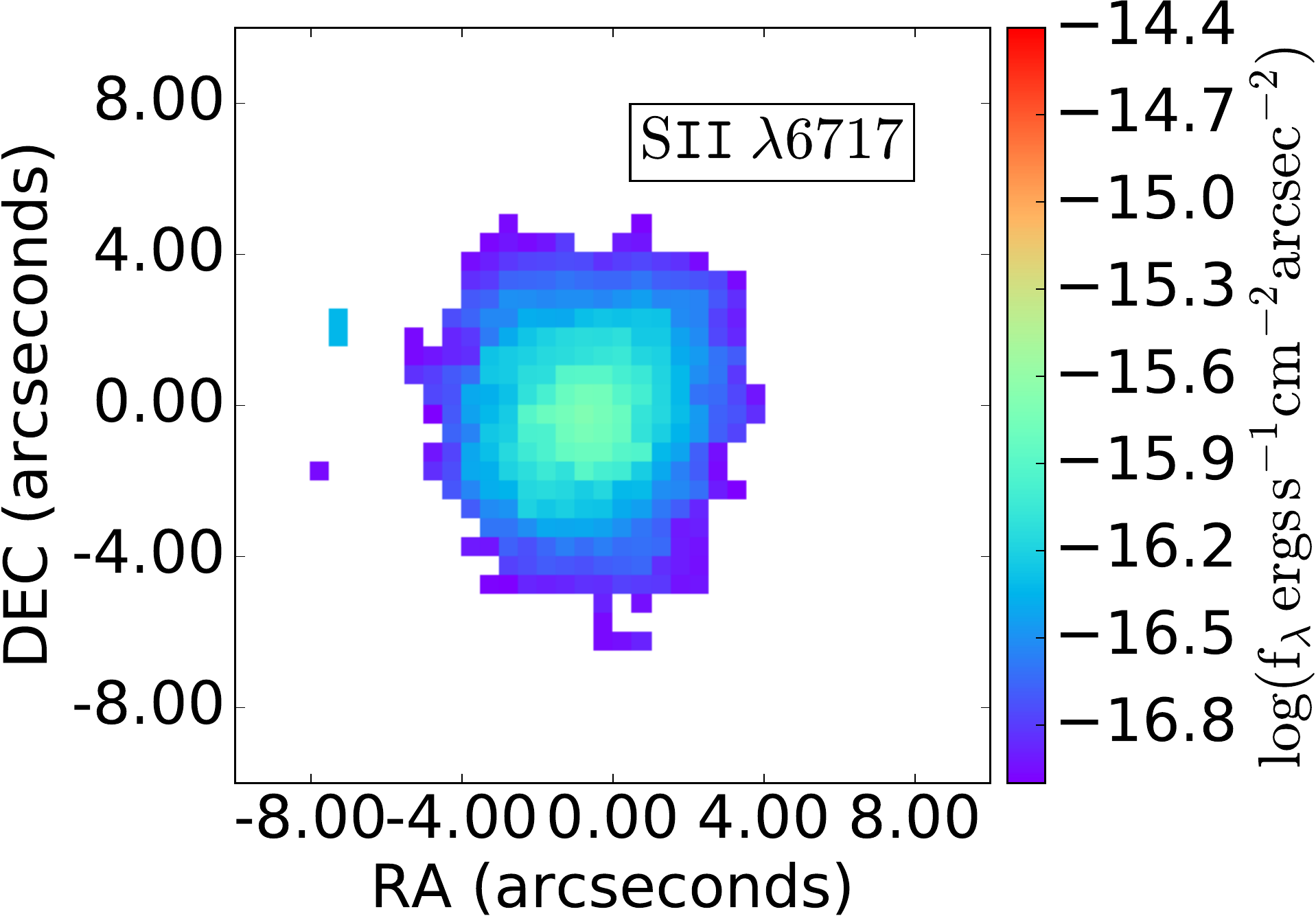} &
			\includegraphics[trim=3.3cm 0.0cm 0.0cm 0.0cm, clip, height=0.185\textheight]{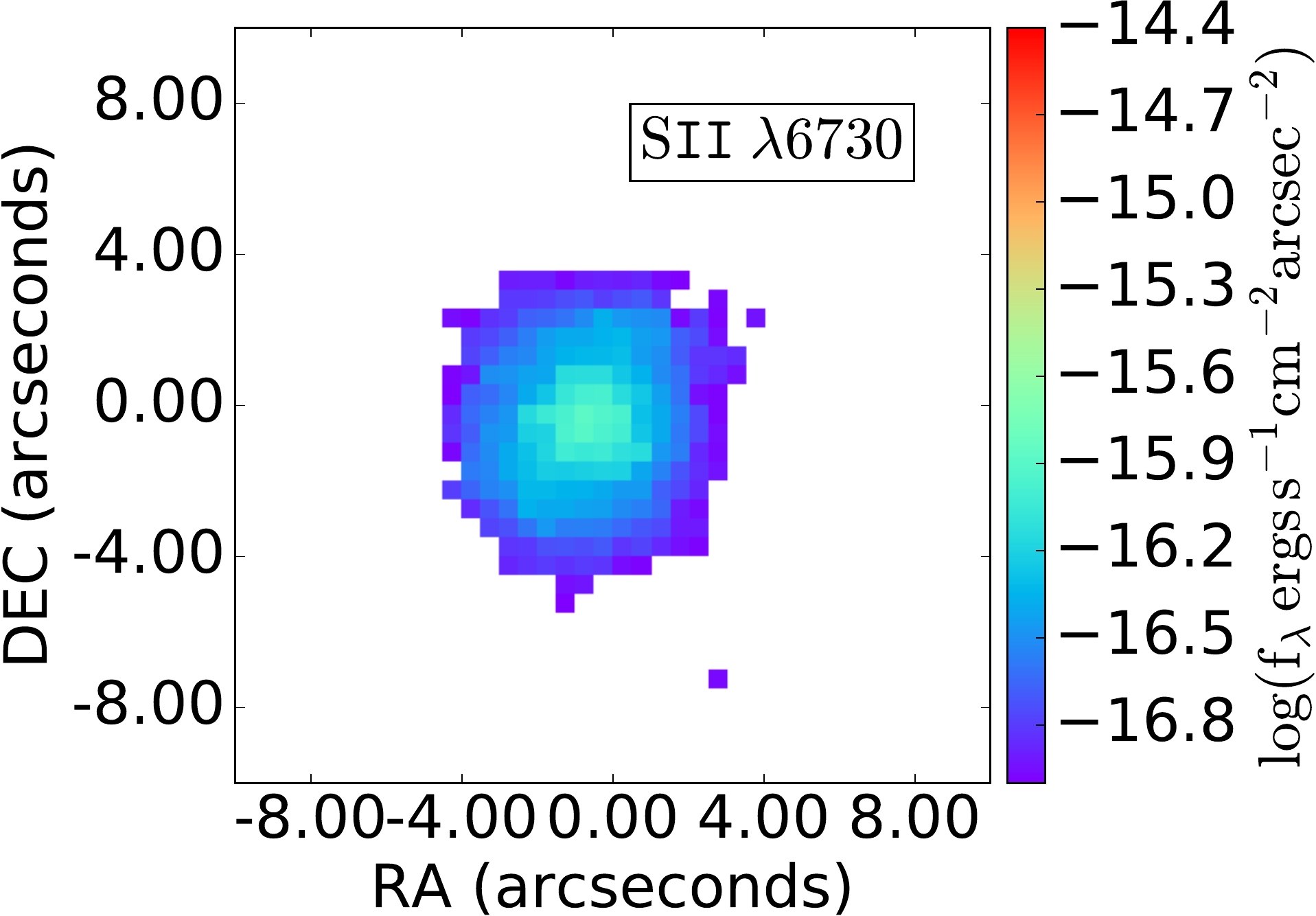}
	\end{tabular}}
	\caption{\textit{Top}: Emission line maps of {\ha} (top left), {\nii} (top right), {\siia} (bottom left) and {\siib} (bottom right) derived by Gaussian fitting (\autoref{sec:fitting}) for a spatial resolution of 2" PSF \citep[corresponding to SAMI;][]{Green:2018aa, Scott:2018aa} and a mean SNR/spaxel of 8 at the scale length of the {\nii} map, indicated by the thin black circle. \textit{Bottom}: Corresponding emission line maps of an observed SAMI galaxy N209807, for comparison. The simulated maps in the top panel have been tailored to the SAMI footprint for ease of comparison.
	} 
	\label{fig:line_maps}
	\end{figure*}

	\subsection{Emission line fitting}\label{sec:fitting}
	Most established IFS data reduction pipelines, e.g. LZIFU for SAMI survey \citep{Ho:2016aa} and Pipe3D for CALIFA survey \citep{Sanchez:2016aa}, are able to perform emission line fitting but are custom-designed for the particular IFU survey. Moreoever, the degree of sophistication that these pipelines operate at for certain steps such as the continuum fitting, is not necessary for our simulation. The focus of our work is to investigate the relative change in emission line ratios with PSF sizes. As such, we perform our own line fitting procedure without resorting to any conventional pipeline, which might require modifying a complex software that was designed for a specific kind of data. 
	
	In order to measure the emission line fluxes, we first fit the continuum and subtract it out. The continuum fitting routine masks out a $\pm$150 \kmps region around all expected spectral features. Then a rolling average is applied on the masked flux, followed by a smooth cubic spline interpolation to derive the stellar continuum. We estimate the continuum uncertainty as the root mean square (RMS) deviation of the fitted continuum from the masked flux. As such, the continuum uncertainty along each spatial pixel is wavelength-independent.
	
	Next we simultaneously fit all neighbouring spectral lines with one Gaussian profile per line using a non-linear least-squares method. A neighbour is defined as any line within $\pm 5$ spectral resolution elements of its adjacent line. For the spectral resolution of 30 \kmps used throughout this paper, the {\nii} and {\ha} lines are well resolved and hence fit indivdually. We constrain the fit to the centroid to a wavelength range $\lambda_0 \pm 3 + \delta z/(1+z)$, where $\lambda_0$ is the line centre wavelength, and $\delta z = 10^{-4}$. We similarly constrain the line width to lie within upper and lower bounds of 300 \kmps and one spectral resolution element, respectively. The amplitudes of the Gaussian fits were allowed to vary freely. We propagate the continuum uncertainty and the noise in the spectra through to the fitting routine. Throughout this paper, we have assumed a spectral resolution of 30 \kmps, implying that all the emission lines are at least marginally resolved. Unlike some observational studies, we do not force the different emission lines to have a common width because we wish to isolate the effect of spatial resolution by reproducing the metallicity gradient with least possible constraints (i.e. maximum freedom) in the spectral fitting step. In principle, invoking contraints from atomic physics and forcing lines to have a common width may allow the use of low SNR spaxels thereby maximally utilising all the data. However, such assumptions do not recognise the fact that there could be multiple {\htrs} along each line of sight. For each spatial pixel, we measure the fluxes for the {\ha}, {\nii} and {\siiab} lines from the fitted Gaussian parameters, along with the corresponding flux uncertainties. Thus, we obtain 2D (spatial) maps for each of these emission lines. \autoref{fig:line_maps} shows example maps for a PSF of 1" and a SNR of 8.
	
	In order to verify that our simulated maps provide a reasonable approximation of real galaxies, we can compare to an observed map from the SAMI galaxy survey \citep{Green:2018aa, Scott:2018aa} selected to have a stellar mass ($M \sim 10^{11}$ $M_\odot$) and star formation rate ($\dot{M}\sim 1$ $M_\odot$ yr$^{-1}$) comparable to the \citetalias{Goldbaum:2016aa} simulation (which is modelled on the Milky Way). The bottom panel of \autoref{fig:line_maps} shows the emission line maps used in our work - {\nii}, {\ha} and {\siiab} - for SAMI galaxy N209807 \citep[SAMI DR2:][]{Scott:2018aa}. A comparison of the two upper and lower sets of panels demonstrates that we are able to reproduce the emission line flux values reasonably well. The limited spatial resolution of SAMI washes out the clumpy structure of the ISM and spiral arms in the massive galaxies. We present a similar low-resolution version of our emission line maps from the simulations in \autoref{fig:line_maps}. We thus demonstrate that our simulation is a reasonable model to test the effects of SNR and spatial resolution on metallicity gradients.	

	\subsection{Metallicity diagnostic}\label{sec:met}
	In the main text we use the \citet[hereafter \citetalias{Dopita:2016aa}]{Dopita:2016aa} diagnostic to derive the metallicity at each pixel; we present results for the alternative \citet[hereafter \citetalias{Kewley:2002fk}]{Kewley:2002fk} diagnostic in \aref{sec:ap}, and show that they are qualitatively similar. We analytically propagate the uncertainties in the line fluxes to obtain the corresponding uncertainty in metallicity. The \citetalias{Dopita:2016aa} calibration uses the {\ha}, {\nii}, {\siia} and {\siib} \,nebular emission lines and the relation
	\begin{equation}
	\log{(\mathrm{O/H})} + 12 = 8.77 + y + 0.45(y + 0.3)^5
	\end{equation}
	where
	\begin{equation}
	y = \log\frac{\ion{N}{ii}}{\ion{S}{ii}} + 0.264\log\frac{\ion{N}{ii}}{\mathrm{H}\alpha}.
	\end{equation}
	The main advantage of the \citetalias{Dopita:2016aa} diagnostic is that reddening corrections are not important because the lines are closely spaced in wavelength. In reality unresolved {\htrs} with different brightness and extinction are a concern for observational studies. This argues for using reddening-insensitive metallicity diagnostics, which is the approach we have followed here. We note, however, that simply using reddening-insensitive line ratios is insufficient to account for the differential extinction across {\htrs}. Accurately quantifying the effect of differential extinction would require simultaneously capturing the structure of {\htrs} and the dusty shells around them which is beyond the scope of this paper. We therefore acknowledge the absence of dust as a potential limitation.
	
	
	\section{Results}\label{sec:results}
	In this section we present our results for the fiducial gas fraction ($f_g=0.2$) simulation using the \citetalias{Dopita:2016aa} metallicity diagnostic, assuming wind parameter $\Omega = 0.5$. In \aref{sec:ap} we present our detailed parameter study to demonstrate that our main result is robust to different values of $f_g$ and $\Omega$, and different choices of metallicity diagnostic. These choices impact the overall accuracy of the inferred metallicity gradient, but the variation with spatial resolution and SNR remains qualitatively unaffected. The full tables and figures associated with \aref{sec:ap} are provided as supplementary online material.

	\begin{figure*}
	\centerline{
		\def\arraystretch{0.1}
		\setlength{\tabcolsep}{0.5pt}
		\begin{tabular}{rl}
			\includegraphics[trim=0.0cm 1.57cm 0.0cm 0.0cm, clip, height=0.274\textheight]{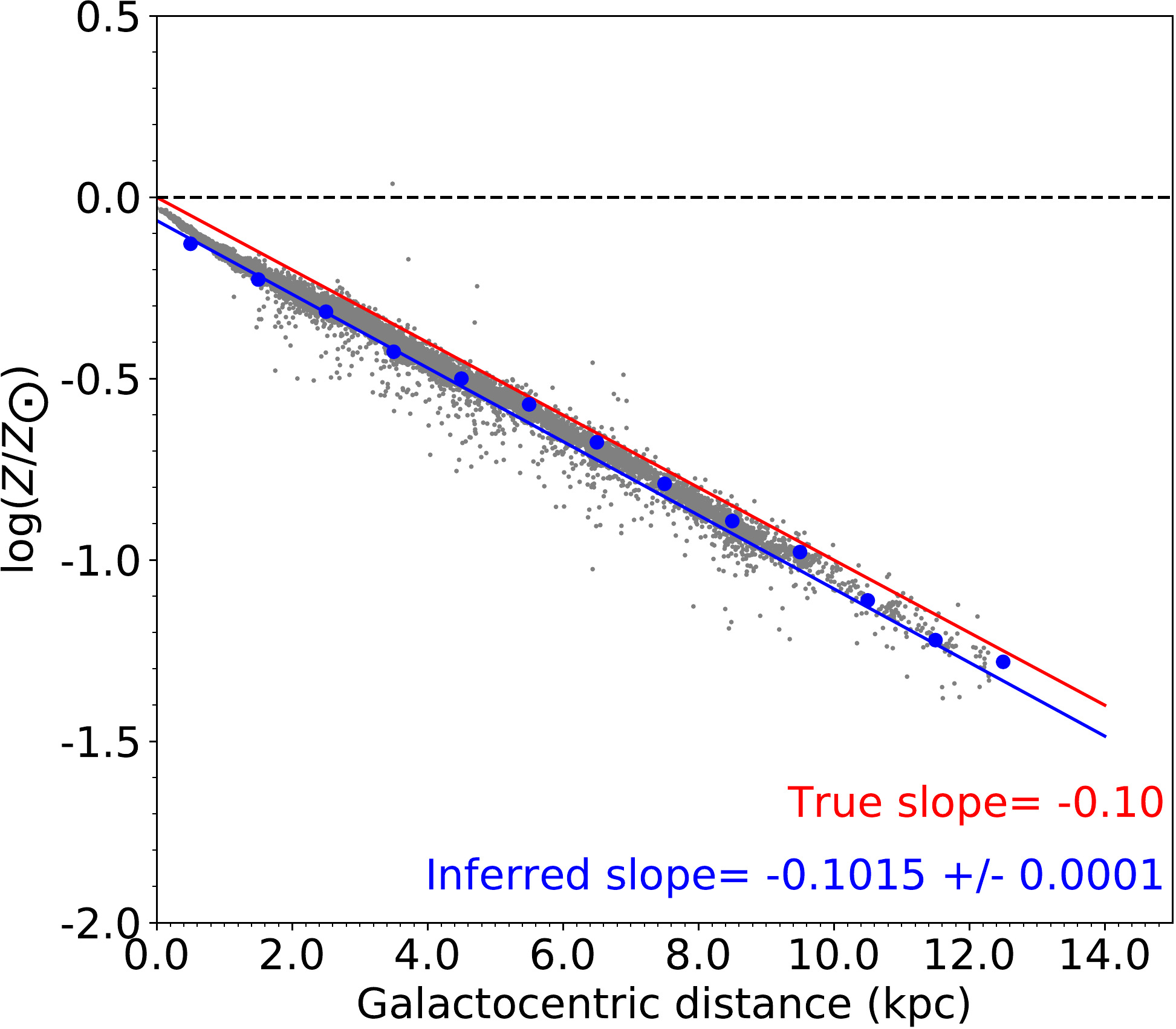}\llap{\makebox[7.0cm][l]{\raisebox{0.7cm}{\includegraphics[trim=3.6cm 2.0cm 4.2cm 0.4cm, clip, height=2.6cm]{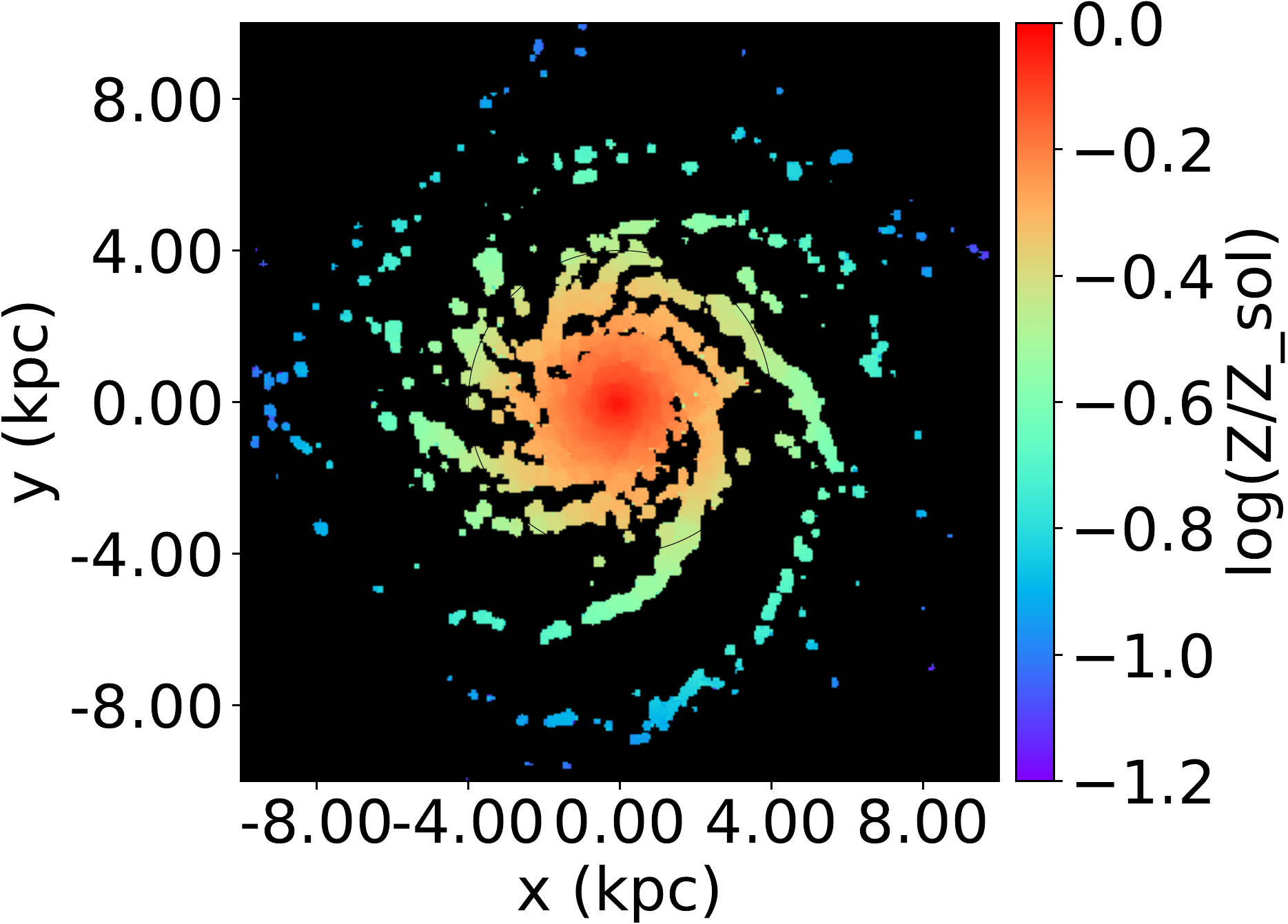}}}}\llap{\makebox[7.0cm][l]{\raisebox{5.5cm}{\includegraphics[trim=0.0cm 0.0cm 0.0cm 0.0cm, clip, height=0.6cm]{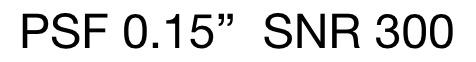}}}} &				
			\includegraphics[trim=2.5cm 1.57cm 0.0cm 0.0cm, clip, height=0.274\textheight]{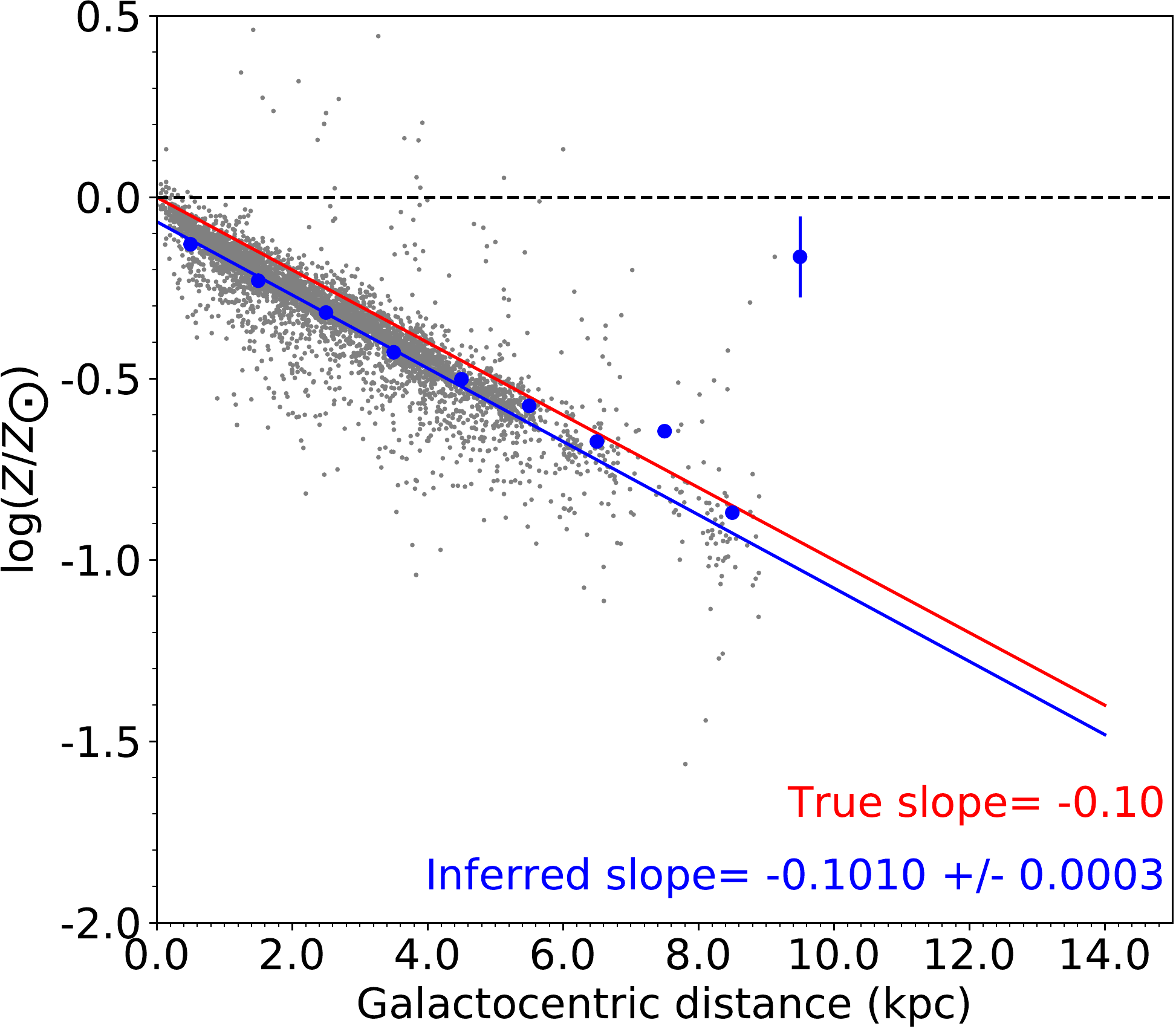}\llap{\makebox[7.0cm][l]{\raisebox{0.7cm}{\includegraphics[trim=3.6cm 2.0cm 4.2cm 0.4cm, clip, height=2.6cm]{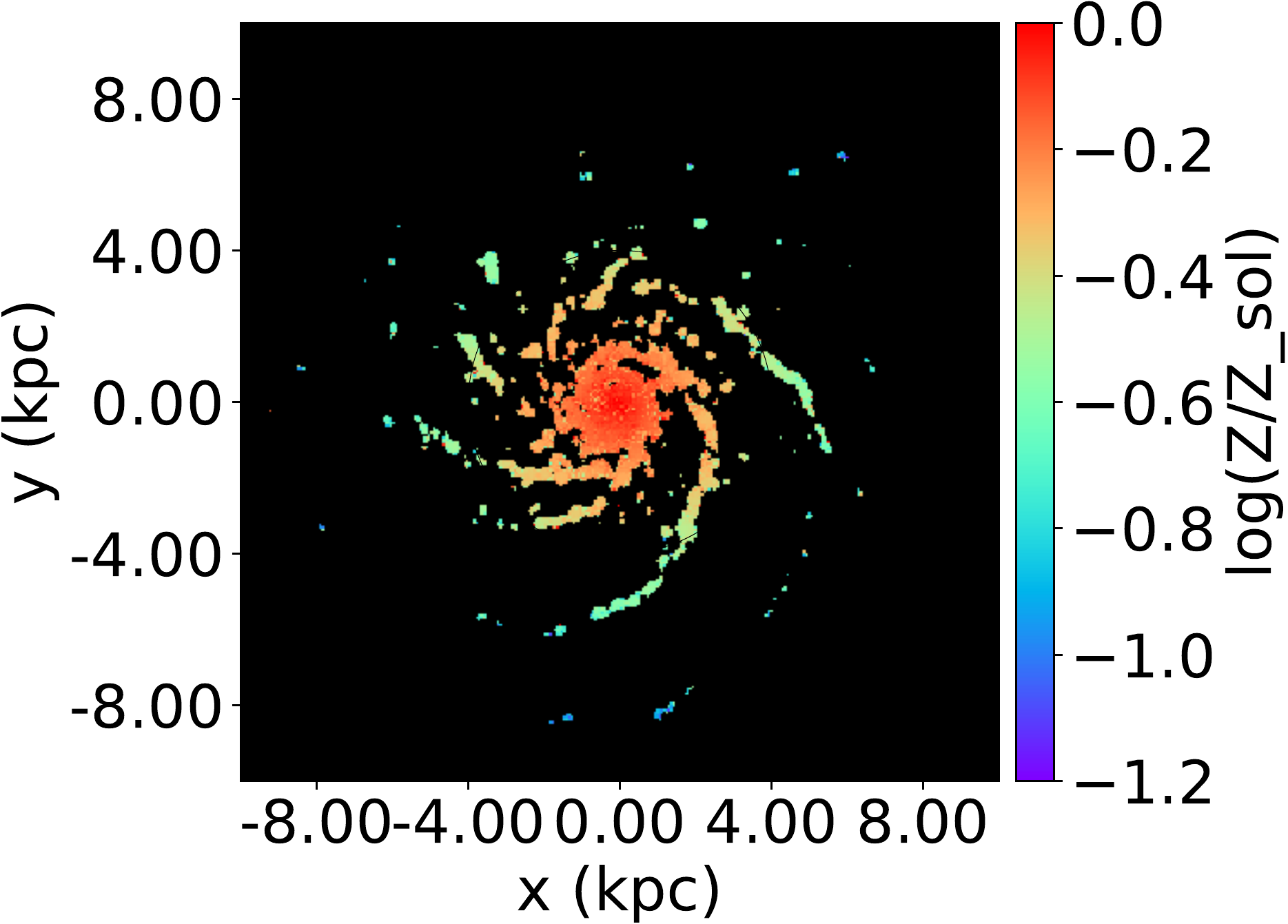}}}}\llap{\makebox[7.0cm][l]{\raisebox{5.5cm}{\includegraphics[trim=0.0cm 0.0cm 0.0cm 0.0cm, clip, height=0.6cm]{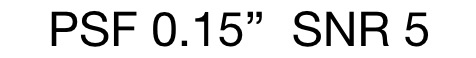}}}} \\
			\\
			\\
			\includegraphics[trim=0.0cm 0.0cm 0.0cm 0.0cm, clip, height=0.305\textheight]{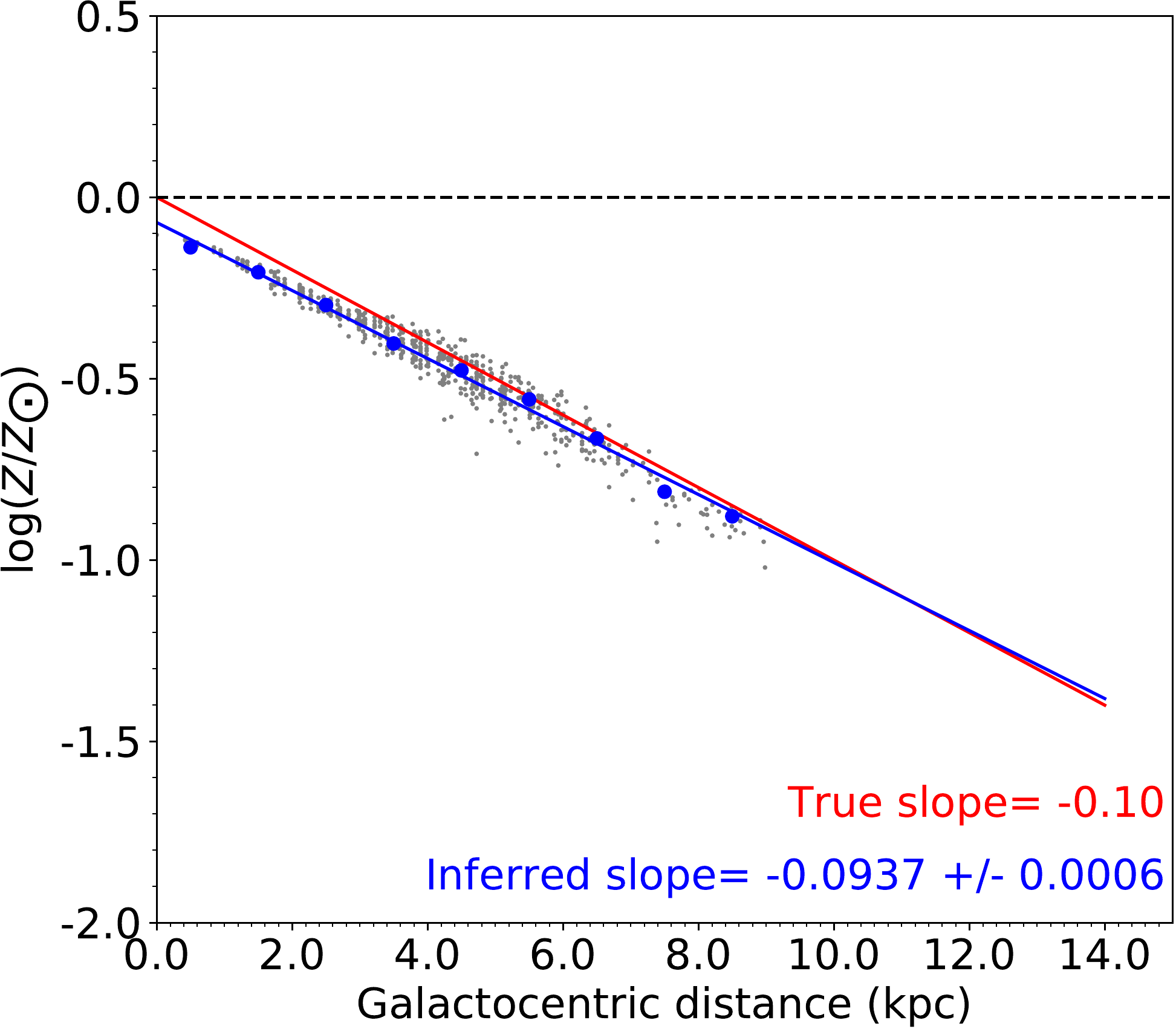}\llap{\makebox[7.1cm][l]{\raisebox{1.4cm}{\includegraphics[trim=3.6cm 2.0cm 4.2cm 0.4cm, clip, height=2.6cm]{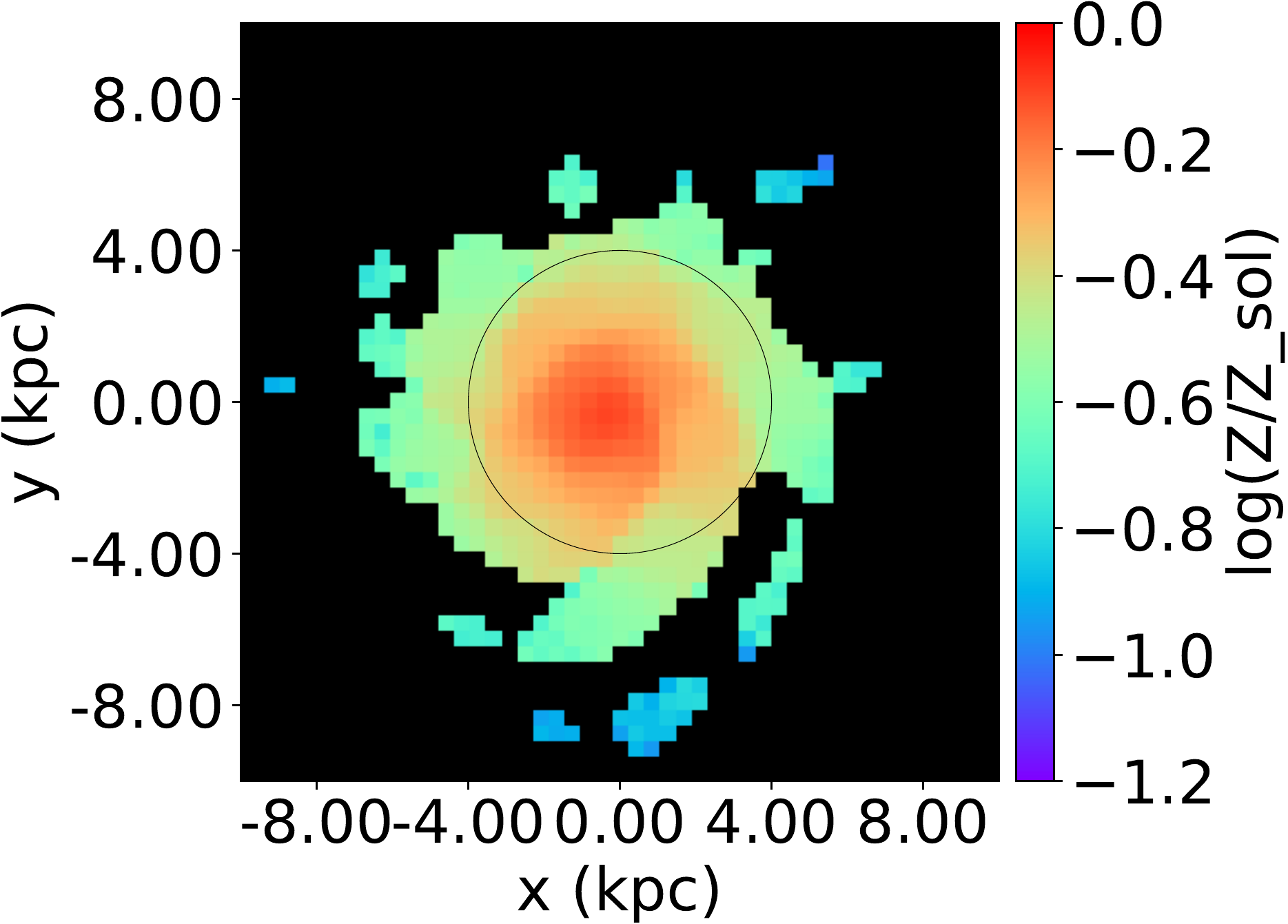}}}}\llap{\makebox[7.2cm][l]{\raisebox{6.2cm}{\includegraphics[trim=0.0cm 0.0cm 0.0cm 0.0cm, clip, height=0.6cm]{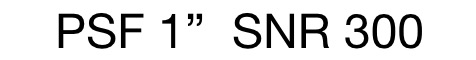}}}} &
			\includegraphics[trim=2.5cm 0.0cm 0.0cm 0.0cm, clip, height=0.305\textheight]{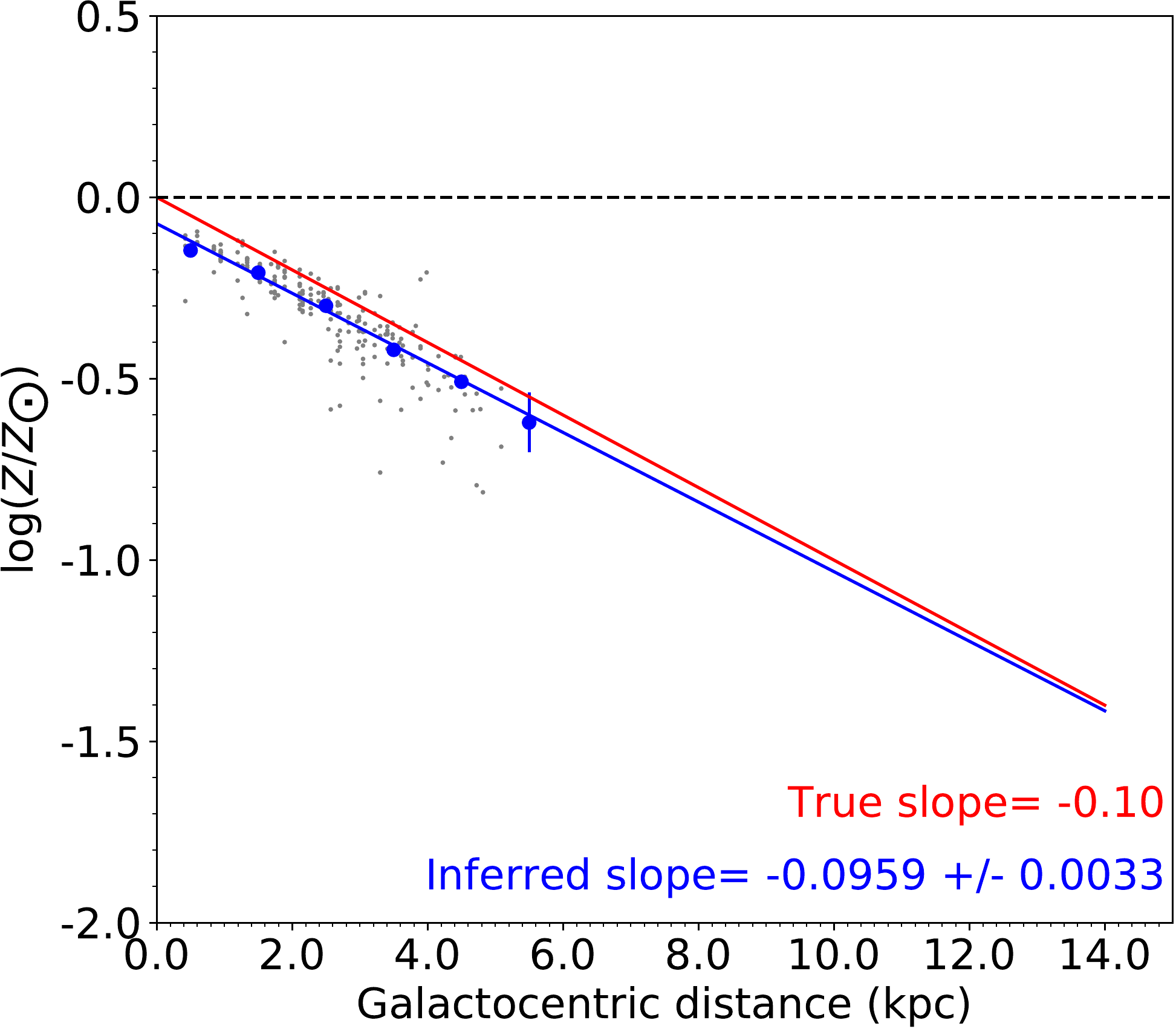}\llap{\makebox[7.1cm][l]{\raisebox{1.4cm}{\includegraphics[trim=3.6cm 2.0cm 4.2cm 0.4cm, clip, height=2.6cm]{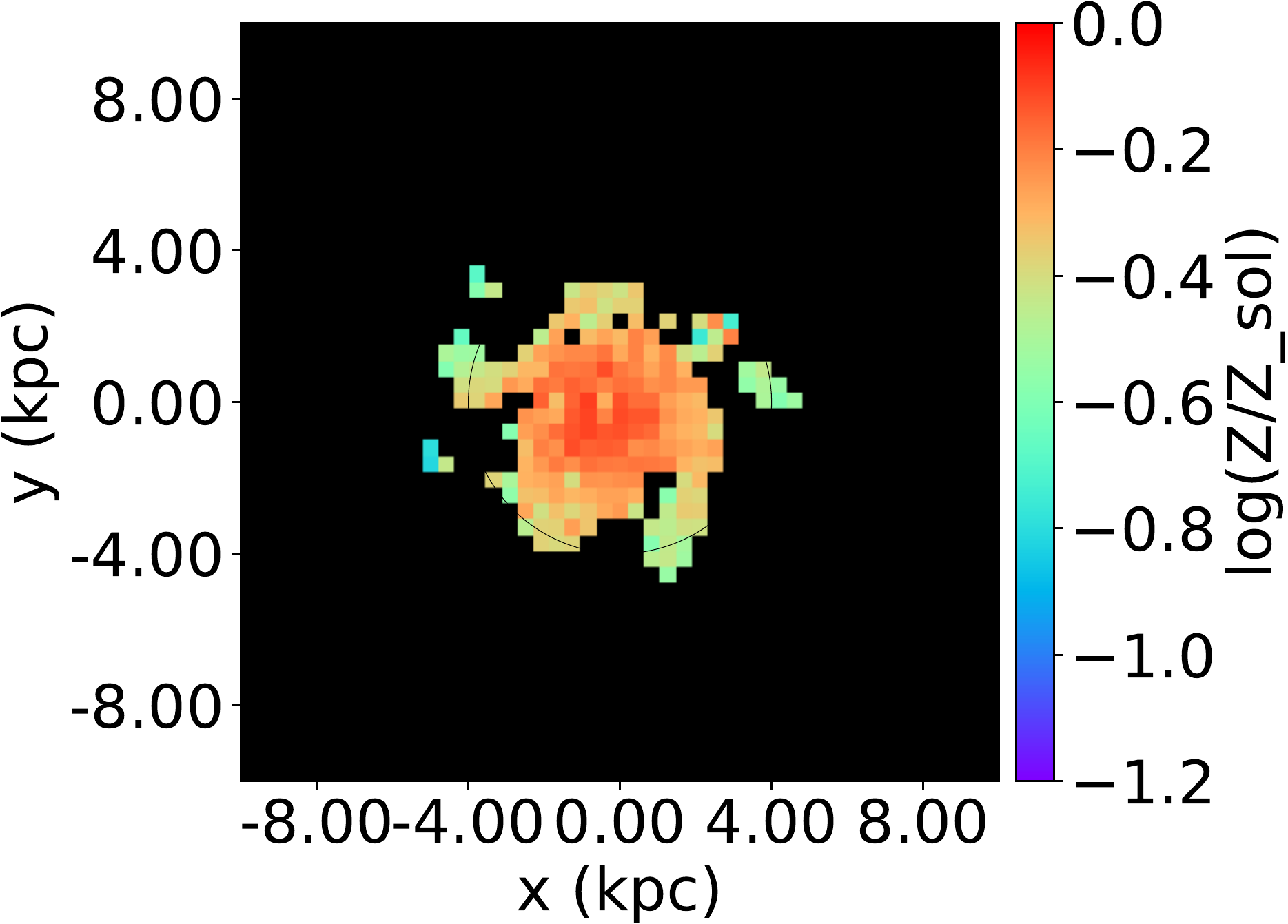}}}}\llap{\makebox[7.2cm][l]{\raisebox{6.2cm}{\includegraphics[trim=0.0cm 0.0cm 0.0cm 0.0cm, clip, height=0.6cm]{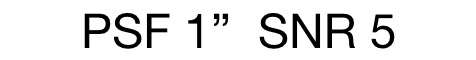}}}}
	\end{tabular}}
	\caption{Sample results for four different combinations of resolution and SNR. In each panel, the grey points denote the metallicity of each spatial pixel as a function of the galactocentric radius; we compute error bars on each of these points, but in the plot we have suppressed them for clarity. In each panel we have discarded pixels with SNR $<$ 3, for each emission line map involved in the metallicity diagnostic. The blue points with error bars are uncertainty-weighted mean metallicities in radial bins of 1 \kpc. The red line denotes the true metallicity as a function of radius, while the blue line is the uncertainty-weighted best linear fit to the grey points. The slopes of the red and blue lines are quoted in each panel as the input and inferred metallicity gradients, in units of dex/kpc. The corresponding 2D metallicity map is shown in the insets. The spatial span of the 2D map is the same as shown in \autoref{fig:enzo}. The colour coding is based on metallicity and ranges from 0 to $-1.2$ in $\log{(Z/Z_\odot)}$ space. From top to bottom, the size of the PSF increases from 0.15" to 1". From left to right the SNR/spaxel deteriorates from 300 to 5. Therefore, the top left panel represents a close-to-ideal case, whereas the bottom right panel is the most realistic scenario.
	} 
	\label{fig:met_maps}
	\end{figure*}

	\begin{figure*}
	\centering
	\includegraphics[width=1.0\linewidth]{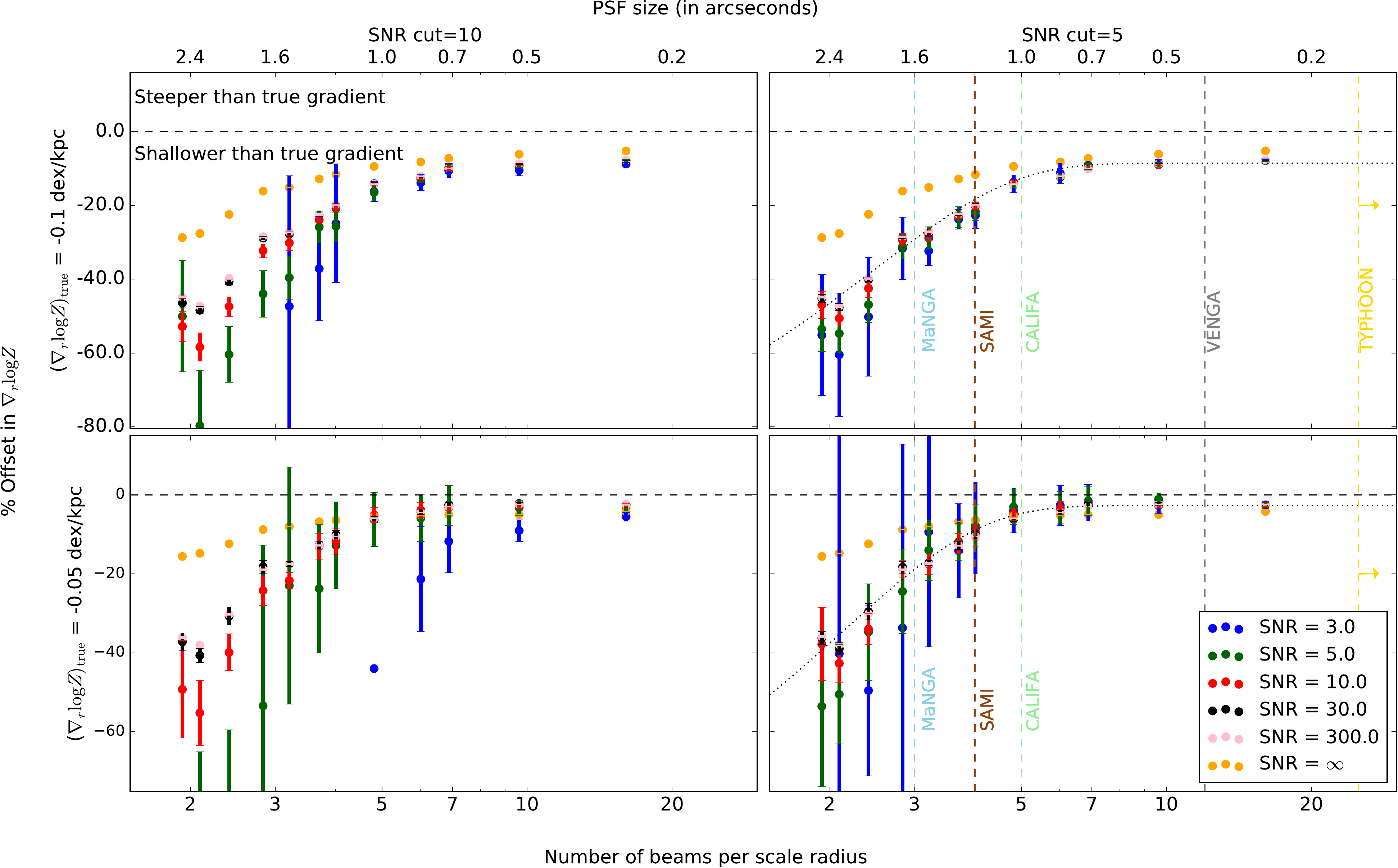}
	\caption{
		Error in inferred metallicity gradient as a function of spatial resolution and SNR. Each point and error bar show the mean and variance of our 10 realisations of the model. The \textit{top} and \textit{bottom} rows correspond to input gradients of $-0.1$ dex/kpc and $-0.05$ dex/kpc respectively. The \textit{left column} shows results for a fit using only pixels with an estimated SNR of 10 or more in each emission line map, whereas the \textit{right column} to a SNR cut-off of 5. The points in each panel show the relative offset (in percentage) between the inferred metallicity gradient and the input gradient, as a function of spatial resolution. The horizontal black dashed line indicates zero error, i.e., inferred metallicity gradient matches true gradient. Points above and below this line signify inferred gradients steeper and shallower than the true gradient, respectively. The coloured points correspond to input SNR values of $3-30$, as indicated in the legend. Blue points on the right column imply datacubes that have a mean SNR = 3 at the scale-length but the measurements have been performed after discarding all pixels with SNR < 5. The black dotted curve is our best-fitting function (see \autoref{sec:quantfit}). Vertical lines on the right column indicate typical spatial resolution levels of some of the current IFU surveys, with lower limits on the resolution marked with an arrow. We compute these typical resolution values by considering our model galaxy (r$_{\mathrm{scale}}$ = 4 {\kpc}) at the mean redshift of the given survey.
	}
	\label{fig:res_metgrad_DD0600}
	\end{figure*}

	\begin{figure*}
	\centerline{
		\def\arraystretch{0.1}
		\setlength{\tabcolsep}{0.5pt}
		\begin{tabular}{rl}
			\includegraphics[trim=0.0cm 1.3cm 2.9cm 0.0cm, clip, height=0.272\textheight]{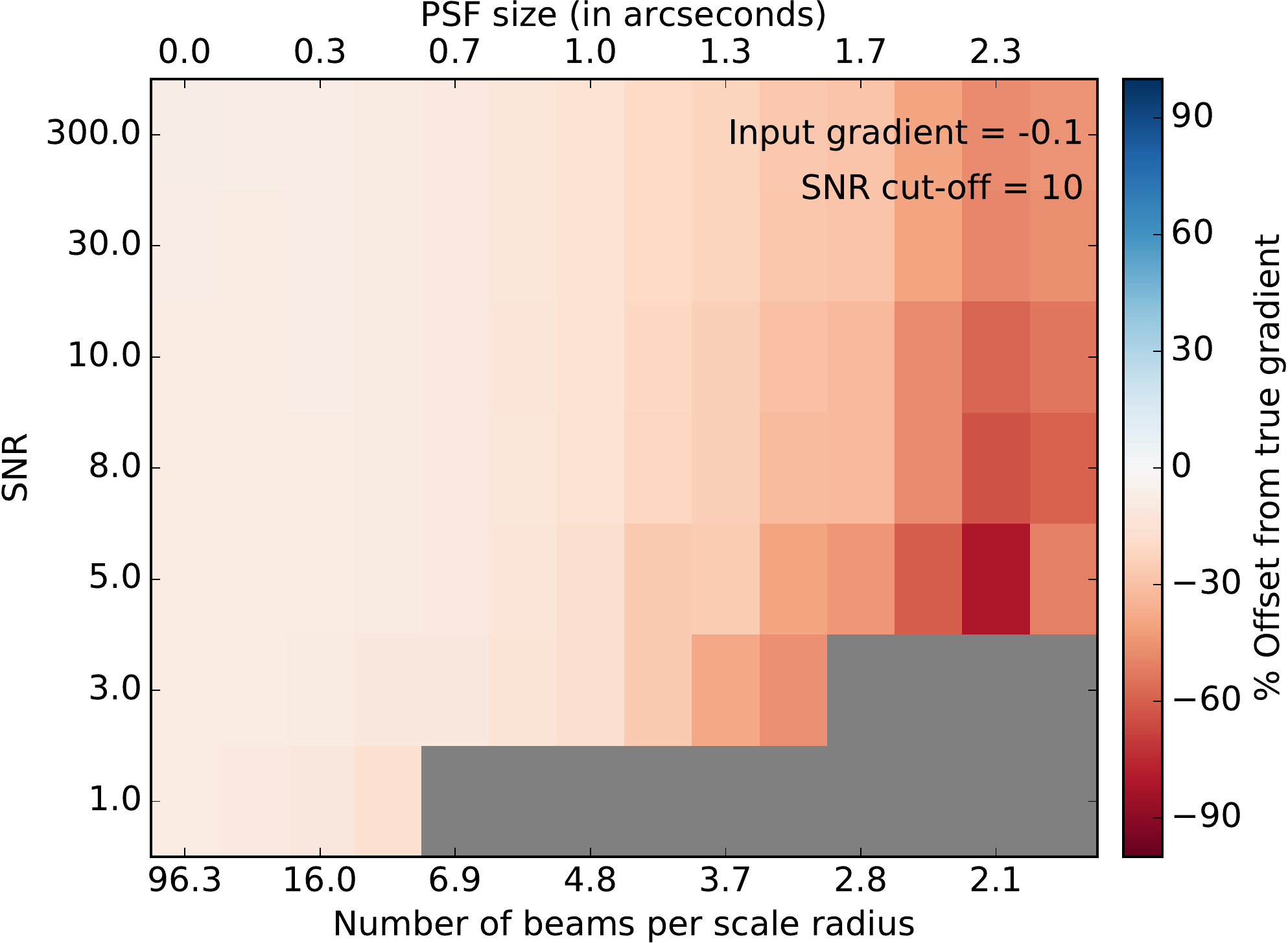} &
			\includegraphics[trim=1.9cm 1.3cm 0.0cm 0.0cm, clip, height=0.272\textheight]{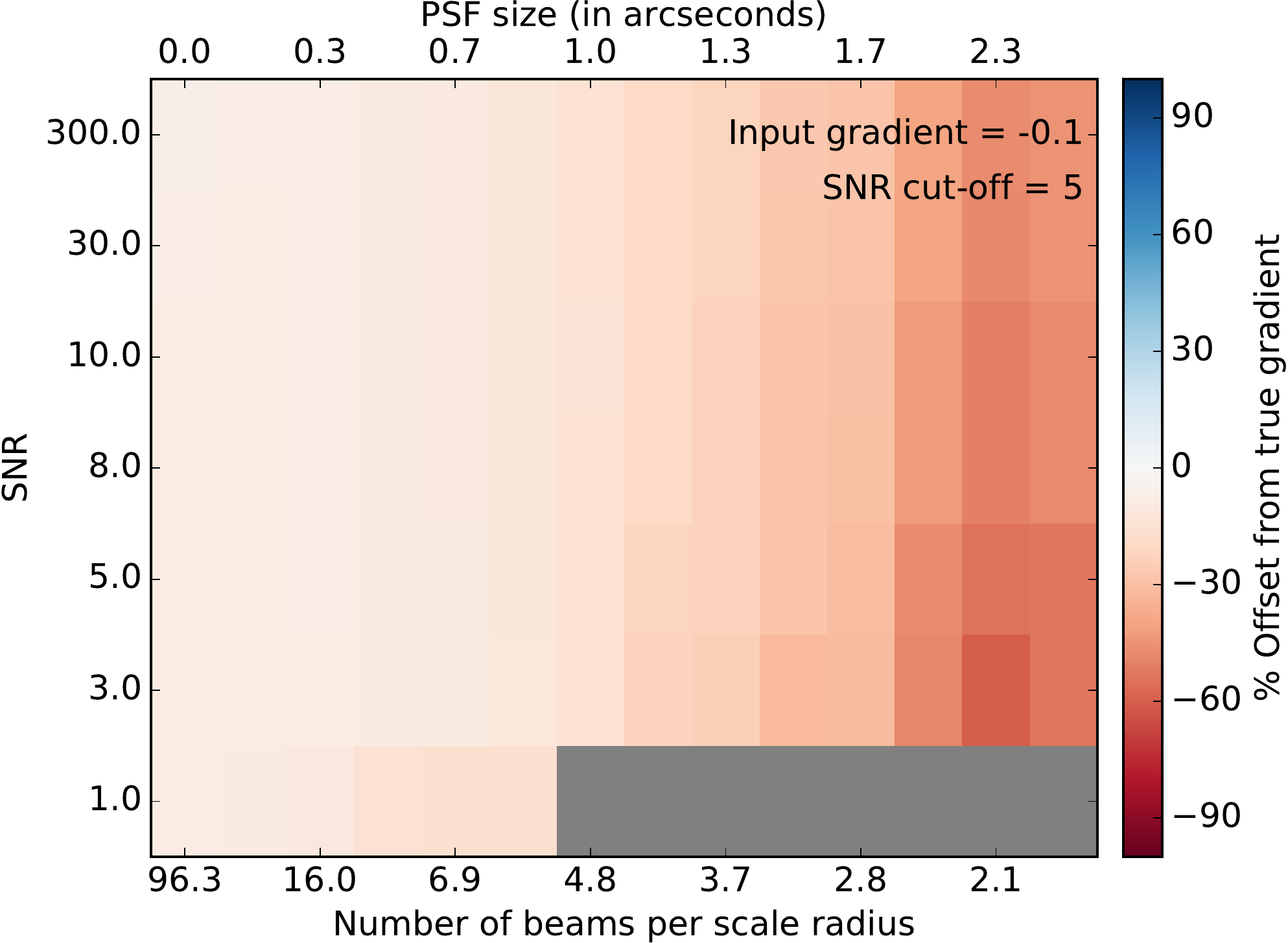} \\
			\includegraphics[trim=0.0cm 0.0cm 2.9cm 1.0cm, clip, height=0.278\textheight]{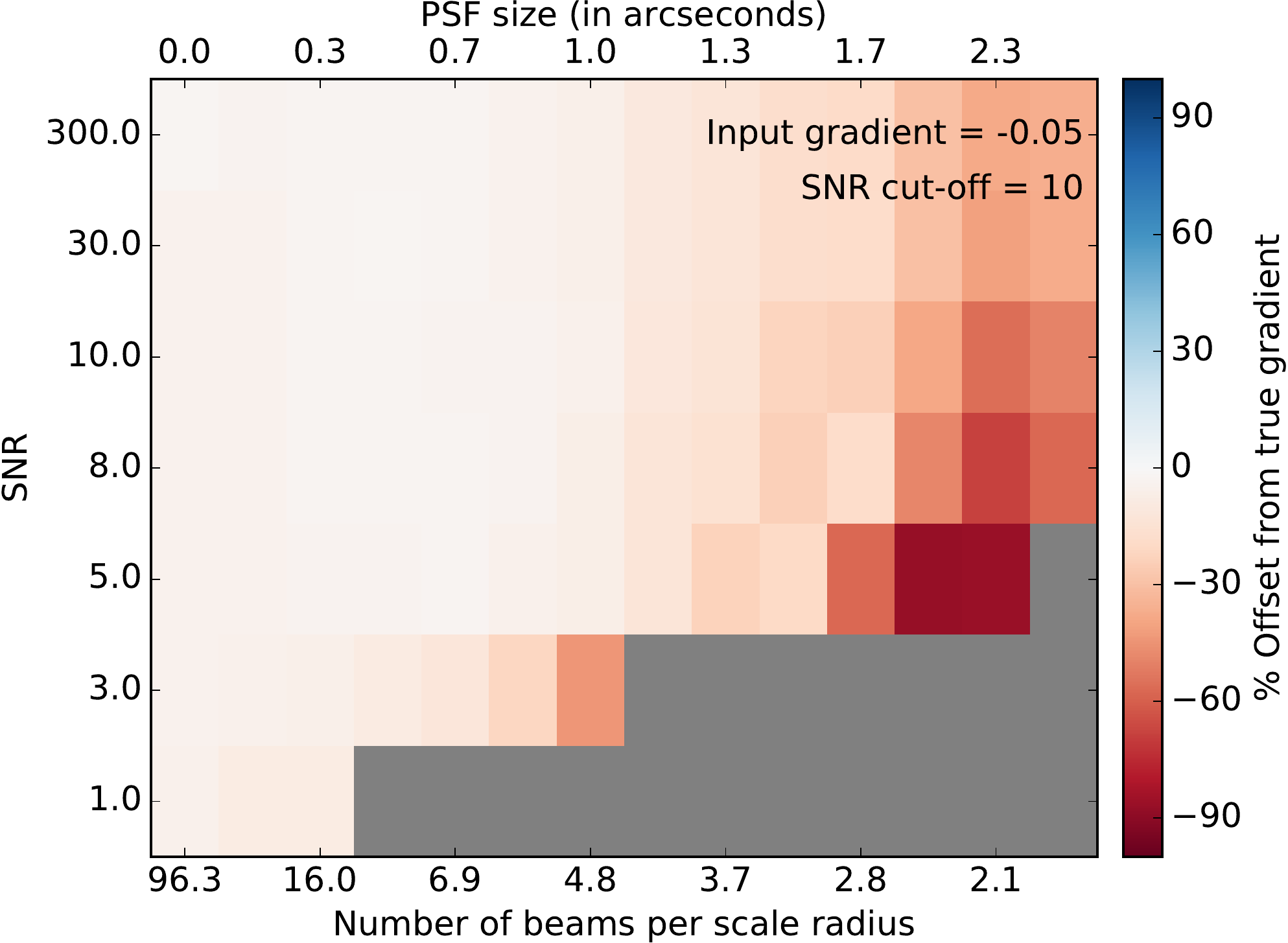} &
			\includegraphics[trim=1.9cm 0.0cm 0.0cm 1.0cm, clip, height=0.278\textheight]{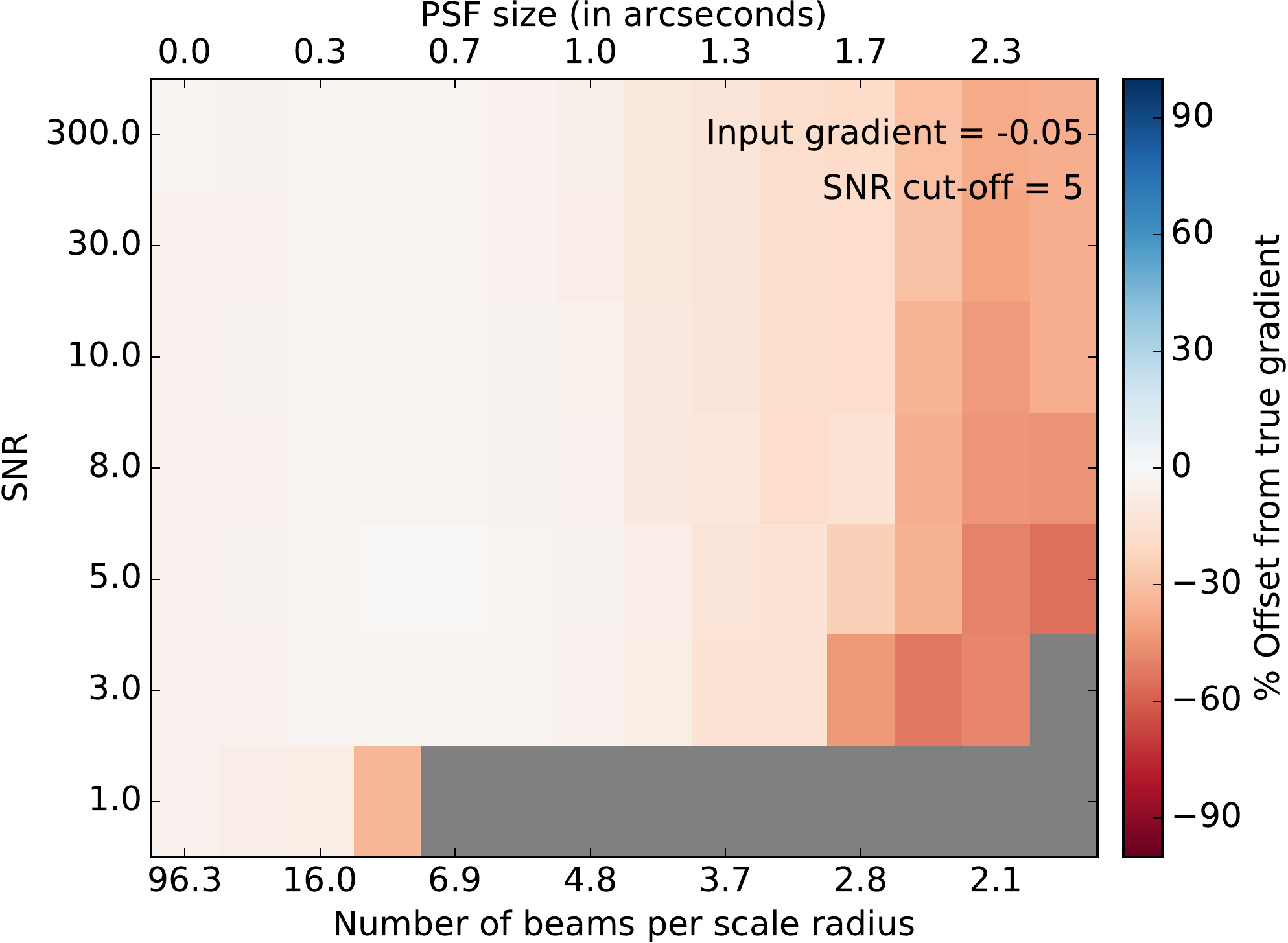}
	\end{tabular}}
	\caption{
		Error in inferred metallicity gradient as a function of resolution and SNR; the results shown here are a superset of those shown in \autoref{fig:res_metgrad_DD0600}. The top and bottom rows correspond to input metallicity gradients of $-0.1$ and $-0.05$ dex/kpc, respectively. The right column shows results for fits using only pixels with SNR of 5 or more in each emission line, while the left column shows results corresponding to SNR cut-off = 10. Grey cells correspond to combinations of resolution and SNR whereby no pixels reached the threshold SNR, and thus no data were available to which to fit the metallicity gradient.
	}
	\label{fig:res_heatmaps_DD0600}
	\end{figure*}

		\begin{table*}{
			\centering
			\begin{tabular}{lllllllllllllll}
\toprule
&\multicolumn{14}{c}{SNR cut-off = 10 }\\ 
\hline 
SNR&\multicolumn{14}{c}{Number of beams per scale radius}\\ 
\hline 
       &   96.26 &    32.09 &    16.04 &     9.63 &     6.88 &     6.02 &     4.81 &      4.01 &     3.70 &     3.21 &     2.83 &     2.41 &      2.09 &     1.93 \\
\midrule
\hline 
&\multicolumn{14}{c}{Input gradient = -0.1 dex/kpc}\\ 
\hline 
   1.0 &  $-$8.1 &   $-$9.5 &  $-$10.2 &  $-$16.1 &        - &        - &        - &         - &        - &        - &        - &        - &         - &        - \\
   3.0 &  $-$7.9 &   $-$8.3 &   $-$8.8 &  $-$10.6 &  $-$10.6 &  $-$13.8 &  $-$16.5 &   $-$26.1 &  $-$38.8 &  $-$45.6 &        - &        - &         - &        - \\
   5.0 &  $-$7.9 &   $-$8.1 &   $-$7.8 &   $-$9.2 &   $-$9.7 &  $-$12.9 &  $-$16.6 &   $-$25.9 &  $-$25.6 &  $-$39.5 &  $-$44.5 &  $-$60.6 &   $-$80.8 &  $-$50.1 \\
   8.0 &  $-$7.9 &   $-$7.9 &   $-$7.8 &   $-$9.0 &   $-$9.9 &  $-$12.0 &  $-$15.1 &   $-$21.3 &  $-$24.6 &  $-$31.7 &  $-$32.6 &  $-$47.4 &   $-$63.6 &  $-$58.8 \\
\bottomrule
\end{tabular}
			\caption{Relative offsets of the inferred gradient with respect to the input gradient for the full parameter space we studied, using the procedure of fitting the gradient using all pixels. Relative offsets in the table are expressed as a percentage of the true value, with the sign chosen so that shallower gradients correspond to negative values. Thus for example if the true gradient is $-0.1$ dex/kpc  and the measured gradient is $-0.09$ dex/kpc, we report a value of $-10$ in the table. The SNR quoted in the first column denotes the output SNR i.e. the SNR measured at the scale length of the synthetic datacube. The output SNR is usually lower than the input SNR, due to the additional uncertainty introducied by continuum subtraction (\autoref{sec:noise}). We fit the gradient using only pixels with SNR $\geq$ 10, in this case.
			}
			\label{tab:res_DD0600}}
	\end{table*}
	
	\begin{table*}{
			\centering
			\begin{tabular}{lllllllllllllll}
\toprule
&\multicolumn{14}{c}{SNR cut-off = 5 }\\ 
\hline 
SNR&\multicolumn{14}{c}{Number of beams per scale radius}\\ 
\hline 
       &   96.26 &    32.09 &    16.04 &     9.63 &     6.88 &     6.02 &     4.81 &     4.01 &     3.70 &     3.21 &      2.83 &     2.41 &     2.09 &     1.93 \\
\midrule
\hline 
&\multicolumn{14}{c}{Input gradient = -0.1 dex/kpc}\\ 
\hline 
   1.0 &  $-$8.0 &   $-$9.0 &  $-$10.3 &  $-$15.5 &  $-$17.1 &  $-$17.1 &        - &        - &        - &        - &         - &        - &        - &        - \\
   3.0 &  $-$7.9 &   $-$7.9 &   $-$7.8 &   $-$8.7 &   $-$9.1 &  $-$11.1 &  $-$14.2 &  $-$23.3 &  $-$23.8 &  $-$32.5 &   $-$31.8 &  $-$48.8 &  $-$60.5 &  $-$53.8 \\
   5.0 &  $-$7.9 &   $-$8.0 &   $-$7.7 &   $-$9.0 &   $-$9.3 &  $-$12.2 &  $-$14.3 &  $-$21.9 &  $-$22.8 &  $-$28.5 &   $-$31.2 &  $-$47.0 &  $-$54.7 &  $-$53.4 \\
   8.0 &  $-$7.9 &   $-$7.9 &   $-$7.8 &   $-$8.9 &   $-$9.6 &  $-$11.8 &  $-$14.1 &  $-$20.1 &  $-$22.7 &  $-$28.4 &   $-$30.2 &  $-$42.4 &  $-$50.5 &  $-$47.3 \\
\bottomrule
\end{tabular}
			\caption{Same as \autoref{tab:res_DD0600} but now fitting the gradient using only pixels with SNR $\geq$ 5. Missing values in the table correspond to cases where the imposed SNR cutoff leaves too few pixels to make it possible to fit a metallicity gradient.
			}
			\label{tab:res_DD0600_snrcut5}}
	\end{table*}
	
	\subsection{Model grid}\label{sec:model_grid}	
	We compute a grid of synthetic IFU datacubes, covering a range of four true metallicity gradients ($-0.1$, $-0.025$, $-0.05$, $-0.01$ dex/kpc), 14 PSF sizes (PSF = [0.05", 0.15", 0.3", 0.5", 0.7", 0.8", 1.0", 1.2", 1.3", 1.5", 1.7", 2.0", 2.3", 2.5"], approximately corresponding to [96.4, 32.1, 16.1, 9.6, 6.9, 6.0, 4.8, 4.0, 3.7, 3.2, 2.8, 2.4, 2.1, 1.9] beams per scale length and [0.04, 0.1, 0.2, 0.4, 0.6, 0.7, 0.8, 1.0, 1.1, 1.2, 1.4, 1.7, 1.9, 2.1] kpc, respectively, sampled at 2 pixels per beam) and seven values of intrinsic SNR of the datacube defined at the scale radius (SNR = [1, 3, 5, 8, 10, 30, 300]). The range of SNR values comfortably covers the observed parameter space for nearby galaxy surveys e.g. MaNGA \citep{Mingozzi:2020aa} and SAMI. We run our models up to large PSF sizes in order to sample the coarse end of the resolution satisfactorily. For each grid point, i.e. for each combination of the above parameters, we compute ten datacubes, the noise being drawn from a different random realisation in each case. Throughout the paper, we report only the mean metallicity gradient measurement of the ten realisations (and the standard deviations as the uncertainties) because the formal uncertainties of the fitted gradient are generally smaller than the variation between different realisations.
	
	\autoref{fig:met_maps} illustrates the next step for a combination of two different PSF sizes (top and bottom) and SNR values (left and right). For each datacube, once we have the metallicity map (\autoref{sec:met}), we fit the pixels (grey points) with a simple linear (in $\log{(Z/Z_{\odot})}$) radial metallicity profile (blue line). All pixels are used for the fitting unless a SNR cut-off is imposed (see below). We propagate the uncertainties in the emission line fluxes through the metallicity diagnostic in order to obtain uncertainties on the metallicity values. The metallicity uncertainty in each pixel is then used to assign pixels relative weights for the radial fit. 
	Propagation of uncertainty is important even in the limiting case of very high SNR, because although the absolute uncertainties in the metallicity of each pixel would be small for very high intrinsic SNR of the data cube, it is the relative uncertainties of the outer-to-inner pixels that affect the fit.

	\subsection{Summary of results}\label{sec:ressum}
	\autoref{fig:res_metgrad_DD0600} shows our main results and Tables~\ref{tab:res_DD0600} and \ref{tab:res_DD0600_snrcut5} quote the values for the full parameter space of our study. Our main finding is that coarser spatial resolution leads to an increasingly shallower metallicity gradient inferred in observations. We return to the question of why lower spatial resolution always has a flattening effect on the inferred metallicity gradient in \autoref{sec:ifu}.

	\autoref{fig:res_metgrad_DD0600} also shows that for a given spatial resolution, the accuracy of the inferred gradient saturates above a certain SNR, i.e. for the limiting case of infinite SNR, there is a best possible metallicity gradient that can be recovered at a given spatial resolution. However, this best possible value is achieved at a SNR $\sim$30 for all practical scenarios. As described before in \autoref{sec:noise}, we define SNR of a datacube as the mean SNR in the \mbox{(0.9 - 1.1)$\,\times\,r_{\mathrm{scale}}$} annulus of the {\nii} emission line map. The value of best possible accuracy depends on the true metallicity gradient and the metallicity diagnostic employed (see \aref{sec:ap} for a discussion) because different diagnostics have varying sensitivities in the metallicity regime of interest.

	We also find that, if one fits the metallicity gradient using all pixels, a lower SNR leads to a shallower inferred metallicity gradient (\autoref{fig:res_metgrad_DD0600}). While correct propagation of uncertainties that down-weights low SNR pixels in the outer parts of galaxies partly counters this effect, at low SNR this correction is imperfect, and the inferred metallicity gradient remains too shallow (right column of \autoref{fig:met_maps}). One way to avoid this problem, which is widely practised by the community, is to discard the noisy pixels based on a threshold SNR cut-off, i.e., fit the metallicity gradient of the galaxy using only those pixels for which the SNR that we infer is above some threshold. When we implement this procedure using a threshold SNR of 5 for each emission line involved in the diagnostic, inferred metallicity gradients are close to the best possible accuracy ($\sim$10-15\,\%) even when the overall SNR is low, as seen in the right column of \autoref{fig:res_metgrad_DD0600}. We select the threshold of 5 because experimentation shows that this is sufficient to reproduce the best possible accuracy for most cases. However, this improvement has a price: for data cubes with low overall SNR, excluding pixels with SNR below 5 from the analysis might not leave enough pixels for a reliable fit, leading to large uncertainties. We can see the interplay of these effects in our model datacubes with the lowest overall SNR (SNR = 3). If we fit the metallicity gradients in these cases using all pixels (i.e., without imposing a SNR cutoff), we infer gradients that are drastically different from the true values and exhibit a non-uniform trend with spatial resolution (blue data in left column of \autoref{fig:res_metgrad_DD0600}). This is because these low SNR (i.e. more noisy) datacubes lead to larger uncertainties during the fitting processes. Discarding pixels with SNR below 5 (right column of \autoref{fig:res_metgrad_DD0600}), brings the results close to the best possible inferred gradient when the spatial resolution is high and thus there are a large number of independent pixels, ensuring that enough will have high SNR to allow reliable estimation of the gradient. However, at low spatial resolution applying a cut in SNR leaves so few pixels that the error in the mean result, and the dispersion between different noise realisations, is actually worse than if no SNR cut had been applied.

	\autoref{fig:res_heatmaps_DD0600} summarises all these effects by showing heat maps for our full parameter space, corresponding to the data in Tables~\ref{tab:res_DD0600} and \ref{tab:res_DD0600_snrcut5}. The colour map denotes the relative offset between the inferred and true gradients, in percentage. Given an observed IFU datacube of some intrinsic SNR and PSF size, one can effectively read off (from Tables~\ref{tab:res_DD0600} and \ref{tab:res_DD0600_snrcut5}, or \autoref{fig:res_heatmaps_DD0600}) by how much would the inferred metallicity gradient be offset from the true value. This is useful for planning large surveys, e.g., given the telescope resolution and source redshift, it is possible to calculate the exposure time required to achieve a certain SNR and thereby achieve a target accuracy in the measured metallicity gradient.
	
	\subsection{Quantifying the effect of spatial resolution}\label{sec:quantfit}
	In order to quantify the effect of the spatial resolution on the inferred metallicity gradient, we fit the data in \autoref{fig:res_metgrad_DD0600} using a least square algorithm, with a function of the form
	\begin{eqnarray}\label{eq:fit}
	y(x) = a\left(\mathrm{erf}\left(\frac{\sqrt{\pi}x}{2b}\right) - 1\right) + c,
	\end{eqnarray}
	where $a$ and $b$ denote the steepness and the ``knee'' of the fit respectively, $c$ is the asymptotic offset of the fit from the true input gradient at high spatial resolution, $x$ is the number of spatial resolution elements per scale radius\add{,} and $y$ is the percentage offset of the inferred gradient from the true value. We choose this functional form because erf($z$) limits to a constant value for $z \rightarrow \infty$ and approximates to a linearly increasing function for $z\ll1$ , which resembles the behaviour of our data. We define the ``knee'' ($b$) as the intersection of the linear (= $a$($x/b$-1) + $c$) and constant (= $c$) parts of the fit function. If the scale length of an observed galaxy is resolved by $b$ beams or fewer, it implies that the inferred metallicity gradient is considerably offset from the most accurate measurement possible, given the diagnostic.

	\begin{table}{
		\centering
		\begin{tabular}{lllllll}
\toprule
Diagnostic & $\Omega$ & f$_g$ & ($\nabla_r \log{Z}$$)_{\rm true}$ &      a &    b &        c \\
\midrule
       D16 &      0.5 &   0.2 &                            $-$0.1 &   89.8 &  3.1 &   $-$8.5 \\
       D16 &      0.5 &   0.2 &                           $-$0.05 &   98.8 &  2.7 &   $-$2.7 \\
       D16 &      0.5 &   0.1 &                            $-$0.1 &   79.9 &  3.3 &   $-$6.7 \\
       D16 &      0.5 &   0.1 &                           $-$0.05 &  182.6 &  1.8 &   $-$3.2 \\
       D16 &      5.0 &   0.2 &                            $-$0.1 &   90.7 &  3.1 &  $-$11.0 \\
       D16 &      5.0 &   0.2 &                           $-$0.05 &   82.6 &  2.8 &   $-$5.8 \\
       D16 &      5.0 &   0.1 &                            $-$0.1 &   70.4 &  3.7 &   $-$7.4 \\
\bottomrule
\end{tabular}
		\caption{
			Best-fit parameters for our model of metallicity gradient error as a function of spatial resolution (equation 23 in the main text), for all choices of diagnostic, and all values of $\Omega$ (wind strength parameter), $f_g$ (simulation gas fraction), and $\nabla_r\log Z$ (true metallicity gradient). Note that the parameter $c$ represents the minimum possible error that can be achieved for a given set of parameters, i.e., it is the value reached in the limit of infinite spatial resolution and SNR.
		}
		\label{tab:quant}}
	\end{table}	

	We fit \autoref{eq:fit} to a data set consisting of all our inferred metallicity gradients at all SNRs and resolutions, derived using a SNR cutoff of 5 (i.e., the points shown in the right column of \autoref{fig:res_metgrad_DD0600}), and obtain best-fit values $a=100.5$, $b=3.1$, $c=-8.9$ and $a=123.0$, $b=2.6$, $c=-2.2$ for true metallicity gradients of $-0.1$ and $-0.05$ dex/kpc, assuming fiducial values for all other parameters. \autoref{tab:quant} gives an overview of the different choices we investigated and the best-fit parameters of the error in inferred gradient as a function of the spatial resolution, for each case. The ``knee'' of the function (parameter $b$) is similar for all cases, indicating that the dependence on resolution is qualitatively unaffected by the different choices of $\Omega$, $f_g$ and diagnostic (see \aref{sec:ap} for a detailed discussion). Using \autoref{eq:fit} and the values from \autoref{tab:quant}, it is possible to estimate the accuracy of the inferred metallicity gradient for a given spatial resolution. However, we caution the reader that our models do not include contributions from the Diffused Ionised Gas (DIG), a point to which we return in \autoref{sec:dig}. As a rough rule of thumb, we find that the scale length should be resolved by more than 4 beams for the inferred metallicity gradient to be close to the most accurate value possible, for a given diagnostic.

	\section{Discussion}\label{sec:disc}
	\subsection{Caveats}
		Here we discuss the physical processes that we have not accounted for in our work and how they could potentially impact our results. 
	
	\subsubsection{Diffused Ionised Gas}\label{sec:dig}
	An important caveat of our work is that we do not account for the emission from the diffuse ionised gas (DIG). The DIG consists of low-density ISM illuminated by the ionising photons that escape dense {\htrs} and appears to be ubiquitous across the ISM of galaxies \citep{Haffner:2009aa}. Poetrodjojo et al. (in prep) show that DIG emission leads to an increase in [\ion{N}{ii}]$\lambda$6584/H$\alpha$ and [\ion{S}{ii}]$\lambda\lambda$6717,30/H$\alpha$ emission line ratios \citep{Blanc:2009aa, Zhang:2017aa}, an effect that is larger at coarser spatial resolutions. As such, DIG is likely to have a considerable impact on the \citetalias{Dopita:2016aa} diagnostic, which makes use of these ratios. However, we have also tested our analysis with the \citet[][hereafter \citetalias{Kewley:2002fk}]{Kewley:2002fk} [\ion{N}{ii}]$\lambda$6584/[\ion{O}{ii}]$\lambda\lambda$3727,29 diagnostic (results in Section A3 of online material) which is less likely to be dominated by DIG effects. We intend to add the effects of DIG emission to our models in future work.
	
	\subsubsection{Dust}\label{sec:dust}
	We do not explicitly account for dust in our model for the ionised gas bubble. The predicted emission line fluxes for each {\htr} that we derive from the photoionisation model grid are effectively the dereddened values. We use the \citetalias{Dopita:2016aa} metallicity diagnostic which is robust to reddening by virtue of using emission lines that are narrowly spaced in wavelength. Using the \citetalias{Dopita:2016aa} indicator ensures that our results will hold, to first order, in real world scenarios where dust is present. For instance, very small Balmer decrements ($\log{(H_{\alpha}/H_{\beta})_{obs}} \lesssim$ 3 $\implies$E(B-V)$\lesssim$0.04) have been observed in high spatial resolution ($\lesssim$100 pc) IFU studies of face-on nearby spiral galaxies (TYPHOON; private communication) which have negligible effect ($\lesssim$0.001 dex) on the \citetalias{Dopita:2016aa} diagnostic.
	
	However, if multiple {\htrs} with different luminosities and reddening values lie along the line of sight, it could potentially lead to a second-order effect in the metallicity diagnostic. The precise magnitude of such an effect is out of the scope of this work but we present a qualitative argument here. For instance, in a scenario where {\htrs} are more embedded when they are younger and have harder spectra, not correcting for differential extinction can potentially lead to the metallicity being overestimated (by missing the young, dust-obscured, low metallicity {\htrs}) in relatively more dusty regions of the galaxy. The resultant effect on the inferred metallicity profile would depend on the spatial distribution of dust in the galaxy. If the dust distribution follows a smooth, radially declining profile, the central metallicity would be overestimated the most, resulting in artificial steepening of the gradient.
	
	We do not model the Balmer absorption because we do not account for the underlying old stellar populations in the galaxy in determining the continuum. For this work, we consider only the young (< 5 {\Myr}) stellar associations which contribute to most of the ionising flux that drives the {\htrs}. Although our continuum subtraction prior to line-fitting ensures that the overall shape of the continuum does not impact the emission line ratios, the lack of the old stellar population and the Balmer decrement potentially underestimates the uncertainties originating from continuum subtraction. Consequently, the uncertainties in the line fluxes and the metallicity gradients are potentially underestimated due to lack of dust.

	\subsubsection{Variation of Wind Strength}\label{sec:omega}

	We parameterise the relative strength of stellar winds versus gas and radiation pressure via the quantity $\Omega$. Quantitatively, $\Omega$ is the ratio of the volume occupied by the inner stellar wind cavity to that occupied by the ionised {\htr} shell. A high $\Omega$ implies a thin shell of ionised gas, whereas a low $\Omega$ value corresponds to an almost spherical ionised bubble with a negligible stellar cavity. The value of $\Omega$ is potentially important for line diagnostics because it changes the ionisation parameter at fixed {\htr} size, and because it changes the amount of spherical divergence that the radiation field undergoes as it propagates through the {\htr}; \citet{Yeh:2012aa} show that the value of $\Omega$ has observable effects on infrared line ratios, for example.
	
	Our fiducial value, $\Omega=0.5$, is primarily motivated by observations. In their analysis of starburst galaxies, \citet{Yeh:2012aa} find that observed infrared line ratios are best explained by $\Omega \sim 0.5$. Similarly, \citet{Lopez:2011aa, Lopez:2014aa} use X-ray observations of Magellanic Cloud {\htrs} to constrain the fractional contribution $f_X$ of the wind bubble to the overall volume-averaged pressure; assuming that the {\htr} interior is isobaric, $f_X$ and $\Omega$ are related simply by $\Omega = f_X / (1-f_X)$ (see the Appendix of \citet{Lopez:2014aa} for further discussion). They find that $f_X \sim 0.1 - 0.5$ in the LMC, again consistent with our fiducial choice; for the SMC, $f_X < 0.1$, likely because the metallicity is too low for the stars to have substantial winds. Finally, in the Milky Way, \citet{Draine:2011b} notes that the nearly-spherical N49 bubble has a maximum surface brightness that is a factor of $\approx 2$ higher than its central surface brightness, a ratio that is too large to be explained by radiation pressure effects alone, but is consistent with what is expected if the ionised gas is confined to a thick shell by a wind bubble. If we approximate the photoionised region as having uniform emissivity, so that the observed surface brightness variations are solely due to variations in the path length through the shell along different lines of sight, then a maximum-to-central surface brightness ratio of 2 corresponds to a shell where the wind bubble occupies the inner $\Omega = 0.28$ of the volume, close to our fiducial value.
	
	Estimates of $\Omega$ from theory are significantly more diverse. One-dimensional simulations often indicate that stellar winds should dominate the pressure of {\htrs}, implying large $\Omega$ \citep[e.g.,][]{Fierlinger:2016aa, Rahner:2017aa, Silich:2018aa}. However, multi-dimensional simulations usually imply small values of $\Omega$, due to efficient leakage of hot gas or mixing-induced radiative cooling \citep[e.g.,][]{Mackey:2015aa, Calura:2015aa, Wareing:2017aa}. The 1D and multi-dimensional simulations are somewhat difficult to compare, because the 1D models explore a much larger range of parameter space, and often find winds to be important primarily in the parts of parameter space -- very massive and dense clusters -- that are most inaccessible to multi-dimensional simulations. Energetic analyses even of large {\htrs} tend to imply that there must be significant loss of wind energy due to breakout or mixing-induced cooling \citep[e.g.,][]{Harper-Clark:2009aa, Rosen:2014aa, Rogers:2014aa, Toala:2018aa}, which would suggest $\Omega \lesssim 1$, but this is hard to quantify. For further discussion of this issue, we refer readers to the recent review by \citet{Krumholz:2019aa}.
	
	Given that there is some uncertainty about the true value of $\Omega$, it is worthwhile to explore how variations in its value could affect our results. We do so by performing two different tests. First, we construct model galaxies in which $\Omega=0.05$ and $5.0$ (compared to our fiducial choice $\Omega=0.5$) independent of radius within the galaxy. Second, we construct a model galaxy in which $\Omega$ varies radially from $5.0$ at the centre to $0.05$ at $2R_e$.
	
	Figures S3 and S4 in the Supplementary Online Material show the results for $\Omega$=0.05 and 5.0, which correspond to a nearly spherical and a thin shell geometry, respectively. Comparing with \autoref{fig:res_metgrad_DD0600} we see no qualitative difference in the results, and nearly identical dependence on resolution. The only significant effect of varying $\Omega$ is to slightly increase the absolute error, so that a few percent error remains even in the limit of high SNR and resolution. The origin of this effect is easy to understand:
	 the {\htr} models used to construct the diagnostic implicitly assume a value of $\Omega$ via the geometry they adopt. While the diagnostics attempt to minimise the effects of geometry, they cannot do so perfectly, leading to a persistent error if the geometry is not exactly what was assumed, even if the measured line strengths are perfect. This variation in the best possible accuracy produces an overall vertical offset to the data depending on $\Omega$. As mentioned before, the vertical offset also depends on other factors, such as the DIG, that has not been accounted for in this work. However, variations in $\Omega$ have only very minor effects on the resolution-dependence of the error. Quantitatively, when we use \autoref{eq:fit} to fit the resolution-dependent error in our recovered metallicity gradients for $\Omega = 0.05$ or $\Omega = 5$, the coefficients $a$ and $b$ that describe the shape and location of the ``knee'' in the curve change by at most $\approx 10\%$ (see \autoref{tab:quant}). We therefore conclude that, even if $\Omega$ takes on a wide range of values, our conclusions about the resolution-dependence of metallicity errors change by at most $\approx 10\%$. Full tables and heatmaps for cases with different $\Omega$ values are  provided as Supplementary Online Material.
	
	 Figure S7 in the Supplementary Online Material compares our model galaxy with $\Omega$ varying with radius to our fiducial case with fixed $\Omega=0.5$. As with the case of varying $\Omega$ independent of radius, we find no qualitative difference between the two cases. We refer the readers to the supplementary material for details of the variable $\Omega$ models.

	\subsection{Dependence on metallicity diagnostic}\label{sec:compdiag}

	Switching between metallicity diagnostics usually introduces an offset in the absolute value of the metallicity determined \citep{Kewley:2008aa}. The absolute magnitude of such offset is not uniform across the metallicity range typically observed in galaxies. Therefore, different diagnostics can potentially introduce slight deviations to the inferred slope. To test the extent of this effect we re-compute our results with the [\ion{N}{ii}]$\lambda$6584/[\ion{O}{ii}]$\lambda\lambda$3727,29 (N2O2) diagnostic following \citetalias{Kewley:2002fk} (as opposed to the \citetalias{Dopita:2016aa} diagnostic used in the main text), keeping everything else unchanged. We note, however, that the absence of a treatment for dust could affect the N2O2 diagnostic due to the inclusion of the [\ion{O}{ii}] lines. In order to use the [\ion{O}{ii}]$\lambda\lambda$3727,29 emission lines for the \citetalias{Kewley:2002fk} diagnostic, we produce an additional set of synthetic datacubes (and corresponding emission line maps) spanning the wavelength range 3670 - 3680 \AA~following the same procedure as described in Sections~\ref{sec:ifu} and \ref{sec:fitting}.
	
	Figures~S8 and S9 (and Tables~S7 and S8) of the Supplementary Online Material (SOM) present the results using the \citetalias{Kewley:2002fk} diagnostic. We find no qualitative difference in our results on comparing Figure S7 of SOM with \ref{fig:res_metgrad_DD0600}. The figures show that the absolute offset of the accuracy of inferred gradient (parameter $c$ in \autoref{tab:quant}) depends on the diagnostic used. However, the accuracy as a function of the spatial resolution, including the `knee' of the functional form (parameter $b$ in \autoref{tab:quant}), remains largely unchanged. The discrepancy in the vertical offset is due to different sensitivities of the different diagnostics in the metallicity regime concerned.

	\subsection{Comparison with previous work}\label{sec:compprev}
	
	Several authors have previously examined the question of how well observations at finite resolution and sensitivity can recover galaxy metallicity gradients. Most recently,
	\citet[][hereafter \citetalias{Carton:2017aa}]{Carton:2017aa} proposed a forward modelling approach to correct for the degrading effects of seeing and spatial binning on metallicity gradients. They find that coarser spatial resolution yields a systematically steeper metallicity gradient (Figure 4 of \citetalias{Carton:2017aa}) which is contrary to what we observe in our models, and to what one would expect based on an intuitive understanding of how line ratio diagnostics operate. Metallicity diagnostics are nonlinear functions of the ratio of a weaker nebular line to a stronger one. Consequently, in a galaxy where metallicity falls with radius, the true radial gradient of the weaker line tends to be steeper than the true radial gradient of the stronger line. Beam-smearing affects the line with the steeper gradient more severely, and so smearing tends to reduce the value of the diagnostic line ratio in the galaxy centres, and raise it in galaxy outskirts. This in turn leads one to infer an artificially shallower gradient. \citetalias{Carton:2017aa} point out that their contrary-to-expectation variation of the inferred gradient could arise from the systematics of their modelling. We can identify three significant differences between their approach and ours that might be responsible for this difference.
	
	First and most simply, their choice of range of the PSF variation corresponds to $\sim 0.25-1$ beams per scale length of the galaxy, which is towards the extreme coarse end of our range of spatial resolutions (\autoref{fig:res_metgrad_DD0600}). Second, \citetalias{Carton:2017aa} use an analytic prescription for the spatial distribution of star formation with a galaxy, while we use a distribution taken from a self-consistent, high-resolution hydrodynamic simulation. This leads to large differences in the morphology of emission: our galaxy has a clumpy, spiral structure similar to that of observed disc galaxies. \citetalias{Carton:2017aa} point out that a clumpy morphology can have significant impact on interpretation of metallicity gradients, but are unable to capture this effect with their method. Third, \citetalias{Carton:2017aa} implement a noise model that depends solely on the emission line fluxes, with no sky contribution. As such, the spatial distribution of SNR within their model galaxy is uniform, which is not necessarily observed in IFU surveys. In addition to the flux-dependence, the noise in observed galaxies is dictated by instrumental noise and contribution from the sky background. The latter two sources of \textit{noise} have no real spatial variation, implying an overall radially declinig \textit{SNR} profile in observed emission line maps. We account for this by implementing a wavelength-dependent noise model that takes into account both the instrumental read noise and the sky noise contribution, in addition to the flux-dependence of noise. As we have shown in this paper, and as \citetalias{Carton:2017aa} have also discussed, accounting for the SNR is crucial to inferring metallicity gradients. We fit the noise-weighted metallicity of each pixel to a smooth radial profile, implying less relative weight on the pixels at the outskirts. But in the \citetalias{Carton:2017aa} noise model the relative weight between pixels would effectively be uniform, which could potentially impact their metallicity gradient inference. Another potential cause for the steepening of metallicity gradient in \citetalias{Carton:2017aa} could be the systematic offsets in their models which tend to infer steeper output gradients for larger negative input gradients (their Figure B1) due to model degenracies stemming from the finite extent of available photionisation grids.
	
	Our work shows better agreement with that of \citet{Yuan:2013aa}, who also find that coarser spatial resolution (or coarser annular binning) leads to flatter inferred gradients. They suggest this flattening is due to selective smearing of the low SNR pixels of the weak [\ion{N}{ii}]$\lambda$6584 line, thereby leading to an overestimate of the \ion{N}{ii} flux in the outskirts of the galaxy, flattening the gradient. Their argument suggests that we should see enhanced flattening for steeper intrinsic gradients, which is entirely consistent with the results shown in \autoref{fig:res_metgrad_DD0600} and the argument made in \autoref{sec:ifu}. However, \citeauthor{Yuan:2013aa} also do not find any appreciable effect of SNR on the inferred gradient, whereas we derive systematically flatter gradients for lower SNR data cubes. This is likely because \citeauthor{Yuan:2013aa} impose a SNR cut-off on the \ion{N}{ii} line, which, as we show in the second column of \autoref{fig:res_metgrad_DD0600}, eliminates the SNR dependence of the inferred gradient.
	
	\subsection{Relevance to current IFU surveys and their interpretations}\label{sec:ifu}

	The vertical lines in the right panel of \autoref{fig:res_metgrad_DD0600} indicate the typical spatial resolution of some of the current IFU surveys. Galaxies within a stellar mass range of $9.8 \lesssim \log{(\mathrm{M}_{\star}/\mathrm{M}_{\bigodot})} \lesssim 10.2$, in both SAMI and MaNGA surveys, have a typical spatial resolution of $\sim$3 beams per scale radius. Such resolution would yield a metallicity gradient $\sim$20 - 30\% shallower than the intrinsic value, modulo metallicity diagnostics used. The magnitude of this offset is highly dependent on the characteristic size of the disk galaxy being probed. The inferred gradient worsens rapidly with change in spatial resolution at the coarse-resolution end. Hence, a galaxy with 2 {\kpc} scale-length probed with SAMI or CALIFA would accommodate only 2 beams across its scale-radius, thereby leading to metallicity gradients that are $\sim$50 - 60 \% shallower. The SDSS-IV ManGA survey \citep{Bundy:2015aa}, with its typical 1 - 2 {\kpc} resolution, resolves the disk scale-lengths with 2 - 3 beams - thus yielding 20 - 40 \% shallower gradients. The TYPHOON/PrISM survey (Seibert et al. in prep.) operates at $<$100 {\pc} spatial resolution, implying $>10$ beams per scale-length of typical disk galaxies. Our work demonstrates that at such fine resolution, the metallicity gradient measurement is limited by the intrinsic scatter in the metallicity diagnostic used, which depends on the intrinsic metallicity gradient of the galaxy. In other words, if the disk scale-length is resolved by $>$6 - 7 beams that would yield the most-accurate-possible metallicity gradient. IFU surveys of nearby galaxies such as the VENGA survey \citep[][$\sim$300 {\pc} resolution]{Blanc:2013aa} and the future MUSE/PHANGS survey (estimated $\sim$50 - 100 {\pc}) have enough spatial resolution to overcome the PSF smearing effects on metallicity gradients.

	The magnitude of the artificial `flattening' of the metallicity gradient could potentially depend on the morphology of the galaxy, which is linked to the stellar mass. We note that our models are based on a Milky Way type, spiral galaxy with a stellar mass of 10$^{10}$ {\msol}. We therefore caution the reader that our quantitative measures will be applicable to large IFU surveys such as SAMI and MaNGA only after the samples have been appropriately trimmed to match our simulated galaxy. We will investigate the effect of morphology in a future work.

	\section{Summary and Conclusions}\label{sec:sum}
	We produce synthetic IFU observations of face-on simulated galaxies in order to investigate how well observations of finite sensitivity and resolution are able to recover metallicity gradients in disc galaxies. Compared to previous investigations of this topic, our work features a significantly more realistic distribution of star formation within the galaxy, a substantially more realistic treatment of sources of noise, and a wider exploration of the parameter space of resolution and signal-to-noise ratio (SNR). Our primary conclusions are as follows.
	\begin{itemize}
		\item Inferred metallicity gradients are generally shallower than the true value, with the error worsening with coarser spatial resolution and higher noise levels. This is because metallicity is a non-linear function of the emission line ratios. A coarser PSF redistributes flux from the centre to the outskirts of the disk - an effect that is more pronounced for the emission lines that have a steeper intrinsic radial profile. This differential outcome of the smearing leads to the emission line ratios, and consequently the metallicity, being overestimated in the peripheral regions thereby flattening the metallicity gradient.
		\item At a given resolution, there is a ``best possible accuracy" for the inferred gradient, no matter how high the SNR. The exact value of the best possible accuracy depends on the resolution, the true gradient and the metallicity diagnostic used. As a rough rule of thumb, however, in a galaxy where the scale length is resolved by 4 telescope beams, the best possible accuracy is a $\sim$ 10--15\,\% error in metallicity gradient.
		\item Achieving the best possible performance at a given resolution requires careful treatment of errors. One must consistently propagate uncertainties all the way from measurement of the errors in individual IFU spaxels, through the steps of measuring line fluxes, converting these to metallicities, and finally fitting the overall gradient. Moreover, it is desirable to exclude pixels with low SNR from fits to the metallicity gradient, and we find that a cut-off at SNR = 5 yields gradient measurements very close to the best possible value. It is conceivable that a more accurate study can be done by performing forward modelling using a full Bayesian MCMC approach so that the weak SNR pixels no longer need to be discarded and we do not lose any data. We plan to investigate this in a future work. The principle motivation of the current work is to emulate the current practice in observational studies, most of which employ strong emission line diagnostics, to infer metallicity gradient for different spatial resolution. 
		\item With proper treatment of uncertainties, the dependence of the error in metallicity gradient on spatial resolution is relatively insensitive to the physical properties of {\htr} or galaxies (strength of stellar wind, gas fraction) or to the choice of metallicity diagnostic. These only affect the absolute error level, not its dependence on resolution.
	\end{itemize}	
	
	In the future we plan to conduct follow up by studies by using the same analysis framework to investigate the effect of the instrumental properties on other observables e.g. star formation maps, ionisation parameter, ISM pressure and other strong emission line ratio diagnostics.
	
	\section*{Acknowledgements}
	We thank Eric Pellegrini and Laura S\'{a}nchez-Menguiano for extensive discussions regarding the {\htr} modeling aspects. We are also thankful to Henry Poetrodjojo for his prompt help regarding TYPHOON and SAMI observations. Parts of this research were conducted by the Australian Research Council Centre of Excellence for All Sky Astrophysics in 3 Dimensions (ASTRO 3D), through project number CE170100013. MRK acknowledges support the Australian Research Council's Discovery Projects funding scheme (grant DP160100695) and from an ARC Future Fellowship (FT180100375). L.J.K. gratefully acknowledges the support of an ARC
	Laureate Fellowship (FL150100113). C.~F.~acknowledges funding provided by the Australian Research Council (Discovery Projects DP150104329 and DP170100603, and Future Fellowship FT180100495), and the Australia-Germany Joint Research Cooperation Scheme (UA-DAAD). We further acknowledge high-performance computing resources provided by the Leibniz Rechenzentrum and the Gauss Centre for Supercomputing (grants~pr32lo, pr48pi and GCS Large-scale project~10391), the Partnership for Advanced Computing in Europe (PRACE grant pr89mu), the Australian National Computational Infrastructure (grants jh2 and ek9), and the Pawsey Supercomputing Centre with funding from the Australian Government and the Government of Western Australia, in the framework of the National Computational Merit Allocation Scheme and the ANU Allocation Scheme. This research has made use of the following: NASA's Astrophysics Data System Bibliographic Services, yt -- an open source analysis tool \citep{Turk:2011aa}, Astropy -- a community-developed core Python package for Astronomy \citep{astropy}, scipy \citep{scipy}, numpy \citep{numpy} and pandas \citep{pandas} packages.
	
	

	\appendix
	\section{Full parameter study}\label{sec:ap}
	In this section we present our analyses for the full parameter space, and for different values $\Omega$, gas fraction ($f_g$) and different diagnostics. \autoref{tab:quant} gives an overview of the different choices we investigated and the best-fit parameters of the error in inferred gradient as a function of the spatial resolution, for each case. The purpose of this section is to demonstrate the robustness of our key results to the different choices of the above parameters. We now discuss the dependence of our result on each of these choices individually. The tables and figures associated with this discussion are available as supplementary online material (SOM). 
	
	\subsection{Dependence on gas fractions}\label{sec:compgf}
	We find no qualitative difference in our results based on the gas fraction of the simulation used, on comparing \autoref{fig:res_metgrad_DD0600} and Figure S1 of the SOM. The difference in gas fraction yields an overall offset of the data. By virtue of a higher gas content, the fiducial gas fraction ($f_g$=0.2) simulation had formed more star particles and consequently more {\htr}s than the low gas fraction ($f_g$=0.1) simulation. This implies the former case has a higher level of overall signal from the emission lines. Thus one would require a longer exposure time to reach the same SNR when observing the $f_g = 0.1$ galaxy than the fiducial, $f_g = 0.2$ case. However, when we compare results at fixed SNR, the value of $f_g$ does not change the qualitative result for the exact error level or the number of beams per scale radius required to achieve maximum accuracy.
		
	\subsection{Dependence on the wind parameter}\label{sec:omega_ap}
	Figures~S3-S6 and Tables~S3-S6 in the SOM denote the resolution-dependent error in metallicity, heatmaps and tabulated values corresponding to Figures \ref{fig:res_metgrad_DD0600} and  \autoref{fig:res_heatmaps_DD0600} in the main text. We have discussed the impact of $\Omega$ in \autoref{sec:omega}.
	
	\section{Merging H~\textsc{ii} regions}\label{sec:mergeHII}
	
	In the main text, we calculate the line emission from {\htrs} in close proximity by merging star particles in every 40 {\pc}$^3$ cell by summing their luminosities and masses, implying that multiple star particles in close proximity would be responsible for driving one large {\htr}. We show that this produces a cluster mass function (CMF) in good agreement with observations. As an alternative approach, in this Appendix we do not merge {\htrs} at all; that is, we treat each $\approx 300$ $M_\odot$ star particle as driving a single {\htr} at its position, regardless of its proximity to any other star particles. The CMF for this case is close to a $\delta$ function at 300 $M_\odot$. We then compute the emission line fluxes and apply the remainder of our processing pipeline as described in the main text. We show the resulting resolution-dependent errors in the metallicity gradient in Figure S10 of the SOM. This may be compared with \autoref{fig:res_metgrad_DD0600}, the corresponding figure for our fiducial treatment. The results are nearly identical, agreeing to $\approx 1\%$ at all resolutions and in call cases. This demonstrates that our treatment of merging has no significant effect on the resolution-dependence of the error that we infer.

	\bibliographystyle{mn2e}
	\bibliography{paper}
	
	\bsp	
	\label{lastpage}

\end{document}